\documentclass[
reprint, 
twocolumns,
superscriptaddress,
nofootinbib,
]{revtex4-2}

\usepackage{amsmath}
\usepackage{amssymb}
\usepackage{booktabs}
\usepackage[dvipsnames]{xcolor}
\usepackage{textcomp}
\usepackage{graphicx}
\graphicspath{ {figures/} } 
\usepackage{dcolumn}
\usepackage{bm}
\usepackage[normalem]{ulem}
\usepackage{dcolumn}
\usepackage{booktabs}
\usepackage{setspace}
\usepackage{tikz}
\usepackage[english]{babel}
\usepackage{natbib}

\usepackage{hyperref}
\hypersetup{
    colorlinks=true,
    linkcolor=blue,
    filecolor=magenta,      
    urlcolor=cyan,
}

\newcommand{\beginsupplement}{%
        \setcounter{table}{0}
        \renewcommand{\thetable}{S\arabic{table}}%
        \setcounter{figure}{0}
        \renewcommand{\thefigure}{S\arabic{figure}}%
}

\newcommand{\pbs}{Department of Psychological and Brain Sciences, Indiana University, Bloomington, IN 47405}
\newcommand{\pns}{Program in Neuroscience, Indiana University, Bloomington, IN 47405}
\newcommand{\sice}{School of Informatics, Computing \& Engineering, Indiana University, Bloomington, IN 47405}

\newcommand{\prob}{\mathcal{P}}
\newcommand{\entropy}{\mathcal{H}}
\newcommand{\mi}{\mathcal{I}}
\newcommand{\pentropy}{\mathcal{H}_{\partial}}
\newcommand{\tc}{\mathcal{T}}
\newcommand{\dtc}{\mathcal{D}}
\newcommand{\oinfo}{\mathcal{O}}
\newcommand{\exx}{\mathcal{E}}

\begin{document}

\author{Thomas F. Varley}
\email{tvarley@indiana.edu}
\affiliation{\pbs} 
\affiliation{\sice}

\author{Maria Pope}
\affiliation{\pns}
\affiliation{\sice}

\author{Maria Grazia Puxeddu}
\affiliation{\pbs}

\author{Joshua Faskowitz} 
\affiliation{\pbs} 

\author{Olaf Sporns}
\affiliation{\pbs} 

\title{Partial entropy decomposition reveals higher-order structures in human brain activity}

\date{\today}

\begin{abstract}

    The standard approach to modeling the human brain as a complex system is with a network, where the basic unit of interaction is a pairwise link between two brain regions. While powerful, this approach is limited by the inability to assess higher-order interactions involving three or more elements directly. In this work, we present a method for capturing higher-order dependencies in discrete data based on partial entropy decomposition (PED). Our approach decomposes the joint entropy of the whole system into a set of strictly non-negative partial entropy atoms that describe the redundant, unique, and synergistic interactions that compose the system's structure. We begin by showing how the PED can provide insights into the mathematical structure of both the FC network itself, as well as established measures of higher-order dependency such as the O-information. When applied to resting state fMRI data, we find robust evidence of higher-order synergies that are largely invisible to standard functional connectivity analyses. This synergistic structure is symmetrical across hemispheres, largely conserved across individual subjects, and is distinct from structural features based on redundancy that have previously dominated FC analyses. Our approach can also be localized in time, allowing a frame-by-frame analysis of how the distributions of redundancies and synergies change over the course of a recording. We find that different ensembles of regions can transiently change from being redundancy-dominated to synergy-dominated, and that the temporal pattern is structured in time. These results provide strong evidence that there exists a large space of unexplored structures in human brain data that have been largely missed by a focus on bivariate network connectivity models. This synergistic ``shadow structures" is dynamic in time and, likely will illuminate new and interesting links between brain and behavior. Beyond brain-specific application, the PED provides a very general approach for understanding higher-order structures in a variety of complex systems. 
    
\end{abstract}

	\maketitle

Since the notion of the ``connectome" was first formalized in neuroscience \cite{sporns_human_2005}, network models of the nervous system have become ubiquitous in the field \cite{sporns_networks_2010,fornito_fundamentals_2016}. In a network model, elements of a complex system (typically neurons or brain regions) are modelled as a graph composed of vertices (or nodes) connected by edges, which denote some kind of connectivity or statistical dependency between them. Arguably the most ubiquitous application of network models to the brain is the ``functional connectivity" (FC) framework \cite{friston_functional_1994,fox_human_2005,fornito_fundamentals_2016}. In whole-brain neuroimaging, FC networks generally define connections as correlations between the associated regional time series (e.g. fMRI BOLD signals, EEG waves, etc). The correlation matrix is then cast as the adjacency matrix of a weighted network, on which a wide number of network measures can be computed \cite{rubinov_complex_2010}.

Despite the widespread adoption of functional connectivity analyses, there remains a little-discussed, but profound limitation inherent to the entire methodology: the only statistical dependencies directly visible to  pairwise correlation are bivariate, and in the most commonly performed network analyses, every edge between pairs $X_i$ and $X_j$ is treated as independent of any other edge. There are no \textit{direct} ways to infer statistical dependencies between three or more variables. ``Higher order" interactions are constructed by aggregating bivariate couplings in analyses such as motifs \cite{sporns_motifs_2004} or community detection \cite{fortunato_community_2010}. One of the largest issues holding back the direct study of higher-order interactions has been the lack of effective, accessible mathematical tools with which such interactions can be recognized \cite{battiston_physics_2021}. Recently, however, work in the field of multivariate information theory has enabled the development of a plethora of different measures and frameworks for capturing statistical dependencies beyond the pairwise correlation \cite{rosas_disentangling_2022}.

The few applications of these techniques to brain data have suggested that higher-order dependencies can encode meaningful bio-markers (such as discriminating between health and pathological states induced by anesthesia or brain injury \cite{luppi_synergistic_2020}) and reflect changes associated with age \cite{gatica_high-order_2021}. Since the space of possible higher-order structures is so much vaster than the space of pairwise dependencies, the development of tools that make these structures accessible opens the doors to a large number of possible studies linking brain activity to cognition and behavior. 

Of the tools that have been applied, one of the most well developed is the \textit{partial information decomposition} \cite{williams_nonnegative_2010,gutknecht_bits_2021} (PID), which reveals that multiple interacting variables can participate in a variety of distinct information-sharing relationships, including redundant, unique, and synergistic modes. Redundant and synergistic information sharing represent two distinct, but related ``types" of higher order interaction:.\textit{Redundancy} refers to information that is ``duplicated" over many elements, so that the same information could be learned by observing $X_1\lor X_2,\lor,\ldots,\lor X_N$. In contrast, \textit{synergy} refers to information that is \textit{only} accessible when considering the joint-states of multiple elements and no simpler combinations of sources. Synergistic information can only be learned by observing $X_1\land\ldots\land X_N$.

Redundant, and synergistic information sharing modes can be combined to create more complex relationships. For example, given three variables $X_1$, $X_2$, and $X_3$, information can be redundantly common to all three, which could be learned by observing $X_1 \lor X_2 \lor X_3$. We can also consider the information redundantly shared by joint states: for example, the information that could be learned by observing $X_1 \lor (X_2 \land X_3)$ (i.e. observing $X_1$ or the joint state of $X_2$ and $X_3$). For a finite set of interacting variables, it is possible to enumerate all possible information-sharing modes, and given a formal definition of ``redundancy", they can be calculated (for details see below).

The identification of redundancy and synergy as possible families of statistical dependence raises questions about how such relationships might be reflected (or missed) by the standard, pairwise correlation-based approach for inferring networks. We propose two criteria by which we might assess the performance of bivariate functional connectivity. The first we call \textit{specificity:} the degree to which a pairwise correlation between some $X_i$ and $X_j$ reports dependencies that are unique to $X_i$ and $X_j$ alone, and not shared with any other edges. In a sense, it reflects how appropriate the ubiquitous assumption that edges are independent is. The second criterion we call \textit{completeness:} whether all of the statistical dependencies present in a data set are accounted for and incorporated into the model, or if predictive structure is ``lost" when restrictive analyses are used. 

We hypothesized that classical functional connectivity would prove to be both non-specific (due to the presence of multivariate redundancies that get repeatedly ``seen" by many pairwise correlations) and incomplete (due to the presence of synergies). To test this hypothesis, we used the a framework derived from the PID: the \textit{partial entropy decomposition} \cite{ince_partial_2017} (PED, explained in detail below) to fully retrieve all components of statistical dependencies in sets of three and four brain regions. As part of this analysis, we propose a measure of redundant entropy applicable to arbitrarily-sized collections, which allows us to fully explore the space of higher order interactions. 

We chose the PED over the PID because the PID requires partitioning the system into predictors and ``targets" (the elements whose behavior we are predicting). This distinction is often artificial, and makes it difficult to analyze the system itself as a structured whole. The PED does not require making a source/target distinction, and serves to generalize the PID to the analysis of whole systems. 

By computing the full PED for \textit{all} triads of 200 brain regions, and a subset of approximately two million tetrads, we can provide a rich and detailed picture of beyond-pairwise dependencies in the brain. Furthermore, by separately considering redundancy and synergy instead of assessing just which one is dominant (as is commonly done \cite{gatica_high-order_2021,varley_multivariate_2022}), we can reveal previously unseen structures in resting state brain activity.

\section{Theory}
\label{sec:theory}
\subsubsection*{A Note on Notation}

In this paper, we will be making reference to multiple different ``kinds" of random variables. In general, we will use uppercase italics to refer to single variables (e.g. $X$). Sets of multiple variables will be denoted in boldface (e.g. $\textbf{X}=\{X_1,\ldots,X_N\}$, with subscript indexing). Specific instances of a variable will be denoted with lower case font: $X=x$. Functions (such as the probability, entropy, and mutual information), will be denoted using caligraphic font. Finally, we will make a distinction between expected values of information-theoretic quantities using upper case function notation (e.g. the Shannon entropy of $X$ is $\entropy(X)$, while the local entropy/surprisal is $h(x)$). For a brief review of local information theory, see the Supplementary Material Section S2. Finally, when referring to the partial entropy function $\pentropy$ (described below), we will use superscript index notation to indicate the full set of variables that contextualizes the individual atom. For example, $\pentropy^{123}(\{1\}\{2\})$ refers to the information redundantly shared by $X_1$ and $X_2$, when both are considered as part of the triad $\textbf{X}=\{X_1,X_2,X_3\}$, while $\pentropy^{12}(\{1\}\{2\})$ refers to the information redundantly shared by $X_1$ and $X_2$ qua themselves.

\subsection{Partial Entropy Decomposition}

The \textit{partial entropy decomposition} (PED) provides a framework with which we can extract \textit{all} of the meaningful ``structure" in a system of interacting random variables \cite{ince_partial_2017}. By ``structure", we are referring to the (possibly higher-order) patterns of information-sharing between elements. Consider a system $\textbf{X} = \{X_1, X_2,\ldots,X_N\}$, comprised of $N$ interacting, discrete random variables: the set of all informative relationships between elements (and ensembles of elements) in \textbf{X} forms its ``structure." We begin by defining the total entropy of \textbf{X} using the Shannon entropy:

\begin{equation}
    \entropy(\textbf{X}) := -\sum_{\textbf{x}\in\bm{\mathfrak{X}}}\prob(\textbf{x})\log_2 \prob(\textbf{x})
\end{equation}

Where \textbf{x} indicates a particular configuration of \textbf{X} and $\bm{\mathfrak{X}}$ is the support set of \textbf{X}. This joint entropy quantifies, on average, how much it is possible to ``know" about \textbf{X} (i.e. how many bits of information would be required, on average, to reduce our uncertainty to zero). The entropy is a summary statistic describing an entire distribution $\prob(X)$:

\begin{equation}
	\entropy(\textbf{X}) = \mathbb{E}[-\log_2 \prob(\textbf{x})]
\end{equation}

Where $-\log_2 \prob(x)$ is the \textit{local entropy} $h(\textbf{x})$. We can intuitively understand the local entropy with the logic of local probability mass exclusions \cite{finn_probability_2018,finn_pointwise_2018}. Suppose that we observe $\textbf{X}=\textbf{x}$. Upon observing \textbf{x}, we can immediately \textit{rule out} the possibility that \textbf{X} is in any state $\neg\textbf{x}$, and by ruling out those possibilities, we exclude all the probability mass associated with $\prob(\textbf{X}=\neg \textbf{x})$. If $\prob(\textbf{x})$ is very low, then upon learning $\textbf{X}=\textbf{x}$, we exclude a large amount of probability mass ($1-\prob(\textbf{x})$), and consequently, $h(\textbf{x})$ is high. Conversely, if $\prob(\textbf{x})$ is large, then only a small amount of probability mass is excluded, and so $h(\textbf{x})$ is low.

\subsubsection{Quantifying Shared Entropy}

The measure $h(\textbf{x})$ is a very crude one: it gives us a single summary statistic that describes the behaviour of the ``whole" without making reference to the structure of the relationships between \textbf{x}'s constituent elements. If $\textbf{X}$ has some non-trivial structure that integrates multiple elements (or ensembles of elements), then we propose that those elements must ``share" entropy. This notion of shared entropy forms the cornerstone of the PED. The way all of the parts of \textbf{X} share entropy forms the ``structure" of the system. In the original proposal of the PED by Ince \cite{ince_partial_2017}, shared entropy ($\entropy_{cs}$) was defined using the local co-information, which treats the entropy of variables as sets and defines the shared entropy using inclusion-exclusion criteria. Unfortunately, as discussed by Finn and Lizier, the set-theoretic interpretation of mutlivariate mutual information is complex, as both the expected and local co-information can be negative \cite{finn_generalised_2020}, and the PED computed using Ince's proposed method can result in negative values that are difficult to interpret.

Here, we propose an alternative way to operationalize the notion of ``redundant entropy" by saying that two variables $X_1,X_2\in\textbf{X}$ share entropy if they induce the same exclusions: i.e. if learning $X_1$ or $X_2$ rules out the same configurations of the whole \cite{finn_probability_2018}. Our goal, then, becomes to determine how the entropy of the whole is parcellated out over (potentially multivariate) sharing modes between parts.

\begin{table}[!ht]
	\begin{tabular}{@{}cccc@{}}
		\toprule
		$P$ &  & \textbf{$X_1$} & \textbf{$X_2$} \\ \midrule
		$P_{00}$                &  & 0              & 0              \\
		$P_{01}$                &  & 0              & 1              \\
		$P_{10}$                &  & 1              & 0              \\
		$P_{11}$                &  & 1              & 1              \\ \bottomrule
	\end{tabular}
	\caption{\textbf{Joint entropy of two discrete random variables that together make up the macro-variable \textbf{X}.}}
	\label{tab:biv}
\end{table}

In our toy system given by Table \ref{tab:biv}, suppose we learn that $X_1 = 0$ OR $X_2 = 0$. Only one global state is excluded: $\textbf{X} = (1,1)$ is incompatible with both possibilities, regardless of which is true. Consequently we are only excluding $P_{11}$ from the overall distribution. We can quantify this ``shared entropy" using the \textit{local entropy of shared exclusions} $h_{sx}$:

\begin{equation}
    h_{sx}^{\textbf{x}}(\{1\}\{2\}) = -\log_2 \prob(x_1 \cup x_2)
\end{equation}

Here, we are adapting the partial entropy notation first introduced by Ince in \cite{ince_measuring_2017}. The function $h_{sx}^{\textbf{x}}(\{1\}\{2\})$ quantifies the total probability mass of $\prob(\textbf{X})$ excluded by learning either $X_1=x_1$ or $X_2=x_2$. Said differently, it is the amount of information that could be learned from either variable alone. Importantly, while it \textit{is} a measure of dependency, it is distinct from the classic mutual information.

We term this function $h_{sx}$ to indicate that it is the shared entropy based on common exclusions (``entropy of shared exclusions") from some set of sources. We also note that the form of $h_{sx}$ is equivalent to the informative part of the local redundancy function derived by Makkeh et al., \cite{makkeh_introducing_2021}, which they term $i_{sx}$. For a discussion of how $h_{sx}$ is related to $i_{sx}$ and the deeper connections between partial \textit{entropy} decomposition and partial \textit{information} decomposition, see Appendix 1. 

So far, we have restricted our examples to the simple case of two variables, $x_1$ and $x_2$, however, we are interested in the general case of information common to arbitrarily large, potentially overlapping subsets of a system that has adopted a particular state \textbf{x}. This requires first enumerating the set of subsets, \textbf{s}, which we will call the set of \textit{sources}. It is equivalent to the power set of $\textbf{x}$, excluding the empty set. For example, if $\textbf{x} = \{x_1, x_2, x_3\}$, then the source set \textbf{s} is equal to:

\begin{equation}
    \textbf{s} = 
	\left\{\!\begin{aligned}
	&\{x_1\}, \{x_2\}, \{x_3\},\\
	&\{x_1,x_2\}, \{x_1,x_3\}, \{x_2,x_3\}, \\
	&\{x_1,x_2,x_3\}
	\end{aligned}\right\}
\end{equation}

We are interested in how collections of sources $\textbf{a} \in \textbf{s}$ might share entropy (i.e. to what extent the exclude the same possible global configurations of \textbf{x}), which allows us to write our redundant entropy function in full generality. For a collection of sources $\{\textbf{a}_1,\ldots,\textbf{a}_k\}$:

\begin{equation}
    h_{sx}(\textbf{a}_1,\ldots,\textbf{a}_k) := \log_2\frac{1}{\prob(\textbf{a}_1\cup\ldots\cup\textbf{a}_k)}
\end{equation}

$h_{sx}$ can be interpreted in terms of logical conjunctions and dysjunctions of variables \cite{gutknecht_bits_2021}. Consider the example: $h_{sx}(\{x_1\}\{x_2,x_3\})$, which quantifies the amount of probability mass about the state of the ``whole" that would be excluded by observing just the part $x_1$ \textbf{or} the joint state of $x_2$ \textbf{and} $x_3$. This relationship between probability mass exclusions on one hand, and formal logic on the other, places $h_{sx}$ on a sound conceptual footing. While initially defined locally, it is possible to compute an expected value $\mathcal{H}_{sx}$ for a joint distribution:

\begin{equation}
    \mathcal{H}_{sx}(\textbf{A}_1,\ldots,\textbf{A}_k) := \mathbb{E}[h_{sx}(\textbf{a}_1,\ldots,\textbf{a}_k)]
\end{equation}

\subsubsection{The Partial Entropy Lattice}

Our function $h_{sx}$ has a number of appealing mathematical properties, which collectively satisfy the set of Axioms initially introduced by Williams \& Beer for the problem of information decomposition \cite{williams_nonnegative_2010} as applied to local information \cite{finn_pointwise_2018,makkeh_introducing_2021}:

\begin{itemize}
    \item[] \textbf{Symmetry}: $h_{sx}$ is invariant under permutation of it's argument: $h_{sx}(\textbf{a}_1,\ldots,\textbf{a}_k)$ = $h_{sx}(\sigma(\textbf{a}_1),\ldots,\sigma(\textbf{a}_k))$
    \item[] \textbf{Monotonicity}: $h_{sx}$ decreases as more sources are added: $h_{sx}(\textbf{a}_1,\ldots,\textbf{a}_k) \leq h_{sx}(\textbf{a}_1,\ldots,\textbf{a}_k, \textbf{a}_{k+1})$
    \item[] \textbf{Self-redundancy}: In the special case of a single source, $h_{sx}$ is equivalent to the classic local Shannon entropy: $h_{sx}(\textbf{a}) = h(\textbf{a})$.
\end{itemize}

For proof of these, see \cite{makkeh_introducing_2021} Appendix A. 
Based on these properties, it is possible to specify the domain of $h_{sx}$ (all non-degenerate combinations of sources) in terms of a partially-ordered lattice structure $\mathfrak{A}$ \cite{williams_nonnegative_2010,finn_pointwise_2018}. Not every combination of sources $\textbf{a}_1\ldots\textbf{a}_k$ is a valid partial entropy atom, only those where no source is a subset of any other:

\begin{equation}
	\mathfrak{A} = \{ \boldsymbol{\alpha} \in \mathbb{P}_1(\textbf{s}) : \forall \textbf{a}_i, \textbf{a}_j \in \boldsymbol{\alpha} , \textbf{a}_i \not \subset \textbf{a}_j \}
\end{equation}

\begin{figure*}
    \centering
    \includegraphics[scale=0.6]{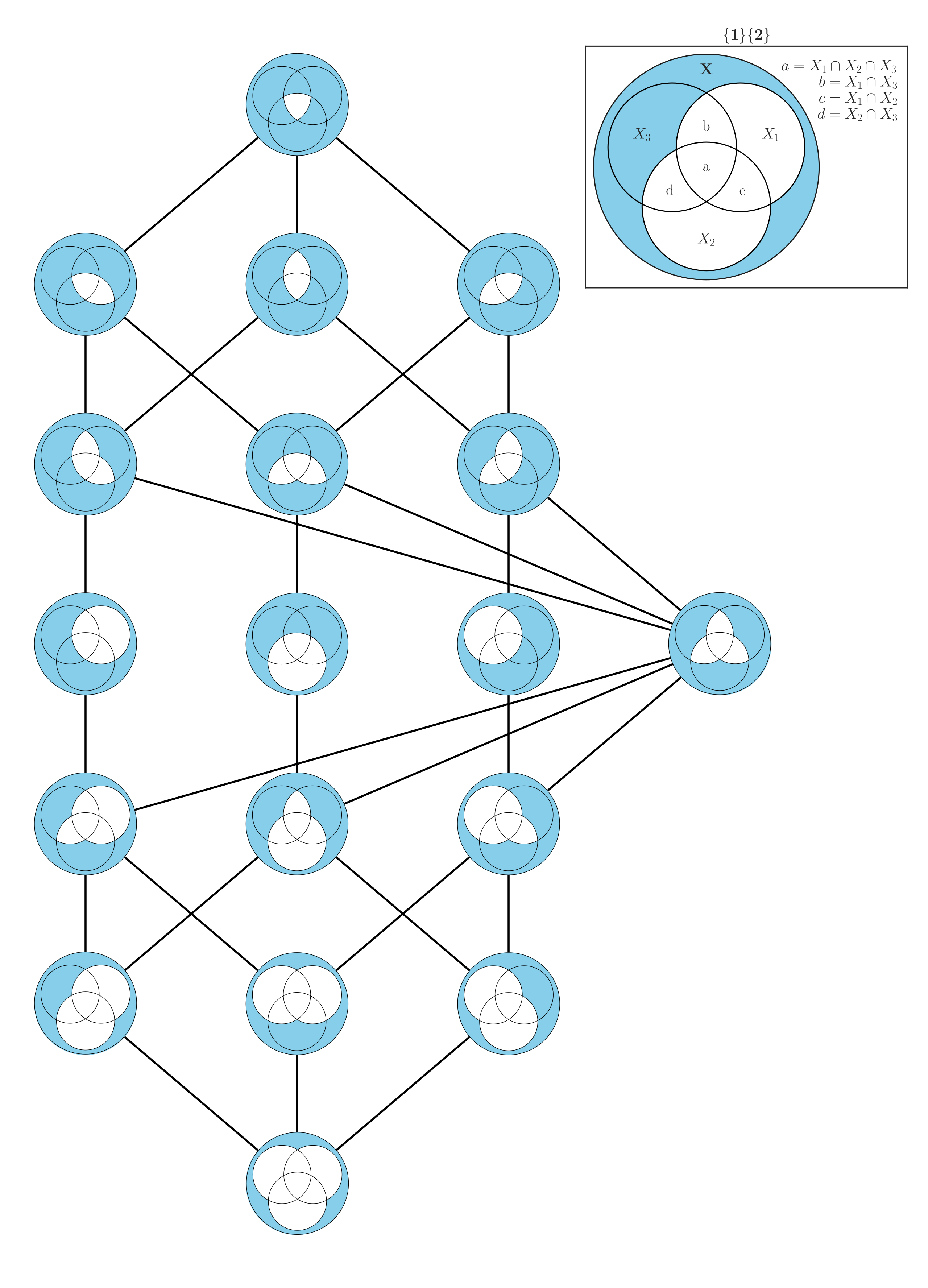}
    \caption{\textbf{The partial entropy lattice.} The lattice of partial entropy atoms induced by the $\mathcal{H}_{sx}$ function. Each vertex of the lattice corresponds to a single PE atom, and the Venn diagram describes the associated structure of probability mass exclusions. The blue area indicates the probability mass from $P(\textbf{x})$ that is excluded by some combination of observations. For example, in the legend, we can see the probability mass excluded by observing $X_1\lor X_2$. The blue area is all of the probability mass one would exclude after learning the state of \textit{either} component alone. The lowest atom is the entropy redundant to all three elements ($\mathcal{H}_{sx}(\{1\}\{2\}\{3\})$), and the dependencies get increasingly synergistic higher on the lattice. }
    \label{fig:lattice}
\end{figure*}

Where $\mathbb{P}_1(\textbf{s})$ indicates the power set of \textbf{s}, excluding the empty set. For an in-depth derivation of the lattice, see \cite{williams_nonnegative_2010,finn_pointwise_2018,gutknecht_bits_2021}, for a visualization of the lattice, see Fig. \ref{fig:lattice}. The value of any element $h_\partial(\bm{\alpha})$ on the lattice can be computed via Mobius inversion:

\begin{equation}
    h_\partial^{\textbf{x}}(\bm{\alpha}) = h_{sx}(\bm{\alpha}) - \sum_{\bm{\beta}\preceq\bm{\alpha}}h_\partial^\textbf{x}(\bm{\beta})
\end{equation}

The result is the entropy specific to a particular $\bm{\alpha}$ \textit{and no simpler combination of sources.} Furthermore, the structure of the lattice and the properties of $h_{sx}$ ensure that $h_\partial^{\textbf{x}}(\bm{\alpha})$ will always be non-negative. We can re-compute the total joint entropy of $\textbf{x}$ as:

\begin{equation}
    h(\textbf{x}) = \sum_{i=1}^{|\mathfrak{A}|}h_\partial^{\textbf{x}}(\bm{\alpha}_i)
    \label{eq:final_sum}
\end{equation}

Like $h_{sx}$, it is also possible to compute an expected value of $h_\partial$ (which will also be strictly non-negative):

\begin{equation}
    \pentropy^{\textbf{X}}(\bm{\alpha}) = \mathbb{E}[h_\partial^\textbf{x}(\bm{\alpha})]
\end{equation}

\subsubsection{Decomposing Marginal and Joint Entropies}

Having defined $h_{sx}$ and the Mobius inversion on the partial entropy lattice, we can now do a complete decomposition of the joint entropy, and its marginal components. For example, consider the bivariate system $\textbf{X}=\{X_1,X_2\}$. We can decompose the joint entropy:

\begin{align}
    \entropy(\textbf{X)} &= \pentropy^{12}(\{1\}\{2\}) + \pentropy^{12}(\{1\}) \\ 
    &+ \pentropy^{12}(\{2\}) + \pentropy^{12}(\{1,2\}) \nonumber
\end{align}

Furthermore, we can decompose the associated marginal entropies in a manner consistent with the partial information decomposition \cite{williams_nonnegative_2010}:

\begin{align}
    \entropy(X_1) = \pentropy^{12}(\{1\}\{2\}) + \pentropy^{12}(\{1\}) \\
    \entropy(X_2) = \pentropy^{12}(\{1\}\{2\}) + \pentropy^{12}(\{2\}) \nonumber
\end{align}

These decompositions can be done for larger ensembles, or more statistical dependencies (see below) and can reveal how higher-order interactions can complicate (and in some cases, compromise) the standard bivariate approaches to functional connectivity.

\subsubsection{Mathematical Analysis of the PED}
\label{sec:res_math}

The partial entropy decomposition reveals a rich and complex structure of statistical dependencies even in small systems. Before considering the empirical results, it is worth discussing how the PED relates to classic measures from information theory and what it reveals about the limitations of bivariate FC measures. 

The first key finding is that the PED provides interesting insights into the nature of bivariate mutual information. Typically, mutual information is conflated with redundancy at the outset (for example, in Venn diagrams), however, when considering the PED of two variables $X_1$ and $X_2$, it becomes clear that:

\begin{equation}
    \mi(X_1;X_2) = \pentropy^{12}(\{1\}\{2\}) - \pentropy^{12}(\{1,2\})
    \label{eq:biv_from_bi}
\end{equation}

This relationship was originally noted by Ince \cite{ince_partial_2017} and later re-derived by Finn and Lizier \cite{finn_generalised_2020}. In a sense, the higher-order information present in the joint-state of ($X_1$ and $X_2$) ``obscures" the lower-order structure. This issue is also inherited by parametric correlation measures based on the Pearson correlation coefficient, since the mutual information is a deterministic function of Pearson's $\rho$ for Gaussian variables \cite{cover_elements_2012}.

When considering the decomposition of local mutual information into informative and misinformative components proposed by Finn and Lizier, it is clear that redundancy corresponds to the informative component of local mutual information, while synergy corresponds to the misinformative component.

We can do a similar analysis extracting the bivariate mutual information from the trivariate PED, which reveals that the bivariate correlation is not \textit{specific}:

\begin{align}
    \mi(X_1;X_2) &= 
    \pentropy^{123}(\{1\}\{2\}\{3\}) + \pentropy^{123}(\{1\}\{2\}) \\
    &- \pentropy^{123}(\{3\}\{1,2\}) - \pentropy^{123}(\{1,2\}\{1,3\}\{2,3\}) \nonumber \\ 
    &- \pentropy^{123}(\{1,2\}\{1,3\}) - \pentropy^{123}(\{1,2\}\{2,3\}) \nonumber \\
    &- \pentropy^{123}(\{1,2\})  \nonumber
    \label{eq:biv_from_tri}
\end{align}

It is clear from Eq. \ref{eq:biv_from_tri} that the \textit{bivariate} mutual information incorporates information that is \textit{triple}-redundant across three variables ($\pentropy^{123}(\{1\}\{2\}\{3\})$), and if one were to take the standard FC approach to a triad (pairwise correlation between all three pairs of elements), that the triple redundancy would be triple counted and erroneously ascribed to three separate interactions. Furthermore, not only does bivariate mutual information double-count redundancy, but it also penalizes higher-order synergies. Any higher-order atom that includes the joint state of $X_1 \land X_2$ counts \textit{against} $\mi(X_1;X_2)$. 

Having established that the presence of higher-order redundancies explicitly precludes bivariate correlation from being specific, we now ask: can we improve the specificity using common statistical methods? One approach aimed at ``controlling" for the context of additional variables in a bivariate correlation analysis is using conditioning or partial correlation. Typically, these analyses are assumed to improve the \textit{specificity} of a pairwise dependency by removing the influence of confounders, however, by decomposing the conditional mutual information between three variables, we can see that conditioning does \textit{not} ensure specificity:

\begin{align}
    \mi(X_1;X_2|X_3) &= \pentropy^{123}(\{1\}\{2\}) \\
    &+ \pentropy^{123}(\{1\}\{2,3\}) + \pentropy^{123}(\{2\}\{1,3\}) \nonumber \\ 
    &+ \pentropy^{123}(\{1,2\}\{1,3\}\{2,3\}) \nonumber \\ 
    &+ \pentropy^{123}(\{1,3\}\{2,3\}) \nonumber \\
    &- \pentropy^{123}(\{1,2\}) - \pentropy^{123}(\{1,2,3\}) \nonumber
    \label{eq:conditional}
\end{align}

The decomposition of $\mi(X_1;X_2|X_3)$ conflates the true pairwise redundancy ($\pentropy^{123}\{1\}\{2\}$) with the a higher-order redundancy involving the joint state of $X_1\land X_3$ and $X_2\land X_3$: $\pentropy^{123}\{1,3\}\{2,3\}$. Furthermore, the conditional mutual information penalizes synergistic entropy shared in the joint state of all three variables ($\pentropy^{123}\{1,2,3\}$). Consequently, we can conclude that the specificity of bivariate functional connectivity \textit{cannot} be salvaged using conditioning or partial correlation. Not only does controlling fail to provide specificity, it also actively compromises completeness, since it brings in higher-order interactions. Given that conditional mutual information and partial correlation are equivalent for Gaussian variables \cite{cliff_assessing_2021}, this issue also affects standard, parametric approaches to conditional connectivity, just as with bivariate mutual information/Pearson correlation.

It is important to understand that these analytic results are \textbf{not} a consequence of the particular form of $h_{sx}$: any shared entropy function that allows for the formation of a partial entropy lattice will produce these same results (many were first derived by Ince, who used a different measure based on the local co-information \cite{ince_partial_2017}).

\subsubsection{Higher-Order Dependency Measures}
\label{sec:res_higher-order}

In addition revealing the structure of commonly-used correlations (bivariate and partial correlations), the PED can also be used to develop intuitions about multivariate generalizations of the mutual information. Many of these generalizations exist, and here we will focus on four: the total correlation \cite{watanabe_information_1960}, the dual total correlation \cite{abdallah_measure_2012}, the O-information \cite{rosas_quantifying_2019,varley_multivariate_2022} (also called the ``enigmatic" information \cite{james_anatomy_2011}) and the S-information \cite{rosas_quantifying_2019} (also called the ``exogenous" information \cite{james_anatomy_2011}). While useful, these measures are often difficult to intuitively understand, and can display surprising behavior. Since they can all be written in terms of sums and differences of joint and marginal entropies, we can use the PED framework to more completely understand them. 

The oldest measure is the total correlation, defined as:

\begin{equation}
    \tc(\textbf{X}) := \sum_{i=1}^{|\textbf{X}|}\entropy(X_i) - \entropy(\textbf{X})
\end{equation}

which is equivalent to the Kullback-Leibler divergence between the true joint distribution $\prob(\textbf{X})$ and the product of the marginals:

\begin{equation}
    \label{eq:tc-dkl}
    \tc(\textbf{X}) = D_{KL}(\prob(\textbf{X}) || \prod_{i=1}^{|\textbf{X}|}\prob(X_i)
\end{equation}

Based on equation \ref{eq:tc-dkl}, we can understand the total correlation as the divergence from the maximum entropy distribution to the true distribution, implying that it might be something like a measure of the ``total" structure of the system (as it's name would suggest). We can decompose the 3-variable case to get a full picture of the structure of the TC:

\begin{align}
    \tc(X_1,X_2,X_3) &= (2\times\{1\}\{2\}\{3\}) \\
    &+\{1\}\{2\}+\{1\}\{3\}+\{2\}\{3\} \nonumber\\
    &-\{1,2\}\{1,3\}\{2,3\} \nonumber\\ 
    &-\{1,2\}\{1,3\}-\{1,2\}\{2,3\}-\{1,3\}\{2,3\} \nonumber\\
    &-\{1,2\}-\{1,3\}-\{2,3\} \nonumber\\ 
    &-\{1,2,3\} \nonumber
\end{align}

We can see that the total correlation is largely a measure of redundancy, sensitive to information shared between single elements, but penalizing higher-order information present in joint states. This can be understood by considering the lattice in Figure \ref{fig:lattice}: each of the $\entropy(X_i)$ terms will only incorporate atoms preceding (or equal to) the unique entropy term $\pentropy^{123}(i)$ - anything that can only be seen by considering the joint-state of \textbf{X} will be negative.

The second generalization of mutual information is the dual total correlation \cite{abdallah_measure_2012}. Defined in terms of entropies by:

\begin{equation}
    \dtc(\textbf{X}) := \entropy(\textbf{X}) - \sum_{i=1}^{|\textbf{X}|}\entropy(X_i | \textbf{X}^{-i})
\end{equation}

where $\textbf{X}^{-i}$ refers to the set of every element of \textbf{X} \textit{excluding} the $i^{th}$. The dual total correlation can be understood as the difference between the total entropy of \textbf{X} and all of the entropy in each element of $X$ that is ``intrinsic" to it and not shared with any other part. When we decompose the three-variable case, we find:

\begin{align}
\dtc(X_1, X_2, X_3) &= \{1\}\{2\}\{3\} \\
& + \{1)\{2\} + \{1\}\{3\} + \{2\}\{3\} \nonumber\\
& + \{1\}\{23\} + \{2\}\{1,3\} + \{3\}\{1,2\} \nonumber \\
&+ \{1,2\}\{1,3\}\{2,3\} \nonumber \\
&- \{1,2\} - \{1,3\} - \{2,3\} - (2\times\{1,2,3\}) \nonumber
\end{align}

This shows that dual total correlation is a much more ``complete" picture of the structure of a system than total correlation. It is sensitive to both shared redundancies and synergies, penalizing only the un-shared, higher-order synergy terms such as $\pentropy^{123}(\{1,2\})$.

The sum of the total correlation and the dual total correlation is the exogenous information \cite{james_anatomy_2011}, also called by the S-information. 

\begin{equation}
    \exx(\textbf{X}) := \tc(\textbf{X}) + \dtc(\textbf{X})
\end{equation}

Prior work has shown the exogenous entropy to be very tightly correlated with the Tononi-Sporns-Edelman complexity \cite{tononi_measure_1994, rosas_quantifying_2019,varley_multivariate_2022}, a measure of global integration/segregation balance. James also showed that the S-information quantified the total information that every element shares with every other element \cite{james_anatomy_2011}. We can see that:

\begin{align}
\exx(X_1, X_2, X_3) &= (3\times\{1\}\{2\}\{3\}) \nonumber \\
&+ 2\times(\{1\}\{2\} + \{1\}\{3\} + \{2\}\{3\})) \nonumber \\
&+ \{1\}\{2,3\} + \{2\}\{1,3\} + \{3\}\{1,2\}  \nonumber \\
&- \{1,2\}\{1,3\} - \{1,2\}\{2,3\} - \{1,3\}\{2,3\} \nonumber \\
& - 2\times(\{1,2\} + \{1,3\} + \{2,3\}) \nonumber \\
&-(3\times\{1,2,3\}) \nonumber
\end{align}

This reveals that S-information to be an unusual measure, in that it counts each redundancy term multiple times (i.e. in the case of three variables, the triple redundancy term appears three times, each double-redundancy term appears twice, etc), and penalizes them likewise when considering unshared synergies.

The final, and arguably most interesting measure is the difference between the total correlation and the dual total correlation is often referred to as the O-information \cite{rosas_quantifying_2019}, and has been hypothesized to give a heuristic measure of the extent to which a given system is dominated by redundant or synergistic interactions:

\begin{equation}
    \oinfo(\textbf{X}) := \tc(\textbf{X}) - \dtc(\textbf{X})
\end{equation}

where $\oinfo(\textbf{X}) > 0$ implies a redundancy-dominated structure and $\oinfo(\textbf{X}) < 0$ implies a synergy dominated one. PED analysis reveals:

\begin{align}
    \oinfo(X_1, X_2, X_3) &= \{1\}\{2\}\{3\} \\
    &- \{1\}\{2,3\} - \{2\}\{1,3\} - \{3\}\{1,2\} \nonumber \\
    &- (2\times\{1,2\}\{1,3\}\{2,3\}) \nonumber \\
    &- \{1,2\}\{1,3\} - \{2,3\}\{1,3\} - \{1,2\}\{2,3\} \nonumber \\
    &+ \{1,2,3\} \nonumber
\end{align}

This shows that the O-information generally satisfies the intuitions proposed by Rosas et al., as it is positively sensitive to the non-pairwise redundancy (in this case just $\pentropy^{123}(\{1\}\{2\}\{3\})$) and negatively sensitive to any higher-order shared information. Curiously, $\oinfo(X_1, X_2, X_3)$ positively counts the highest, un-shared synergy atom ($\pentropy^{123}(\{1,2,3\})$. Conceivably, it may be possible for a set of three variables with \textit{no redundancy} to return a positive O-information, although whether this can actually occur is an area of future research. 

For three-element systems, the O-information is also equivalent to the co-information \cite{rosas_quantifying_2019}, which forms the base of the original redundant entropy function $\entropy_{cs}$ proposed by Ince \cite{ince_partial_2017}. From this we can see that, at least for three variables, co-information is not a pure measure of redundancy, conflating the true redundancy and the highest synergy term, as well as penalizing other higher-order modes of information-sharing. A similar argument was made by Williams and Beer using the mutual information-based interpretation of co-information \cite{williams_nonnegative_2010}. While the O-information and co-information diverge for $N>3$, we anticipate that the behavior of the co-information will remain similarly complex at higher $N$.
These results reveal how the PED framework can provide clarity to the often-murky world of multivariate information theory.

\subsubsection{Novel Higher-Order Measures}

From these PED atoms, we can construct a novel measures of higher-order dependence that extends beyond TC, DTC, O-Information and S-Information. 

When considering higher-order redundancy, we are interested in all of those atoms that duplicate information over three or more individual elements. We define this as the \textit{redundant structure}. For a three element system:

\begin{equation}
    \mathcal{S}_{R}(X_1,X_2,X_3) = \{1\}\{2\}\{3\}
\end{equation}

For a four-element system:

\begin{align}
    \mathcal{S}_{R}(X_1,X_2,X_3,X_4) &= \{1\}\{2\}\{3\}\{4\} \\ 
    &+ \{1\}\{2\}\{3\} +  \{1\}\{2\}\{4\} \nonumber \\ &+ \{1\}\{3\}\{4\} + \{2\}\{3\}\{4\} \nonumber
\end{align}

And so on for larger systems. 

We can also define an analogous measure of synergistic structure: all those atoms representing information shared over the joint state of two or more elements. For example, for a three element system:

\begin{align}
    \mathcal{S}_{S}(X_1,X_2,X_3) &= \{1\}\{2,3\} + \{2\}\{1,3\} + \{3\}\{1,2\} \nonumber \\
    &+ \{1,2\}\{1,3\}\{2,3\} \nonumber \\
    &+ \{1,2\}\{1,3\} + \{2,3\}\{1,3\} \nonumber \\&+ \{1,2\}\{2,3\}
\end{align}

For three element systems, the difference $\mathcal{S}_{R} - \mathcal{S}_{S}$ is analagous to a ``corrected" O-information: the atom $\{1,2\}\{1,3\}\{2,3\}$ is only counted once and the confounding triple synergy $\{1,2,3\}$ is not included. 
Finally, we can define a measure of total (integrated) structure (i.e. all shared information) as the sum of all atoms composed of multiple sources:

\begin{align}
    \mathcal{S} = \sum_{\boldsymbol{\alpha} \in \boldsymbol{\mathfrak{A}}}\boldsymbol{\alpha} \iff |\boldsymbol{\alpha}| > 1
\end{align}

\subsection{Applications to the Brain}

The mathematical structure of the PED is domain agnostic: any complex system composed of discrete random variables is amenable to this kind of information-theoretic analysis. In this paper, we focus on data collected from the human brain with functional magnetic resonance imaging (fMRI). For detailed methods, see the Materials \& Methods section (\ref{sec:mnm}, but in brief, data from ninety five human subjects resting quietly was recorded as part of the Human Connectome Project \cite{van_essen_wu-minn_2013}. All of the scans were concatenated and each channel binarized about the mean \cite{sporns_dynamic_2021} to create multidimensional, binary time series. We then computed the full PED for all triads, and approximatley two million tetrads, to compare to the standard, bivariate functional connectivity network (computed with mutual information). 

By looking at the redundant and synergistic structures, and relating them to the standard FC, we can explore how higher-order dependencies are represented in bivariate networks, as well as what brain regions participate in more redundancy- or synergy-dominated ensembles. 

\section{Results}

\subsection{PED Reveals the Limitations of Bivariate Networks}

\begin{figure*}
    \centering
    \includegraphics[scale=0.5]{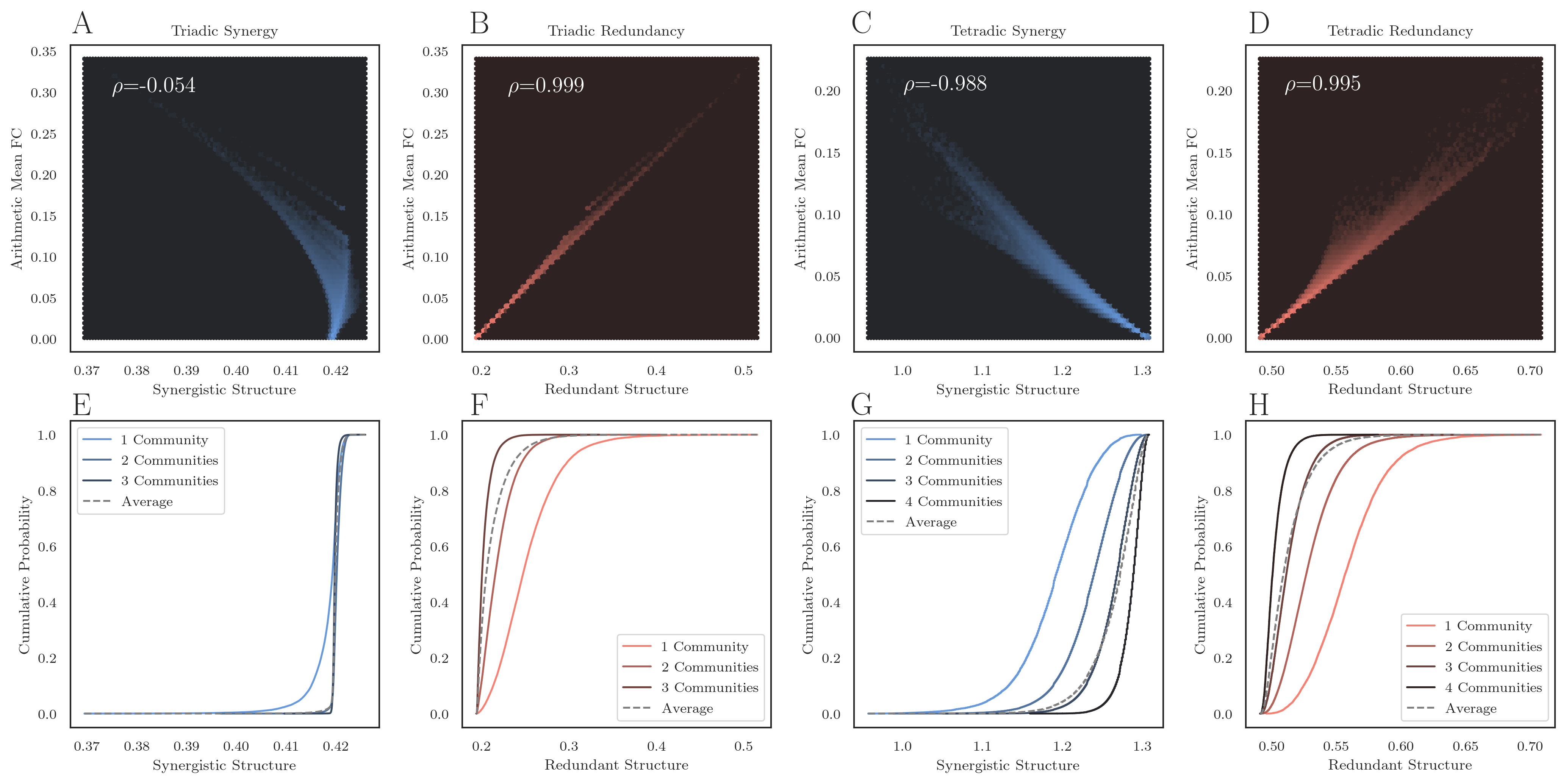}
    \caption{\textbf{The limits of bivariate functional connectivity. A.} In triads, bivariate functional connectivity is largely independent of synergistic structure, and \textbf{B,} is very positively correlated with redundant structure. \textbf{C.} In tetrads, bivariate functional connectivity is strongly negatively correlated with synergistic structure and \textbf{D,} is strongly correlated with redundant structure. \textbf{E-F.} Triads that have all elements within one FC community have significantly less synergistic structure than those that have elements with two communities, while for redundnat structure, there was a clear pattern that the more FC communities a triad straddled, the lower it's overall redundant structure. \textbf{G-H.} The same pattern was even more pronounced in tetrads: as the number of FC communities a tetrad straddled increased, the expected synergistic structure climbed, while expected redundant structure fell.}
    \label{fig:communities}
\end{figure*}

We now discuss how the PED relates multivariate measures of bivariate network structure commonly used in the functional connectivity literature. These measures describe statistical dependencies between ensembles of regions, but mediated by the topology of bivariate connections. We hypothesized that this emergence from bivariate dependencies would render them largely insensitive to synergies, which in turn would mean that such measures do not solve the issue of incompleteness in functional connectivity. 

Following \cite{onnela_intensity_2005}, we compared the redundant and synergistic structure of triads and tetrads to a measure of subgraph strength: the arithmetic mean of all edges in the subgraph. We found that the arithmetic mean FC density was positively correlated with redundancy for triads ($\rho=0.999$, $p<10^{-20}$) and tetrads ($\rho = 0.995$, $p<10^{-20}$), indicating that information duplicated over many brain regions contributes to multiple edges, leading to double-counting.  In contrast, for triads, arithmetic mean FC density was largely independent of synergistic structure ($\rho=-0.05$, $p<10^{-20}$), but for tetrads they were strongly anticorrelated ($\rho=-0.988$, $p<10^{-20}$).

In addition to subgraph structure, another common method of assessing polyadic interactions in networks is via community detection \cite{fortunato_community_2010}. Using the multi-resolution consensus clustering algorithm \cite{jeub_multiresolution_2018}, we clustered the bivariate functional connectivity matrix into non-overlapping communities. We then looked at the distributions of higher-order redundant and synergistic structure for triads and tetrads that spanned different numbers of consensus communities. We found that triads where all nodes were members of one community had significantly less synergy than triads that spanned two or three communities (Kolmogorov-Smirnov two sample test, $D=0.44$, $p<10^{-20}$). The pattern was more pronounced when considering tetrads: tetrads that all belonged to one community had lower synergy than those that spanned two communities ($D=0.45$, $p<10^{-20})$, who in turn had lower synergy than those that spanned three communities ($D=0.37$, $p<10^{-20}$). In Figure \ref{fig:communities} (top row), we show cumulative probability density plots for the distribution of synergies for triads and tetrads that spanned one, two, three, and four FC communities, where it is clear that participation in increasingly diverse communities is associated with greater synergistic structure. In contrast, redundant structure was higher in triads that were all members of a small number of communities. Triads that spanned three communities had lower redundancy than triads that spanned two communities ($D=0.48$, $p<10^{-20}$), which in turn had lower redundancy than those that were all members of one community ($D=0.47$, $p<10^{-20}$) (see Fig. \ref{fig:communities}, bottom row). 
These results, coupled with the mathematical analysis of the PED discussed in Section \ref{sec:theory} provide strong theoretical and empirical evidence that bivariate, correlation-based FC measures are largely sensitive to redundant information duplicated over many individual brain regions, but largely insensitive to (or even anti-correlated with) higher-order synergies involving the joint state of multiple regions. These results imply the possibility that there is a vast space of neural dynamics and structures that have not previously been captured in FC analyses.

\subsubsection{PED with $\mathcal{H}_{sx}$ is consistent with O-information}

To test whether the PED using the $\mathcal{H}_{sx}$ redundancy function was consistent with other, information-theoretic measures of redundancy and synergy, we compared the average redundant and synergistic structures (as revealed by PED), to the O-information. We hypothesized that redundant structure would be positively correlated with O-information (as $\mathcal{O} > 0$ implies redundancy dominance) and that synergistic structure would be negatively correlated, for the same reason.

For both triads and tetrads, our hypothesis was bourne out. The Pearson correlation between O-information and redundant structure was significantly positive for both triads ($\rho=0.72$, $p<10^{-20}$) and tetrads ($\rho=0.82$, $p<10^{-20}$). Conversely, the Pearson correlation between the O-information and the synergistic structure was significantly negative (triads: $\rho=-0.7$, $p<10^{-20}$, tetrads: $\rho=-0.72$, $p<10^{-20}$). These results show that the structures revealed by the PED are consistent with other, non-decomposition-based inference methods and serves to validate the overall framework. 

Interestingly, when comparing the triadic O-information to the corrected triadic O-information (which does not double-count $\pentropy^{123}(\{1,2\}\{1,3\}\{2,3\})$ and does not add back in the atom $\pentropy^{123}(\{1,2,3\})$), we can see that the addition of $\pentropy^{123}(\{1,2,3\})$ can lead to erroneous conclusions. Of all those triads that had a negative corrected O-information (i.e. had a greater synergistic structure than redundant structure), $61.7\%$ had a positive O-information, which could only be attributable to the presence of the triple-synergy being (mis)interpreted as redundancy and overwhelming the true difference. This suggests that, for small systems, the O-information may not provide an unbiased estimator of redundancy/synergy balance. 

\subsection{Characterizing Higher-Order Brain Structures}

Having established the presence of beyond-pairwise redundancies and synergies in brain data, and shown that standard, network-based approaches show an incomplete picture of the overall architecture, we now describe the distribution of redundancies and synergies across the human brain. 

\begin{figure*}
    \centering
    \includegraphics[scale=0.5]{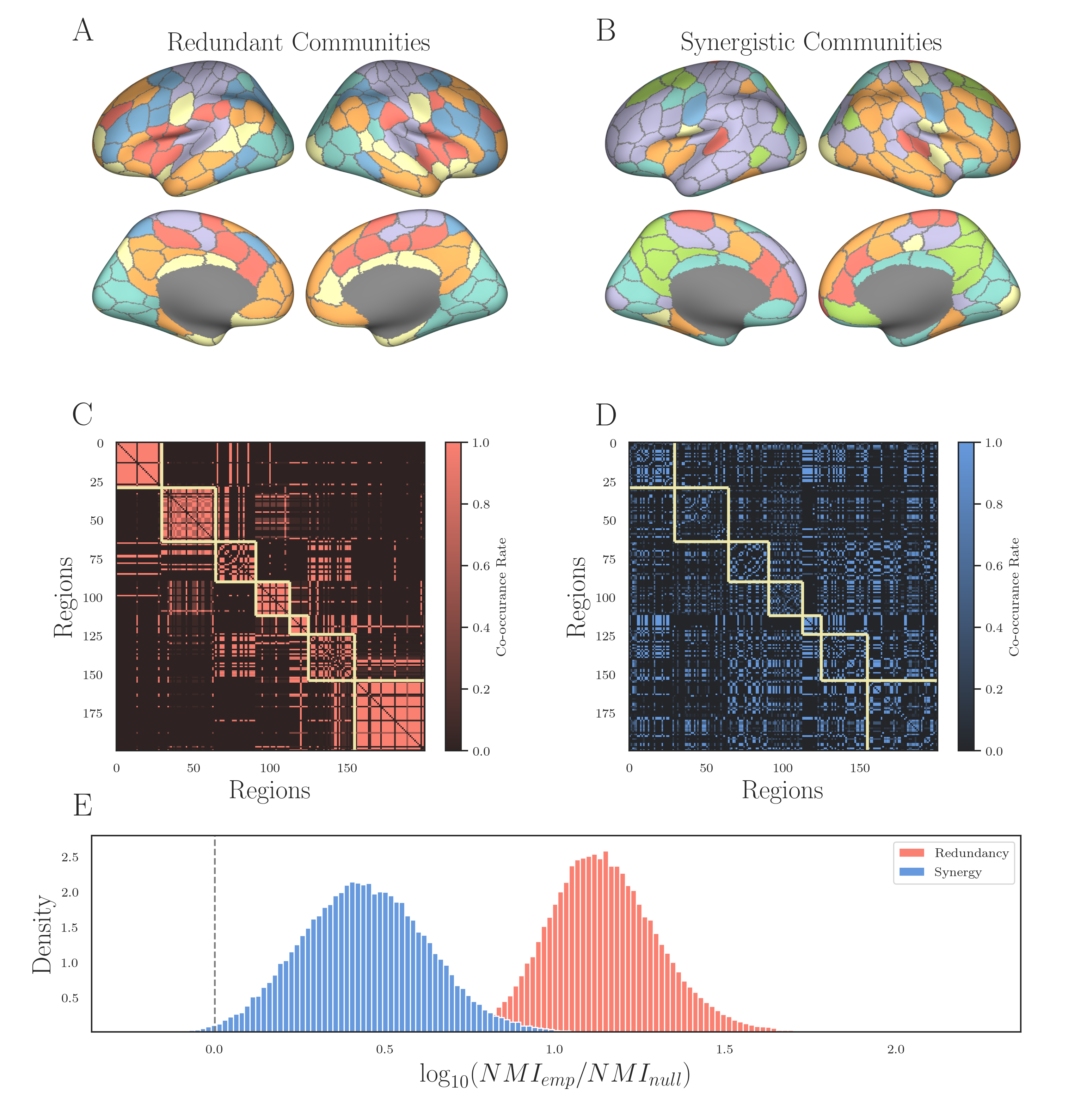}
    \caption{\textbf{Redundant and synergistic hypergraph community structure. A-B.} Surface plots of the two communities structures: on the left is the redundant structure and on the right is the synergistic structure. We can see that both patterns are largely symmetrical for both information-sharing modes, although the synergistic structure has two large, lateralized communities. \textbf{C-D.} The co-classification matrices for redundant structure (left) and the synergistic structure (right). The higher the value of a pair, the more frequently the hypergraph modularity maximization \cite{kumar_new_2020} assigns those two regions to the same hyper-community. The yellow squares indicate the seven canonical Yeo functional networks \cite{yeo_organization_2011}, and we can see that the higher-order redundant structure matches the bivariate Yeo systems well (despite consisting of information shared redundantly across three nodes). In contrast, the synergistic structure largely fails to match the canonical network structure at all. \textbf{E.} For each of the 95 subjects and for each of the 1000 permutation nulls used to significance test the NMI between subject-level community structure and the master level structure, we computed the log-ratio of the empirical NMI to the null NMI. For redundancy, there was not a single null, over any subject, that was greater than the associated empirical NMI. For the case of the synergy, only 0.6\% of nulls were greater than their associated empirical NMI.} 
    \label{fig:community_parc}
\end{figure*}

We began by applying a higher-order generalization of the standard community detection approach using a hypergraph modularity maximization algorithm \cite{kumar_new_2020}. this algorithm partitions collections of (potentially overlapping) sets of nodes called \textit{hyperedges} into communities that have a high degree of internal integration and a lower degree of between-community integration. We selected all those triads that had a greater synergistic structure than any of the one million maximum entropy null triads (see Materials and Methods), which yielded a set of 3,746 unique triads. From these, we constructed an unweighted hypergraph with 200 nodes and 3,746 hyperedges (casting each triad as a hyperedge incident on three nodes). We then performed 1,000 trials of the hypergraph clustering algorithm proposed by Kumar et al., \cite{kumar_new_2020}, from which we built a consensus matrix that tracked how frequently two brain regions $X_i$ and $X_j$ were assigned to the same hyper-community. We repeated the process for the 3,746 maximally redundant triads to create two partitions: a synergistic structure and a redundant structure. 

In Figure \ref{fig:community_parc} we show surface plots of the resulting communities computed from the concatenated time series comprising all ninety-five subjects and all 4 runs. The redundant structure (left) is very similar to the canonical seven Yeo systems \cite{yeo_organization_2011}: we can see a well-developed DMN (orange), a distinct visual system (sky blue), a somato-motor strip (violet), and a fronto-parietal network (dark blue). In contrast, when considering the synergistic structure (right), a strikingly different pattern is apparent. Synergistic connectivity appears more lateralized over left and right hemispheres (orange and violet communities respectively), although there is a high degree of symmetry along the cortical midline comprised of apparently novel communities. These include a synergistic coupling between visual and limbic regions (sky blue), as well a occipital subset of the DMN (green) and a curious, symmetrical set of regions combining somato-motor and DMN regions (red). 

These results show two things: the first is further confirmation that the canonical structures studied in an FC framework can be interpreted as reflecting primarily patterns of redundant information. The second is that higher-order synergies are structured in non-random ways, combining multiple brain regions into integrated systems that are usually thought to be independent when considering just correlation-based analyses. If the synergistic structure were reflecting mere noise, then we would not expect the high-degree of symmetry and structure we observe. 

To test whether the patterns we observed were consistent across individuals, we re-ran the entire pipeline (PED of all triads, hypergraph clustering of redundant and synergistic triads, etc) for each of the 95 subjects seperately. Then, for each subject, we computed the normalized mutual information (NMI) \cite{rubinov_complex_2010} between the subject-level partition and the relevant master partition (redundancy or synergy) created from the concatenated time series of all four scans from each of the ninety-five subjects. We significance tested each comparison with a permutation null model. For each null, we permuted the subject-level community assignment vector of nodes, recomputing the NMI between the master partition and a shuffled subject-level partition (1,000 permutations). In the case of the redundant partition, we found that that no subjects ever had a shuffled null that was greater than the empirical NMI: all had significant NMI ($0.52\pm0.07$). In the case of the synergistic partition, 91 of the 95 subjects showed significant NMI ($0.1\pm0.03$, $p<0.05$, Benjamini-Hochberg FDR corrected). These results suggest that both structures (redundant and synergistic) are broadly conserved across individuals, however, it appears that the synergistic partitions are generally more variable between subjects than the redundant partition (which hews closer to the master partition constructed by combining the data from all subjects). 
When we computed the normalized mutual information of all the subject level redundancy partitions to the canonical Yeo systems, we found a high degree of correlation (NMI = 0.6196$\pm$0.0117, $p<10^{-20}$). The same analysis with the subject level synergy partitions found a much lower degree of concordance (NMI =  0.2290$\pm$0.0117, $p<10^{-20}$). 

\subsubsection{Redundancy-synergy gradient \& time-resolved analysis}

\begin{figure*}
    \centering
    \includegraphics[scale=0.4]{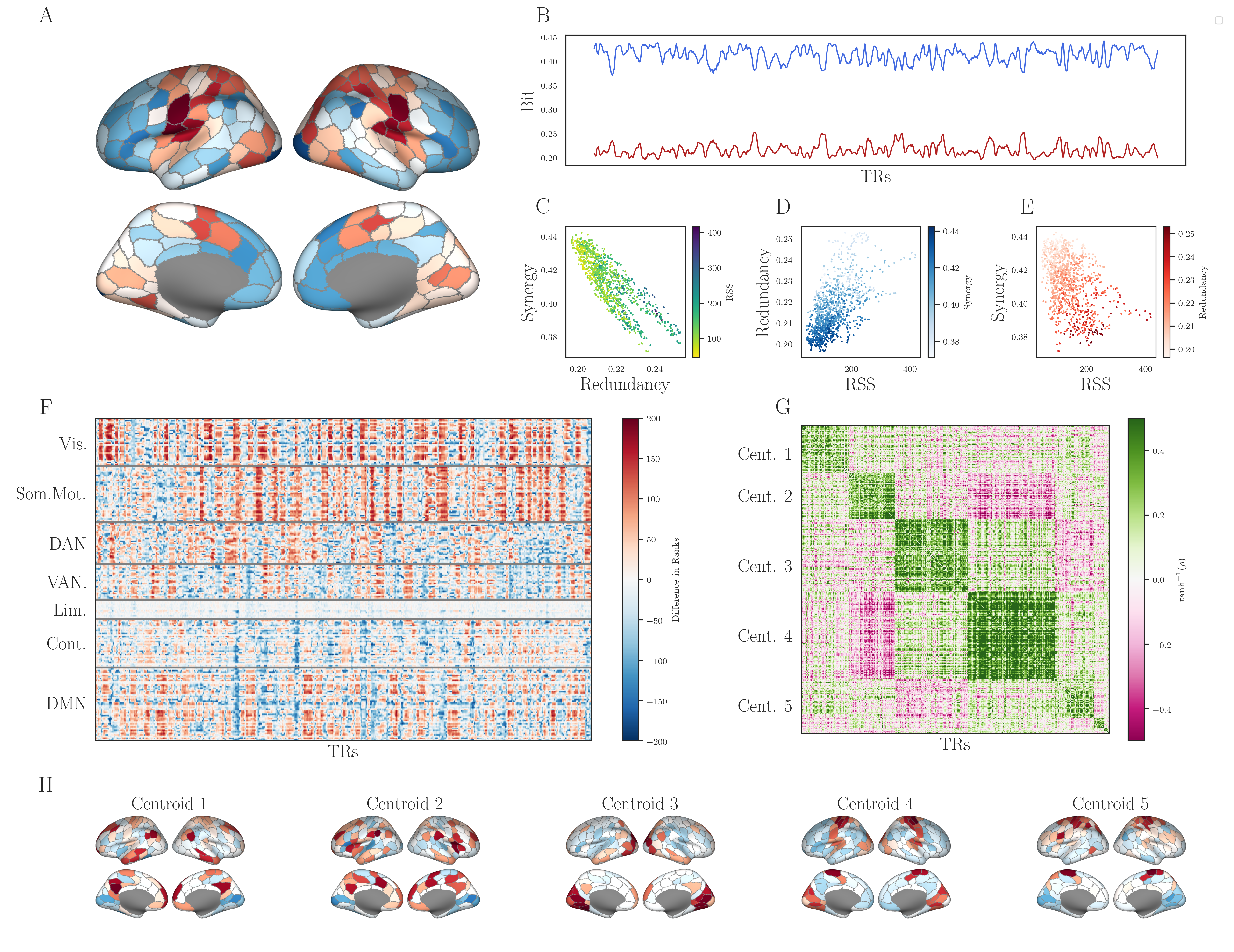}
    \caption{\textbf{Time-resolved analysis. A.} Surface plots for the distributions of relative synergies and relative redundancies across the human brain. These results match prior work by Luppi et al., \cite{luppi_synergistic_2022}, with primary sensory and motor cortex being relatively redundant, while multi-modal association areas being relatively synergistic. \textbf{B.} Over the course of one subject's scan (1100 TRs), the total redundant and synergistic structure varies over time, although never so much that the curves cross (i.e. there is never more redundant structure than synergistic structure present). \textbf{C.} Instantaneous redundant and synergistic structure are anti-correlated ($\rho=-0.83$, $p<10^{-50})$. \textbf{D.} Redundancy is positively correlated with the amplitude of bivariate co-fluctuations ($\rho=0.6$, $p<10^{-50}$) and \textbf{E.} synergy is negatively correlated with co-fluctuation amplitude ($\rho=-0.43$, $p<10^{-50}$). \textbf{F.} For each TR, we show the difference in the rank-redundancy and rank-synergy for each node (red indicates a higher rank-redundancy than rank-synergy and vice versa for blue). When nodes are stratified by Yeo system \cite{yeo_organization_2011} (grey, horizontal lines), it is clear that different systems alternate between high-redundancy and high-synergy configurations in different ways. \textbf{G.} For every pair of columns in Panel F. we compute the Pearson correlation between them to construct a time $\times$ time similarity matrix, which we then clustered using the MRCC algorithm \cite{jeub_multiresolution_2018}. Note that rows and columns are not in time order, but rather, re-ordered to reveal the state-structure of the time series. \textbf{H.} Five example states (centroids of each community show in Panel G.) projected onto the cortical surface. It is clear that the instantaneous pattern of relative synergies and redundancies varies from the average structure presented in Panel A. For example, in States 3 and 4, the visual system is highly redundant (as in the average), however in state 5, the visual system is synergistic.}
    \label{fig:time_resolved}
\end{figure*}

Thus far, we have analyzed higher-order redundancy and synergy separately. To understand how they interact, we began by replicating the analysis of Luppi et al., \cite{luppi_synergistic_2022}. We counted how many times each brain region appeared in the set of 3,746 most synergistic and 3,746 most redundant triads. We then ranked each node to create two vectors which rank how frequently each region participates in high-redundancy and high-synergy configurations. By subtracting those two rank vectors, we get a measure of \textit{relative} redundancy/synergy dominance. A value greater than zero indicates that a region's relative redundancy (compared to all other regions) is greater than its relative synergy (compared to all other regions), and vice versa. 

By projecting the rank-differences onto the cortical surface (Fig. \ref{fig:time_resolved}A), we recover the same gradient-like pattern first reported by Luppi et al., with relatively redundant regions located in primary sensory and motor cortex, and relatively synergistic regions located in multi-modal and executive cortex. This replication is noteworthy, as Luppi et al., used an entirely different method of computing synergy (based on the information flow from past to future in pairs of brain regions), while we are looking at generalizations of static FC for which dynamic order does not matter. The fact that the same gradient appears when using both analytical methods strongly suggests it is a robust feature of brain activity.

\begin{figure*}
    \centering
    \includegraphics[scale=0.6]{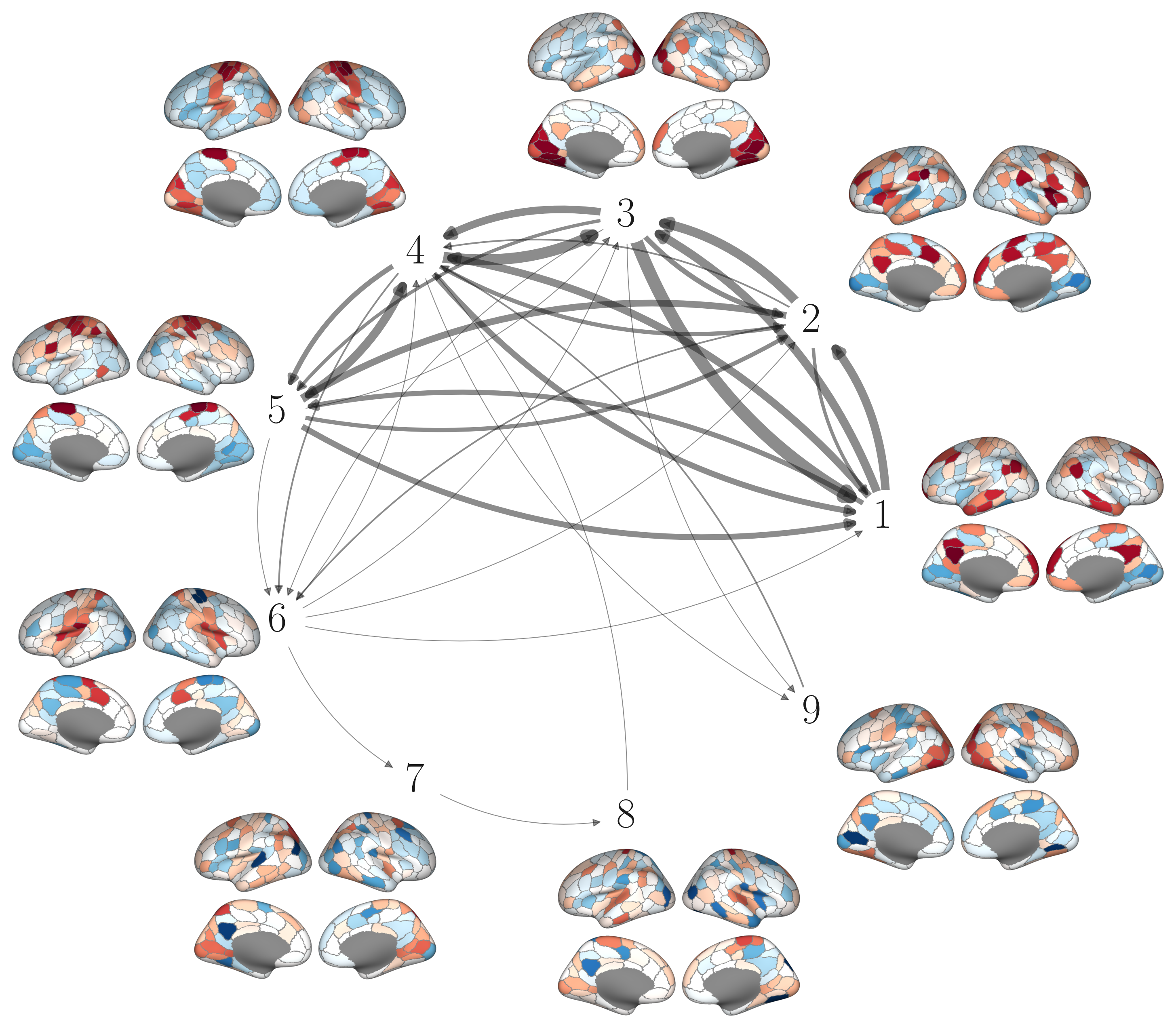}
    \caption{\textbf{State-to-state transitions.} For each of the nine distinct states, we can see how many times each state transitions another (self-loops are not shown for visual clarity). We can see that the various states have meaningful differences between each-other (e.g. the visual system or the somato-motor systems both transition from redundancy- to synergy-dominated configurations over time), however, within a state, the patterns are largely symmetrical across hemispheres. }
    \label{fig:transmat}
\end{figure*}

A limitation of the analysis by Luppi et al. is the restriction that only \textit{average} values of synergy and redundancy are accessible: the results describe expected values over all TRs and obscure any local variability. The PED analysis using $h_{sx}$ can be localized (see Sec. \ref{sec:theory}) to individual frames. This allows us to see how the redundant and synergistic structure fluctuate over the course of a resting state scan, and how the distributions of relative synergies and redundancies vary over the cortex. Figure \ref{fig:time_resolved}B shows how the redundant and synergistic structure fluctuate over the course of 1100 TRs taken from a single subject (for scans concatenated). This allows us to probe the information structure of previously identified patterns in frame-wise dynamics. Analysis of instantaneous pairwise co-fluctuations (also called ``edge time series") reveals a highly structured pattern, with periods of relative disintegration interspersed with high co-fluctuation ``events" \cite{esfahlani_high-amplitude_2020,betzel_hierarchical_2022}. The distribution of these co-fluctuations reflect various factors of cognition \cite{tanner_synchronous_2022}, generative structure \cite{pope_modular_2021}, functional network organization \cite{sporns_dynamic_2021}, and individual differences \cite{betzel_individualized_2022}. By correlating the instantaneous average whole-brain redundant and synergistic structures with instantaneous whole-brain co-fluctuation amplitude (RSS), we can get an understanding of the ``informational structure" of high-RSS ``events." We found that redundancy is positively correlated with co-fluctuation RSS ($\rho=0.6$, $p<10^{-50}$) and synergy is negatively correlated with co-fluctuation amplitude ($\rho=-0.43$, $p<10^{-50}$). Given that synergy is known to drive bivariate functional connectivity \cite{esfahlani_high-amplitude_2020}, this is again consistent with the hypothesis that FC patterns largely reflect redundancy and are insensitive to higher-order synergies.

With full PED analysis completed for every frame, it is possible to compute the instantaneous distribution of relative redundancies and synergies across the cortex for every TR. The resulting multidimensional time-series can be seen in Fig. \ref{fig:time_resolved}F. When sorted by Yeo systems \cite{yeo_organization_2011}, we can see that different systems show distinct relative redundancy/synergy profiles. The nodes in the somato-motor system had the highest median value ($22.0\pm73$), followed by the visual system ($14.0\pm80$), indicating that they were, on-average relatively more redundant than synergistic. In contrast, the ventral attentional system had the lowest median value ($-11.0\pm66$), indicating a relatively synergistic dynamic. Other systems seemed largely balanced: with median values near zero but a wide spread between them, such as the dorsal attention network ($1.0\pm70$), fronto-parietal control system ($-5.0\pm56$), and the DMN ($-2.0\pm67$). These are systems that transiently shift from largely redundancy-dominated to synergy-dominated regimes in equal measure. Finally, the limbic system had small values and relatively little spread ($-5.0\pm18$), indicating a system that never achieved either extreme. 

We then correlated every TR against every other frame to construct a weighted, signed recurrence network \cite{varley_network_2022}, which we could then cluster using the MRCC algorithm \cite{jeub_multiresolution_2018} (Fig. \ref{fig:time_resolved}G). This allowed us to assign every TR to one of nine discrete ``states", each of which can be represented by its centroid (for five examples see Fig \ref{fig:time_resolved}H). We can see that these states are generally symmetrical, but show markedly different patterns relative redundancy and synergy across the cortex, and some systems can change valance entirely. For example, in states three and four the visual system is highly redundant (consistent with the average behavior), while in state five the same regions are more synergy-dominated. In the same vein, the somato-motor strip is highly redundant in state 4, but slightly synergy-biased in state 3. This shows that the dynamics of information processing are variable in time, with different areas of cortex transiently becoming more redundant or more synergistic in concert. 

The sequence of states occupied at each TR is a discrete time series which we can analyze as a finite-state machine (for visualization, see Figure \ref{fig:transmat}). Shannon temporal mutual information found that the present state was significantly predictive of the future state (1.59 bit, $p<10^{-50}$), and that the transitions between states were generally more deterministic \cite{hoel_quantifying_2013,varley_topological_2021} (2.29 bit $p<10^{-50}$) than would be expected by chance. While the sample size is small (1099 transitions), these results suggest that the transition between states is structured in non-random ways.  

\section{Discussion}
In this paper, we have explored a novel framework for extracting higher-order dependencies from data and applied it to fMRI rcordings. We found that the human brain is rich in beyond-pairwise, synergistic structures, as well as redundant information copied over many brain regions. Based on a partial entropy decomposition framework \cite{ince_partial_2017,finn_generalised_2020} our method returns strictly non-negative values, does not require grouping elements into ``sources" and ``targets", and is localizable, permitting a time-resolved analysis of the system's dynamics. 

Prior work on the partial entropy decomposition has analytically shown that the bivariate mutual information between two elements incorporates non-local information that is redundantly present over more than two elements \cite{ince_partial_2017,finn_generalised_2020}. This means that classic approaches to functional connectivity are \textit{non-specific}: the link between two elements does not reflect information uniquely shared by those two but double (or triple-counts) higher-order redundancies distributed over the system. We verified this empirically by comparing the distribution of higher-order (beyond pairwise) redundancies to a bivariate correlation network and found that the redundancies closely matched the classic network structure. 

These non-local redundancies shed new light on a well-documented feature of bivariate functional connectivity networks: the transitivity of correlation \cite{zalesky_use_2012}. In functional connectivity networks, if $X_i$ and $X_j$ are correlated, as well as $X_j$ and $X_k$, then there is a much higher-than expected chance that $X_i$ and $X_k$ are correlated (even though this is not theoretically necessary \cite{langford_is_2001}). Since the Pearson correlation related the mutual information under Gaussian assumptions \cite{cover_elements_2012}, we claim that the observed transitivity of functional connectivity is a consequence of previously-unrecognized, non-local redundancies copied over ensembles of nodes. This hypothesis is consistent with our findings that redundancies correlate with key features of functional network topology, including subgraph density and community structure. 

In addition to higher-order redundancies, we also found strong evidence of higher order synergies: information present in the joint states of multiple brain regions and only accessible when considering ``wholes" rather than just ``parts." These synergies appear to be structured in part by the physical brain (for example, being largely symmetric across hemispheres), but also don't readily correspond to the standard functional connectivity networks previously explored in the literature. Since synergiestic structures appear to be largely anti-correlated with the standard bivariate network structures, it is plausible that these synergistic systems represent a novel organization of human brain activity. 

These higher-order interactions represent a vast space of largely unexplored, but potentially significant aspects of brain activity. One possible avenue of study is how higher-order synergies reflect individual differences \cite{cutts_uncovering_2022,betzel_individualized_2022} and subject identifiability \cite{jo_subject_2021}. The finding that the synergistic community structure was more variable across subjects than the redundant structure suggests that synergistic dependencies may reflect more unique, individualized differences, while the redundant structure (reflected in the functional connectivity) represents a more conserved architecture. This is consistent with recent theoretical work linking synergy to individuality \cite{krakauer_information_2020}, as well as empirical findings that the evolution of humans is associated with an enrichment of synergistic cortical structures \cite{luppi_synergistic_2022}. The ability to expand beyond pairwise network models of the brain into the much richer space of beyond-pairwise structures offers a the opportunity to explore previously inaccessible relationships between brain activity, cognition, and behavior. 

Since normal cognitive functioning requires the coordination of many different brain regions \cite{barttfeld_signature_2015,demertzi_human_2019,shine_dynamics_2016}, and pathological states are associated with the dis-integrated dynamics \cite{ahmed_neuronal_2016,damoiseaux_functional_2012,luppi_consciousness-specific_2019}, it is reasonable to assume that alterations to higher-order, synergistic coordination may also reflect clinically significant changes in cognition and health. Recent work has already indicated that changes in bivariate synergy track loss of consciousness under anesthesia and following traumatic and anoxic brain injury \cite{luppi_synergistic_2020} suggesting that higher-order dependencies can encode clinically significant biomarkers. We hypothesize that beyond-pairwise synergies in particular may be worth exploring in the context of recognizing early signs of Alzheimer's and other neurodegenerative diseases, as synergy requires the coordination of many regions simultaneously and may begin to show signs of fragmentation earlier than standard, functional connectivity-based patterns (which are dominated by non-local redundancies may obscure early fragmentation of the system). 

Finally, the localizable nature of the $\mathcal{H}_{sx}$ partial entropy function allows us a high degree of temporal precision when analyzing brain dynamics. The standard approach to time-varying connectivity is using a sliding-windows analysis, however, this approach blurs temporal features and obscures higher-frequency events \cite{zamani_esfahlani_edge-centric_2022}. By being able to localize the redundancies and synergies in time, we can see that there is a complex interplay between both ``types" of integration. When considering expected values, we find a distribution of redundancies and synergies that replicates the findings of Luppi et al., \cite{luppi_synergistic_2022}, however, when we localize the analysis in time, we find a high degree of variability between frames. It appears that there are not consistently ``redundant" or ``synergistic" brain regions (or ensembles), but rather, various brain regions can transiently participate in highly synergistic or highly redundant behaviors at different times. The structure of these dynamics appears to be non-random (based on the structure of the state-transition matrix), however, the significance of the various combinations of redundancy and synergy remains a topic for much future work. The fact that some systems (such as the visual system) can be either redundancy- or synergy-dominated at different times complicates the notion of a ``synergistic core". Instead, there may be a ``synergistic landscape" of configurations that the system traverses, with different configurations of brain regions transiently serving as the core and providing a flexible architecture for neural computation in response to different demands. 

This analysis does have some limitations, however. The most significant is that the size of the partial entropy lattice grows explosively as the size of the system increases: a system with only eight elements will have a lattice with 5.6$\times10^{22}$ unique partial entropy atoms. While our aggregated measures of redundant and synergistic structure can summarize the dependencies in a principled way, simply computing that many atoms is computationally prohibitive. In this paper, we took a large system of 200 nodes, and calculated every triad and a large number of tetrads, however, this also quickly runs into combinatorial difficulties, as the number of possible groups of size $k$ one can make from $N$ elements grows with the binomial coefficient. Heuristic measures such as the O-information can help, although as we have seen, this measure can conflate redundancy and synergy in sometimes surprising ways. One possible avenue of future work could be to leverage optimization algorithms to find small, tractable subsets of systems that show interesting redundant or synergistic structure, as was done in \cite{novelli_large-scale_2019,wollstadt_rigorous_2021,varley_multivariate_2022}. Alternately, coarse-graining approaches that can reduce the dimensionality of the system while preserving the informational or causal structure may allow the analysis of a compressed version of the system small enough to be tractable \cite{hoel_quantifying_2013,varley_emergence_2022}.

In the context of this study, the use of fMRI BOLD data presents some inherent limitations, such as a small number of samples (TRs) from which to infer probability distributions, and the necessity of binarizing a slow, continuous signal. Generalizing the logic of shared probability mass exclusions remains an area of on-going work \cite{schick-poland_partial_2021}, although for the time being, the $h_{sx}$ function requires discrete random variables. BOLD itself is also fundamentally a proxy measure of brain activity based on oxygenated blood flow and not a direct measure of neural activity. Applying this work to electrophysciological data (M/EEG, which can be discretized in principled ways to enable information-theoretic analysis \cite{varley_differential_2020}), and naturally discrete spiking neural data \cite{newman_revealing_2022}, will help deepen our understanding of how higher-order interactions contribute to cognition and behavior. The applicability of the PED to multiple scales of analysis highlights one of the foundational \textit{strengths} of the approach (and information-theoretic frameworks more broadly): being based on the fundamental logic of inferences under conditions of uncertainty, the PED can be applied to a large number of complex systems (beyond just the brain), or to multiple scales within a single system to provide a detailed, and holistic picture of the system's structure. 

\section{Conclusions}
In this work, we have shown how the joint entropy of a complex system can be decomposed into atomic components of redundancy and synergy, which reveal higher-order, beyond-pairwise dependencies in the structure of the system. When applied to human brain data, this partial entropy decomposition framework reveals previously unrecognized, higher-order structures in the human brain. We find that the well-known patterns of functional connectivity networks largely reflect redundant information copied over many brain regions. In contrast, the synergies for a kind of ``shadow structure" that is largely independent from, or anticorrelated with, the bivariate network and has consequently remained less well explored. The patterns of redundancy and synergy over the cortex are dynamic across time, with different ensembles of brain regions transiently forming redundancy- or synergy-dominated structures. This space of beyond-pairwise dynamics is likely rich in previously unidentified links between brain activity and cognition. The PED can also be applied to problems beyond neuroscience and may provide a general tool with which higher-order structure can be studied in any complex system. 

\section{Materials \& Methods}
\label{sec:mnm}

\subsection{Human Connectome Project fMRI Data}
The data used in this study was taken from a set of 100 unrelated subjects included in the Human Connectome Project (HCP) \cite{van_essen_wu-minn_2013}. Refs \cite{van_essen_wu-minn_2013,glasser_minimal_2013} provide a detailed description of the acquisition and preprocessing of this data, which have been used in many previous studies\cite{sporns_dynamic_2021,pope_modular_2021}. Briefly, all subjects gave informed consent to protocols approved by the Washington University Institutional Review Board. Data was collected with a Siemens 3T Connectom Skyra using a head coil with 32 channels. Functional data analysed here was acquired during resting state with a gradient-echo echo-planar imaging (EPI) sequence. Collection occurred over four scans on two separate days (scan duration: 14:33 min; eyes open). The main acquisition parameters included TR = 720 ms, TE = 33.1 ms, flip angle of 52°, 2 mm isotropic voxel resolution, and a multiband factor of 8. Resting state data was mapped to a 200-node parcellation scheme \cite{schaefer_local-global_2018} covering the entire cerebral cortex. 

Considerations for subject inclusion were established before the study and are as follows. The mean and mean absolute deviation of the relative root mean square (RMS) motion throughout any of the four resting scans were calculated. Subjects that exceeded 1.5 times the interquartile range in the adverse direction for two or more measures they were excluded. This resulted in the exclusion of four subjects, and an additional subject due to a software error during diffusion MRI processing. The included subjects had demographic characteristics of: 56\% female, mean age = 29.29 $\pm$ 3.66, age range = 22-36 years.

\subsubsection{Preprocessing}
The minimal preprocessing of HCP rs-fMRI data can be found described in detail in ref. \cite{glasser_minimal_2013}. Five main steps were followed: 1) susceptibility, distortion, and motion correction; 2) registration to subject-specific T1-weighted data; 3) bias and intensity normalization; 4) projection onto the 32k\_fs\_LR mesh; and 5) alignment to common space with a multimodal surface registration (81). This pipeline produced an ICA+FIX time series in the CIFTI grayordinate coordinate system. We included two additional preprocessing steps: 6) global signal regression and 7) detrending and band pass filtering (0.008 to 0.08 Hz) \cite{parkes_evaluation_2018}. We discarded the first and last 50 frames of each time series after confound regression and filtering to produce final scans with length 13.2 min (1,100 frames). All four scans from 95 subjects were then z-scored and concatenated to give a final time-series of 200 brain regions and 418,000 time points. 

\subsubsection{Discretizing BOLD Signals}

Unfortunately, the $\mathcal{H}_{sx}$ measure is only well-defined for discrete random variables. Consequently, we discretized our data by binarizing the z-scored time series: setting any value greater than zero to one and any value less than zero to zero. Prior work has established that transforming BOLD signals into binary point processes preserves the majority of the total correlation structure \cite{tagliazucchi_criticality_2012,sporns_dynamic_2021}, so we are confident that our analysis is robust, especially considering the large number of samples. 

We chose to binarize around the z-score (as opposed to alternative point-processing techniques such as local maxima), as the z-score ensures that each individual channel is generally maximally entropic (i.e. $\prob(X_i = 1) \approx \prob(X_i = 0) \approx 1/2$). This ensures that every individual channel has approximately the same entropy, and so deviations from maximum entropy at the level of the entire triad or tetrad can \textit{only} emerge from correlations between two or more channels, rather than being influenced by biases at the channel-level. The choice to binarize about the mean also links this work to previous work on decomposing functional connectivity into discrete partitions \cite{sporns_dynamic_2021}.

\subsection{Statistical Analyses}

\subsubsection{Triads \& tetrads}
In standard FC analysis, it is typical to compute the pairwise correlation between all pairs of brain regions, resulting $\binom{N}{2}$ unique pairs. For this analysis, we computed all triads of brain regions, resulting in $\binom{200}{3} = 1,313,400$ unique triples. For each triad, we computed the joint entropy, and performed the full partial entropy decomposition to compute each of the eighteen partial entropy atoms. Finally, each of the atoms was normalized by the total joint entropy, to give a measure of how much each atom contributes to the whole entropy. This allows us to directly compare triads that have different joint entropies. 

It was not feasible to brute-force all possible tetrads, which is a set of approximately sixty-four million. Instead, we randomly sub-sampled sets of four randomly, collecting 1954000 tetrads ($\approx3\%$ of the total space) and analyzing them. 

\subsubsection{Bivariate functional connectivity networks}

To directly compare the PED framework to the standard, correlation-based FC network framework, we constructed single, representative FC network by computing the pairwise mutual information between every pair of regions in the fMRI scan (as was done in \cite{pope_modular_2021}). 

\begin{equation}
    \mi(X;Y) = \entropy(X) + \entropy(Y) - \entropy(X,Y)
\end{equation}

\subsubsection{Subgraph Analysis}
Since we are interested in how the bivariate FC framework reflects (or fails to reflect) higher-order redundancies and synergies, we also compute a battery of structure metrics on matching subgraphs taken from the FC network. Formally presented by Onnela et al., \cite{onnela_intensity_2005}, we consider arithmetic mean of the subgraph connectivity:

\begin{equation}
    \mathcal{G}_{\mathfrak{A}}(\textbf{X}) = \frac{\sum_{i\not= j} \mi(X_i;X_j)}{|\textbf{X}|^{2} - |\textbf{X}|}
\end{equation}

For a given triad of tetrad \textbf{X}, we compared the mean FC density to the various redundant and synergistic information-sharing structures of \textbf{X}.

\subsubsection{Community Detection on Bivariate Matrices}

Multi-resolution consensus clustering \cite{jeub_multiresolution_2018} was used to detect network communities in the functional connectivity matrix across multiple scales. The algorithm proceeds in three main stages. In the first stage, modularity maximization using the Louvain method is performed for 1,000 different values of the resolution parameter, $\gamma$. This produced a range of $\gamma$ values that resulted with partitions having between 2 and $N$ communities. The second stage consisted of a more fine-grained sweep (10,000 steps) over the $\gamma$ values defined in the first stage of the process. We aggregate the partitions produced by this sweep into a node-by-node co-classification matrix storing how frequently nodes are partitioned into the same community. A null model with expected values of co-classification based on the size and number of communities was subtracted from the co-classification matrix \cite{jeub_multiresolution_2018}. Finally, in the third stage, the null-adjusted co-classification matrix was clustered again using consensus clustering with 100 repetitions and a consensus threshold $\tau$ of 0 \cite{lancichinetti_consensus_2012}. The resulting partition was used for analyses.

We assessed the similarity between single-subject partitions and consensus partitions using Normalized Mutual Information (NMI). Each partition can be formalized as a vector of integers of dimension $N$ whose entries denote the nodes’ allegiance to communities. NMI estimates the similarity between two partitions by counting co-occurrences in the two vectors.

We computed NMI between each one of the 95 single-subject partitions and the consensus partition, in both cases of redundancy and synergy hypergraphs. We assessed the significance of NMI values by comparing them with a null case obtained by randomly shuffling 1000 times communities labels in the single-subject partitions. The $p$-values of the statistical test, calculated as the fraction of null-case NMI greater than the actual NMI, have been corrected with a Benjamini-Hochberg test.

\subsubsection{Null Model}

To ensure that the statistical dependencies we were observing reflect non-trivial interactions, we significance-tested triads and tetrads against a null distribution composed of one million, maximum entropy null models. We constructed sets of totally independent, maximum entropy binary time series and computed the PED on each set of three or four null channels. From this, we can construct distributions of the expected null structure and expected synergistic structure against which to compare triads and tetrads.  

\subsubsection{Hypergraph Community Detection}

Each of the triads can be thought of as a hyper-edge on a 3-uniform hypergraph of 200 nodes. For the synergistic structure, we selected only those hyperedges who had a \textit{greater} synergistic structure than any of the one million maximum-entropy nulls that formed our null distribution. This resulted in a hypergraph with 200 hundred nodes and 3,746 regular hyper-edges. We used the same criteria to build a redundant structure hypergraph using the top 3,746 most redundant hyperedges. 

Both hypergraphs were clustered using the \texttt{HyperNetX} package (available on Github: \url{https://github.com/pnnl/HyperNetX}) implementation of the hyper-modularity optimization by Kumar and Vaidyanathan et al., \cite{kumar_new_2020}.

Briefly, the algorithm by Kumar and Vaidyanathan et al., takes a modularity maximization approach to partitioning the vertices of a hypergraph into non-overlapping communities. In dyadic networks, the modularity function compares the distribution of within- and between-community edges to the expected distribution based on a degree-preserving, configuration null model \cite{newman_modularity_2006}. In the case of hypergraphs, a hyper-configuration model can be used instead. A generalized modularity metric can then be used as an objective function in a Louvain-based, modularity maximization search.

\subsubsection{Temporal Structure}
To explore the temporal structure of the state-transition series, we used the active information storage \cite{wibral_measuring_2014,lizier_framework_2014} (a measure of how predictable is the future given the past) and the determinism \cite{hoel_quantifying_2013,varley_topological_2021}, (a measure of how constrained the future is given the past). For a one dimensional, discrete random variable \textit{X} that evolves through time, we can compute the information that the past $X_{t-1}$ discloses about the future $X_{t}$ with the mutual information:

\begin{equation}
    AIS(X) = I(X_{t-1};X_{t})
\end{equation}

This measure quantifies the degree to which knowing the past reduces our uncertainty about the future. This term can be further decomposed into two components: the determinism and the degeneracy \cite{hoel_quantifying_2013}:

\begin{equation}
    I(X_{t-1};X_{t}) = Det(X) - Deg(X)
\end{equation}

Where determinism is:

\begin{equation}
    Det(X) = \log_2(N) - H(X_{t}|X_{t-1})
\end{equation}

And degeneracy is:

\begin{equation}
    Deg(X) = \log_2(N)-H(X_{t})
\end{equation}

The determinism quantifies how reliably a given past state $x_{t-1}$ leads to a single future state $x_{t}$. If $P(x_{t}|x_{t-1})\approx1$, then we say that $x_{t-1}$ \textit{deterministically} leads to $x_{t}$. 

We significance tested both the active information storage and the determinism by comparing the empirical values to an ensemble of ten thousand randomly permuted nulls generated by shuffling the time series. Since the degeneracy is unchanged by permutation of the temporal structure (since the marginal entropy $H(X_{t})$ is the same), any changes in active information storage produced by shuffling must be driven by changes in the determinism.

\subsection{Software}

All partial information/entropy decompositions were done using the \texttt{SxPID} package released with \cite{makkeh_introducing_2021} and can be accessed on Github: \url{https://github.com/Abzinger/SxPID}. All scripts required to reproduce this analysis will be attached as supplementary material to the final published work. 

\bibliography{partial_ent_decomp.bib}

\begin{thebibliography}{71}%
\makeatletter
\providecommand \@ifxundefined [1]{%
 \@ifx{#1\undefined}
}%
\providecommand \@ifnum [1]{%
 \ifnum #1\expandafter \@firstoftwo
 \else \expandafter \@secondoftwo
 \fi
}%
\providecommand \@ifx [1]{%
 \ifx #1\expandafter \@firstoftwo
 \else \expandafter \@secondoftwo
 \fi
}%
\providecommand \natexlab [1]{#1}%
\providecommand \enquote  [1]{``#1''}%
\providecommand \bibnamefont  [1]{#1}%
\providecommand \bibfnamefont [1]{#1}%
\providecommand \citenamefont [1]{#1}%
\providecommand \href@noop [0]{\@secondoftwo}%
\providecommand \href [0]{\begingroup \@sanitize@url \@href}%
\providecommand \@href[1]{\@@startlink{#1}\@@href}%
\providecommand \@@href[1]{\endgroup#1\@@endlink}%
\providecommand \@sanitize@url [0]{\catcode `\\12\catcode `\$12\catcode
  `\&12\catcode `\#12\catcode `\^12\catcode `\_12\catcode `\%12\relax}%
\providecommand \@@startlink[1]{}%
\providecommand \@@endlink[0]{}%
\providecommand \url  [0]{\begingroup\@sanitize@url \@url }%
\providecommand \@url [1]{\endgroup\@href {#1}{\urlprefix }}%
\providecommand \urlprefix  [0]{URL }%
\providecommand \Eprint [0]{\href }%
\providecommand \doibase [0]{https://doi.org/}%
\providecommand \selectlanguage [0]{\@gobble}%
\providecommand \bibinfo  [0]{\@secondoftwo}%
\providecommand \bibfield  [0]{\@secondoftwo}%
\providecommand \translation [1]{[#1]}%
\providecommand \BibitemOpen [0]{}%
\providecommand \bibitemStop [0]{}%
\providecommand \bibitemNoStop [0]{.\EOS\space}%
\providecommand \EOS [0]{\spacefactor3000\relax}%
\providecommand \BibitemShut  [1]{\csname bibitem#1\endcsname}%
\let\auto@bib@innerbib\@empty
\bibitem [{\citenamefont {Sporns}\ \emph {et~al.}(2005)\citenamefont {Sporns},
  \citenamefont {Tononi},\ and\ \citenamefont {Kötter}}]{sporns_human_2005}%
  \BibitemOpen
  \bibfield  {author} {\bibinfo {author} {\bibfnamefont {O.}~\bibnamefont
  {Sporns}}, \bibinfo {author} {\bibfnamefont {G.}~\bibnamefont {Tononi}},\
  and\ \bibinfo {author} {\bibfnamefont {R.}~\bibnamefont {Kötter}},\
  }\bibfield  {title} {\bibinfo {title} {The {Human} {Connectome}: {A}
  {Structural} {Description} of the {Human} {Brain}},\ }\bibfield  {journal}
  {\bibinfo  {journal} {PLoS Computational Biology}\ }\textbf {\bibinfo
  {volume} {1}},\ \href {https://doi.org/10.1371/journal.pcbi.0010042}
  {10.1371/journal.pcbi.0010042} (\bibinfo {year} {2005}),\ \bibinfo {note}
  {number: 4}\BibitemShut {NoStop}%
\bibitem [{\citenamefont {Sporns}(2010)}]{sporns_networks_2010}%
  \BibitemOpen
  \bibfield  {author} {\bibinfo {author} {\bibfnamefont {O.}~\bibnamefont
  {Sporns}},\ }\href@noop {} {\emph {\bibinfo {title} {Networks of the
  {Brain}}}}\ (\bibinfo  {publisher} {MIT Press},\ \bibinfo {year}
  {2010})\BibitemShut {NoStop}%
\bibitem [{\citenamefont {Fornito}\ \emph {et~al.}(2016)\citenamefont
  {Fornito}, \citenamefont {Zalesky},\ and\ \citenamefont
  {Bullmore}}]{fornito_fundamentals_2016}%
  \BibitemOpen
  \bibfield  {author} {\bibinfo {author} {\bibfnamefont {A.}~\bibnamefont
  {Fornito}}, \bibinfo {author} {\bibfnamefont {A.}~\bibnamefont {Zalesky}},\
  and\ \bibinfo {author} {\bibfnamefont {E.}~\bibnamefont {Bullmore}},\ }\href
  {https://doi.org/10.1016/C2012-0-06036-X} {\emph {\bibinfo {title}
  {Fundamentals of {Brain} {Network} {Analysis}}}}\ (\bibinfo  {publisher}
  {Elsevier},\ \bibinfo {year} {2016})\BibitemShut {NoStop}%
\bibitem [{\citenamefont {Friston}(1994)}]{friston_functional_1994}%
  \BibitemOpen
  \bibfield  {author} {\bibinfo {author} {\bibfnamefont {K.~J.}\ \bibnamefont
  {Friston}},\ }\bibfield  {title} {\bibinfo {title} {Functional and effective
  connectivity in neuroimaging: {A} synthesis},\ }\href
  {https://doi.org/10.1002/hbm.460020107} {\bibfield  {journal} {\bibinfo
  {journal} {Human Brain Mapping}\ }\textbf {\bibinfo {volume} {2}},\ \bibinfo
  {pages} {56} (\bibinfo {year} {1994})}\BibitemShut {NoStop}%
\bibitem [{\citenamefont {Fox}\ \emph {et~al.}(2005)\citenamefont {Fox},
  \citenamefont {Snyder}, \citenamefont {Vincent}, \citenamefont {Corbetta},
  \citenamefont {Essen},\ and\ \citenamefont {Raichle}}]{fox_human_2005}%
  \BibitemOpen
  \bibfield  {author} {\bibinfo {author} {\bibfnamefont {M.~D.}\ \bibnamefont
  {Fox}}, \bibinfo {author} {\bibfnamefont {A.~Z.}\ \bibnamefont {Snyder}},
  \bibinfo {author} {\bibfnamefont {J.~L.}\ \bibnamefont {Vincent}}, \bibinfo
  {author} {\bibfnamefont {M.}~\bibnamefont {Corbetta}}, \bibinfo {author}
  {\bibfnamefont {D.~C.~V.}\ \bibnamefont {Essen}},\ and\ \bibinfo {author}
  {\bibfnamefont {M.~E.}\ \bibnamefont {Raichle}},\ }\bibfield  {title}
  {\bibinfo {title} {The human brain is intrinsically organized into dynamic,
  anticorrelated functional networks},\ }\href
  {https://doi.org/10.1073/pnas.0504136102} {\bibfield  {journal} {\bibinfo
  {journal} {Proceedings of the National Academy of Sciences}\ }\textbf
  {\bibinfo {volume} {102}},\ \bibinfo {pages} {9673} (\bibinfo {year}
  {2005})},\ \bibinfo {note} {number: 27 Publisher: National Academy of
  Sciences Section: Biological Sciences}\BibitemShut {NoStop}%
\bibitem [{\citenamefont {Rubinov}\ and\ \citenamefont
  {Sporns}(2010)}]{rubinov_complex_2010}%
  \BibitemOpen
  \bibfield  {author} {\bibinfo {author} {\bibfnamefont {M.}~\bibnamefont
  {Rubinov}}\ and\ \bibinfo {author} {\bibfnamefont {O.}~\bibnamefont
  {Sporns}},\ }\bibfield  {title} {\bibinfo {title} {Complex network measures
  of brain connectivity: {Uses} and interpretations},\ }\href
  {https://doi.org/10.1016/j.neuroimage.2009.10.003} {\bibfield  {journal}
  {\bibinfo  {journal} {NeuroImage}\ }\bibinfo {series} {Computational {Models}
  of the {Brain}},\ \textbf {\bibinfo {volume} {52}},\ \bibinfo {pages} {1059}
  (\bibinfo {year} {2010})},\ \bibinfo {note} {number: 3}\BibitemShut {NoStop}%
\bibitem [{\citenamefont {Sporns}\ and\ \citenamefont
  {Kötter}(2004)}]{sporns_motifs_2004}%
  \BibitemOpen
  \bibfield  {author} {\bibinfo {author} {\bibfnamefont {O.}~\bibnamefont
  {Sporns}}\ and\ \bibinfo {author} {\bibfnamefont {R.}~\bibnamefont
  {Kötter}},\ }\bibfield  {title} {\bibinfo {title} {Motifs in {Brain}
  {Networks}},\ }\href {https://doi.org/10.1371/journal.pbio.0020369}
  {\bibfield  {journal} {\bibinfo  {journal} {PLOS Biology}\ }\textbf {\bibinfo
  {volume} {2}},\ \bibinfo {pages} {e369} (\bibinfo {year} {2004})}\BibitemShut
  {NoStop}%
\bibitem [{\citenamefont {Fortunato}(2010)}]{fortunato_community_2010}%
  \BibitemOpen
  \bibfield  {author} {\bibinfo {author} {\bibfnamefont {S.}~\bibnamefont
  {Fortunato}},\ }\bibfield  {title} {\bibinfo {title} {Community detection in
  graphs},\ }\href {https://doi.org/10.1016/j.physrep.2009.11.002} {\bibfield
  {journal} {\bibinfo  {journal} {Physics Reports}\ }\textbf {\bibinfo {volume}
  {486}},\ \bibinfo {pages} {75} (\bibinfo {year} {2010})},\ \bibinfo {note}
  {number: 3}\BibitemShut {NoStop}%
\bibitem [{\citenamefont {Battiston}\ \emph {et~al.}(2021)\citenamefont
  {Battiston}, \citenamefont {Amico}, \citenamefont {Barrat}, \citenamefont
  {Bianconi}, \citenamefont {Ferraz~de Arruda}, \citenamefont {Franceschiello},
  \citenamefont {Iacopini}, \citenamefont {Kéfi}, \citenamefont {Latora},
  \citenamefont {Moreno}, \citenamefont {Murray}, \citenamefont {Peixoto},
  \citenamefont {Vaccarino},\ and\ \citenamefont
  {Petri}}]{battiston_physics_2021}%
  \BibitemOpen
  \bibfield  {author} {\bibinfo {author} {\bibfnamefont {F.}~\bibnamefont
  {Battiston}}, \bibinfo {author} {\bibfnamefont {E.}~\bibnamefont {Amico}},
  \bibinfo {author} {\bibfnamefont {A.}~\bibnamefont {Barrat}}, \bibinfo
  {author} {\bibfnamefont {G.}~\bibnamefont {Bianconi}}, \bibinfo {author}
  {\bibfnamefont {G.}~\bibnamefont {Ferraz~de Arruda}}, \bibinfo {author}
  {\bibfnamefont {B.}~\bibnamefont {Franceschiello}}, \bibinfo {author}
  {\bibfnamefont {I.}~\bibnamefont {Iacopini}}, \bibinfo {author}
  {\bibfnamefont {S.}~\bibnamefont {Kéfi}}, \bibinfo {author} {\bibfnamefont
  {V.}~\bibnamefont {Latora}}, \bibinfo {author} {\bibfnamefont
  {Y.}~\bibnamefont {Moreno}}, \bibinfo {author} {\bibfnamefont {M.~M.}\
  \bibnamefont {Murray}}, \bibinfo {author} {\bibfnamefont {T.~P.}\
  \bibnamefont {Peixoto}}, \bibinfo {author} {\bibfnamefont {F.}~\bibnamefont
  {Vaccarino}},\ and\ \bibinfo {author} {\bibfnamefont {G.}~\bibnamefont
  {Petri}},\ }\bibfield  {title} {\bibinfo {title} {The physics of higher-order
  interactions in complex systems},\ }\href
  {https://doi.org/10.1038/s41567-021-01371-4} {\bibfield  {journal} {\bibinfo
  {journal} {Nature Physics}\ ,\ \bibinfo {pages} {1}} (\bibinfo {year}
  {2021})}\BibitemShut {NoStop}%
\bibitem [{\citenamefont {Rosas}\ \emph {et~al.}(2022)\citenamefont {Rosas},
  \citenamefont {Mediano}, \citenamefont {Luppi}, \citenamefont {Varley},
  \citenamefont {Lizier}, \citenamefont {Stramaglia}, \citenamefont {Jensen},\
  and\ \citenamefont {Marinazzo}}]{rosas_disentangling_2022}%
  \BibitemOpen
  \bibfield  {author} {\bibinfo {author} {\bibfnamefont {F.~E.}\ \bibnamefont
  {Rosas}}, \bibinfo {author} {\bibfnamefont {P.~A.~M.}\ \bibnamefont
  {Mediano}}, \bibinfo {author} {\bibfnamefont {A.~I.}\ \bibnamefont {Luppi}},
  \bibinfo {author} {\bibfnamefont {T.~F.}\ \bibnamefont {Varley}}, \bibinfo
  {author} {\bibfnamefont {J.~T.}\ \bibnamefont {Lizier}}, \bibinfo {author}
  {\bibfnamefont {S.}~\bibnamefont {Stramaglia}}, \bibinfo {author}
  {\bibfnamefont {H.~J.}\ \bibnamefont {Jensen}},\ and\ \bibinfo {author}
  {\bibfnamefont {D.}~\bibnamefont {Marinazzo}},\ }\bibfield  {title} {\bibinfo
  {title} {Disentangling high-order mechanisms and high-order behaviours in
  complex systems},\ }\href {https://doi.org/10.1038/s41567-022-01548-5}
  {\bibfield  {journal} {\bibinfo  {journal} {Nature Physics}\ ,\ \bibinfo
  {pages} {1}} (\bibinfo {year} {2022})},\ \bibinfo {note} {publisher: Nature
  Publishing Group}\BibitemShut {NoStop}%
\bibitem [{\citenamefont {Luppi}\ \emph {et~al.}(2020)\citenamefont {Luppi},
  \citenamefont {Mediano}, \citenamefont {Rosas}, \citenamefont {Allanson},
  \citenamefont {Pickard}, \citenamefont {Carhart-Harris}, \citenamefont
  {Williams}, \citenamefont {Craig}, \citenamefont {Finoia}, \citenamefont
  {Owen}, \citenamefont {Naci}, \citenamefont {Menon}, \citenamefont {Bor},\
  and\ \citenamefont {Stamatakis}}]{luppi_synergistic_2020}%
  \BibitemOpen
  \bibfield  {author} {\bibinfo {author} {\bibfnamefont {A.~I.}\ \bibnamefont
  {Luppi}}, \bibinfo {author} {\bibfnamefont {P.~A.~M.}\ \bibnamefont
  {Mediano}}, \bibinfo {author} {\bibfnamefont {F.~E.}\ \bibnamefont {Rosas}},
  \bibinfo {author} {\bibfnamefont {J.}~\bibnamefont {Allanson}}, \bibinfo
  {author} {\bibfnamefont {J.~D.}\ \bibnamefont {Pickard}}, \bibinfo {author}
  {\bibfnamefont {R.~L.}\ \bibnamefont {Carhart-Harris}}, \bibinfo {author}
  {\bibfnamefont {G.~B.}\ \bibnamefont {Williams}}, \bibinfo {author}
  {\bibfnamefont {M.~M.}\ \bibnamefont {Craig}}, \bibinfo {author}
  {\bibfnamefont {P.}~\bibnamefont {Finoia}}, \bibinfo {author} {\bibfnamefont
  {A.~M.}\ \bibnamefont {Owen}}, \bibinfo {author} {\bibfnamefont
  {L.}~\bibnamefont {Naci}}, \bibinfo {author} {\bibfnamefont {D.~K.}\
  \bibnamefont {Menon}}, \bibinfo {author} {\bibfnamefont {D.}~\bibnamefont
  {Bor}},\ and\ \bibinfo {author} {\bibfnamefont {E.~A.}\ \bibnamefont
  {Stamatakis}},\ }\bibfield  {title} {\bibinfo {title} {A {Synergistic}
  {Workspace} for {Human} {Consciousness} {Revealed} by {Integrated}
  {Information} {Decomposition}},\ }\href
  {https://doi.org/10.1101/2020.11.25.398081} {\bibfield  {journal} {\bibinfo
  {journal} {bioRxiv}\ ,\ \bibinfo {pages} {2020.11.25.398081}} (\bibinfo
  {year} {2020})},\ \bibinfo {note} {publisher: Cold Spring Harbor Laboratory
  Section: New Results}\BibitemShut {NoStop}%
\bibitem [{\citenamefont {Gatica}\ \emph {et~al.}(2021)\citenamefont {Gatica},
  \citenamefont {Cofré}, \citenamefont {Mediano}, \citenamefont {Rosas},
  \citenamefont {Orio}, \citenamefont {Diez}, \citenamefont {Swinnen},\ and\
  \citenamefont {Cortes}}]{gatica_high-order_2021}%
  \BibitemOpen
  \bibfield  {author} {\bibinfo {author} {\bibfnamefont {M.}~\bibnamefont
  {Gatica}}, \bibinfo {author} {\bibfnamefont {R.}~\bibnamefont {Cofré}},
  \bibinfo {author} {\bibfnamefont {P.~A.}\ \bibnamefont {Mediano}}, \bibinfo
  {author} {\bibfnamefont {F.~E.}\ \bibnamefont {Rosas}}, \bibinfo {author}
  {\bibfnamefont {P.}~\bibnamefont {Orio}}, \bibinfo {author} {\bibfnamefont
  {I.}~\bibnamefont {Diez}}, \bibinfo {author} {\bibfnamefont {S.~P.}\
  \bibnamefont {Swinnen}},\ and\ \bibinfo {author} {\bibfnamefont {J.~M.}\
  \bibnamefont {Cortes}},\ }\bibfield  {title} {\bibinfo {title} {High-{Order}
  {Interdependencies} in the {Aging} {Brain}},\ }\bibfield  {journal} {\bibinfo
   {journal} {Brain Connectivity}\ }\href
  {https://doi.org/10.1089/brain.2020.0982} {10.1089/brain.2020.0982} (\bibinfo
  {year} {2021}),\ \bibinfo {note} {publisher: Mary Ann Liebert, Inc.,
  publishers}\BibitemShut {NoStop}%
\bibitem [{\citenamefont {Williams}\ and\ \citenamefont
  {Beer}(2010)}]{williams_nonnegative_2010}%
  \BibitemOpen
  \bibfield  {author} {\bibinfo {author} {\bibfnamefont {P.~L.}\ \bibnamefont
  {Williams}}\ and\ \bibinfo {author} {\bibfnamefont {R.~D.}\ \bibnamefont
  {Beer}},\ }\bibfield  {title} {\bibinfo {title} {Nonnegative {Decomposition}
  of {Multivariate} {Information}},\ }\href {http://arxiv.org/abs/1004.2515}
  {\bibfield  {journal} {\bibinfo  {journal} {arXiv:1004.2515 [math-ph,
  physics:physics, q-bio]}\ } (\bibinfo {year} {2010})},\ \bibinfo {note}
  {arXiv: 1004.2515}\BibitemShut {NoStop}%
\bibitem [{\citenamefont {Gutknecht}\ \emph {et~al.}(2021)\citenamefont
  {Gutknecht}, \citenamefont {Wibral},\ and\ \citenamefont
  {Makkeh}}]{gutknecht_bits_2021}%
  \BibitemOpen
  \bibfield  {author} {\bibinfo {author} {\bibfnamefont {A.~J.}\ \bibnamefont
  {Gutknecht}}, \bibinfo {author} {\bibfnamefont {M.}~\bibnamefont {Wibral}},\
  and\ \bibinfo {author} {\bibfnamefont {A.}~\bibnamefont {Makkeh}},\
  }\bibfield  {title} {\bibinfo {title} {Bits and pieces: understanding
  information decomposition from part-whole relationships and formal logic},\
  }\href {https://doi.org/10.1098/rspa.2021.0110} {\bibfield  {journal}
  {\bibinfo  {journal} {Proceedings of the Royal Society A: Mathematical,
  Physical and Engineering Sciences}\ }\textbf {\bibinfo {volume} {477}},\
  \bibinfo {pages} {20210110} (\bibinfo {year} {2021})},\ \bibinfo {note}
  {publisher: Royal Society}\BibitemShut {NoStop}%
\bibitem [{\citenamefont {Ince}(2017{\natexlab{a}})}]{ince_partial_2017}%
  \BibitemOpen
  \bibfield  {author} {\bibinfo {author} {\bibfnamefont {R.~A.~A.}\
  \bibnamefont {Ince}},\ }\bibfield  {title} {\bibinfo {title} {The {Partial}
  {Entropy} {Decomposition}: {Decomposing} multivariate entropy and mutual
  information via pointwise common surprisal},\ }\href
  {http://arxiv.org/abs/1702.01591} {\bibfield  {journal} {\bibinfo  {journal}
  {arXiv:1702.01591 [cs, math, q-bio, stat]}\ } (\bibinfo {year}
  {2017}{\natexlab{a}})},\ \bibinfo {note} {arXiv: 1702.01591}\BibitemShut
  {NoStop}%
\bibitem [{\citenamefont {Varley}\ \emph {et~al.}(2022)\citenamefont {Varley},
  \citenamefont {Pope}, \citenamefont {Faskowitz},\ and\ \citenamefont
  {Sporns}}]{varley_multivariate_2022}%
  \BibitemOpen
  \bibfield  {author} {\bibinfo {author} {\bibfnamefont {T.~F.}\ \bibnamefont
  {Varley}}, \bibinfo {author} {\bibfnamefont {M.}~\bibnamefont {Pope}},
  \bibinfo {author} {\bibfnamefont {J.}~\bibnamefont {Faskowitz}},\ and\
  \bibinfo {author} {\bibfnamefont {O.}~\bibnamefont {Sporns}},\ }\href
  {https://doi.org/10.48550/arXiv.2206.06477} {\bibinfo {title} {Multivariate
  {Information} {Theory} {Uncovers} {Synergistic} {Subsystems} of the {Human}
  {Cerebral} {Cortex}}} (\bibinfo {year} {2022}),\ \bibinfo {note} {number:
  arXiv:2206.06477 arXiv:2206.06477 [cs, math, q-bio]}\BibitemShut {NoStop}%
\bibitem [{\citenamefont {Finn}\ and\ \citenamefont
  {Lizier}(2018{\natexlab{a}})}]{finn_probability_2018}%
  \BibitemOpen
  \bibfield  {author} {\bibinfo {author} {\bibfnamefont {C.}~\bibnamefont
  {Finn}}\ and\ \bibinfo {author} {\bibfnamefont {J.~T.}\ \bibnamefont
  {Lizier}},\ }\bibfield  {title} {\bibinfo {title} {Probability {Mass}
  {Exclusions} and the {Directed} {Components} of {Mutual} {Information}},\
  }\href {https://doi.org/10.3390/e20110826} {\bibfield  {journal} {\bibinfo
  {journal} {Entropy}\ }\textbf {\bibinfo {volume} {20}},\ \bibinfo {pages}
  {826} (\bibinfo {year} {2018}{\natexlab{a}})},\ \bibinfo {note} {number: 11
  Publisher: Multidisciplinary Digital Publishing Institute}\BibitemShut
  {NoStop}%
\bibitem [{\citenamefont {Finn}\ and\ \citenamefont
  {Lizier}(2018{\natexlab{b}})}]{finn_pointwise_2018}%
  \BibitemOpen
  \bibfield  {author} {\bibinfo {author} {\bibfnamefont {C.}~\bibnamefont
  {Finn}}\ and\ \bibinfo {author} {\bibfnamefont {J.~T.}\ \bibnamefont
  {Lizier}},\ }\bibfield  {title} {\bibinfo {title} {Pointwise {Partial}
  {Information} {Decomposition} {Using} the {Specificity} and {Ambiguity}
  {Lattices}},\ }\href {https://doi.org/10.3390/e20040297} {\bibfield
  {journal} {\bibinfo  {journal} {Entropy}\ }\textbf {\bibinfo {volume} {20}},\
  \bibinfo {pages} {297} (\bibinfo {year} {2018}{\natexlab{b}})},\ \bibinfo
  {note} {number: 4 Publisher: Multidisciplinary Digital Publishing
  Institute}\BibitemShut {NoStop}%
\bibitem [{\citenamefont {Finn}\ and\ \citenamefont
  {Lizier}(2020)}]{finn_generalised_2020}%
  \BibitemOpen
  \bibfield  {author} {\bibinfo {author} {\bibfnamefont {C.}~\bibnamefont
  {Finn}}\ and\ \bibinfo {author} {\bibfnamefont {J.~T.}\ \bibnamefont
  {Lizier}},\ }\bibfield  {title} {\bibinfo {title} {Generalised {Measures} of
  {Multivariate} {Information} {Content}},\ }\href
  {https://doi.org/10.3390/e22020216} {\bibfield  {journal} {\bibinfo
  {journal} {Entropy}\ }\textbf {\bibinfo {volume} {22}},\ \bibinfo {pages}
  {216} (\bibinfo {year} {2020})},\ \bibinfo {note} {number: 2 Publisher:
  Multidisciplinary Digital Publishing Institute}\BibitemShut {NoStop}%
\bibitem [{\citenamefont {Ince}(2017{\natexlab{b}})}]{ince_measuring_2017}%
  \BibitemOpen
  \bibfield  {author} {\bibinfo {author} {\bibfnamefont {R.~A.~A.}\
  \bibnamefont {Ince}},\ }\bibfield  {title} {\bibinfo {title} {Measuring
  {Multivariate} {Redundant} {Information} with {Pointwise} {Common} {Change}
  in {Surprisal}},\ }\href {https://doi.org/10.3390/e19070318} {\bibfield
  {journal} {\bibinfo  {journal} {Entropy}\ }\textbf {\bibinfo {volume} {19}},\
  \bibinfo {pages} {318} (\bibinfo {year} {2017}{\natexlab{b}})},\ \bibinfo
  {note} {number: 7 Publisher: Multidisciplinary Digital Publishing
  Institute}\BibitemShut {NoStop}%
\bibitem [{\citenamefont {Makkeh}\ \emph {et~al.}(2021)\citenamefont {Makkeh},
  \citenamefont {Gutknecht},\ and\ \citenamefont
  {Wibral}}]{makkeh_introducing_2021}%
  \BibitemOpen
  \bibfield  {author} {\bibinfo {author} {\bibfnamefont {A.}~\bibnamefont
  {Makkeh}}, \bibinfo {author} {\bibfnamefont {A.~J.}\ \bibnamefont
  {Gutknecht}},\ and\ \bibinfo {author} {\bibfnamefont {M.}~\bibnamefont
  {Wibral}},\ }\bibfield  {title} {\bibinfo {title} {Introducing a
  differentiable measure of pointwise shared information},\ }\href
  {https://doi.org/10.1103/PhysRevE.103.032149} {\bibfield  {journal} {\bibinfo
   {journal} {Physical Review E}\ }\textbf {\bibinfo {volume} {103}},\ \bibinfo
  {pages} {032149} (\bibinfo {year} {2021})},\ \bibinfo {note} {publisher:
  American Physical Society}\BibitemShut {NoStop}%
\bibitem [{\citenamefont {Cover}\ and\ \citenamefont
  {Thomas}(2012)}]{cover_elements_2012}%
  \BibitemOpen
  \bibfield  {author} {\bibinfo {author} {\bibfnamefont {T.~M.}\ \bibnamefont
  {Cover}}\ and\ \bibinfo {author} {\bibfnamefont {J.~A.}\ \bibnamefont
  {Thomas}},\ }\href@noop {} {\emph {\bibinfo {title} {Elements of
  {Information} {Theory}}}}\ (\bibinfo  {publisher} {John Wiley \& Sons},\
  \bibinfo {year} {2012})\BibitemShut {NoStop}%
\bibitem [{\citenamefont {Cliff}\ \emph {et~al.}(2021)\citenamefont {Cliff},
  \citenamefont {Novelli}, \citenamefont {Fulcher}, \citenamefont {Shine},\
  and\ \citenamefont {Lizier}}]{cliff_assessing_2021}%
  \BibitemOpen
  \bibfield  {author} {\bibinfo {author} {\bibfnamefont {O.~M.}\ \bibnamefont
  {Cliff}}, \bibinfo {author} {\bibfnamefont {L.}~\bibnamefont {Novelli}},
  \bibinfo {author} {\bibfnamefont {B.~D.}\ \bibnamefont {Fulcher}}, \bibinfo
  {author} {\bibfnamefont {J.~M.}\ \bibnamefont {Shine}},\ and\ \bibinfo
  {author} {\bibfnamefont {J.~T.}\ \bibnamefont {Lizier}},\ }\bibfield  {title}
  {\bibinfo {title} {Assessing the significance of directed and multivariate
  measures of linear dependence between time series},\ }\href
  {https://doi.org/10.1103/PhysRevResearch.3.013145} {\bibfield  {journal}
  {\bibinfo  {journal} {Physical Review Research}\ }\textbf {\bibinfo {volume}
  {3}},\ \bibinfo {pages} {013145} (\bibinfo {year} {2021})}\BibitemShut
  {NoStop}%
\bibitem [{\citenamefont {Watanabe}(1960)}]{watanabe_information_1960}%
  \BibitemOpen
  \bibfield  {author} {\bibinfo {author} {\bibfnamefont {S.}~\bibnamefont
  {Watanabe}},\ }\bibfield  {title} {\bibinfo {title} {Information
  {Theoretical} {Analysis} of {Multivariate} {Correlation}},\ }\href
  {https://doi.org/10.1147/rd.41.0066} {\bibfield  {journal} {\bibinfo
  {journal} {IBM Journal of Research and Development}\ }\textbf {\bibinfo
  {volume} {4}},\ \bibinfo {pages} {66} (\bibinfo {year} {1960})},\ \bibinfo
  {note} {number: 1}\BibitemShut {NoStop}%
\bibitem [{\citenamefont {Abdallah}\ and\ \citenamefont
  {Plumbley}(2012)}]{abdallah_measure_2012}%
  \BibitemOpen
  \bibfield  {author} {\bibinfo {author} {\bibfnamefont {S.~A.}\ \bibnamefont
  {Abdallah}}\ and\ \bibinfo {author} {\bibfnamefont {M.~D.}\ \bibnamefont
  {Plumbley}},\ }\bibfield  {title} {\bibinfo {title} {A measure of statistical
  complexity based on predictive information with application to finite spin
  systems},\ }\href {https://doi.org/10.1016/j.physleta.2011.10.066} {\bibfield
   {journal} {\bibinfo  {journal} {Physics Letters A}\ }\textbf {\bibinfo
  {volume} {376}},\ \bibinfo {pages} {275} (\bibinfo {year}
  {2012})}\BibitemShut {NoStop}%
\bibitem [{\citenamefont {Rosas}\ \emph {et~al.}(2019)\citenamefont {Rosas},
  \citenamefont {Mediano}, \citenamefont {Gastpar},\ and\ \citenamefont
  {Jensen}}]{rosas_quantifying_2019}%
  \BibitemOpen
  \bibfield  {author} {\bibinfo {author} {\bibfnamefont {F.}~\bibnamefont
  {Rosas}}, \bibinfo {author} {\bibfnamefont {P.~A.~M.}\ \bibnamefont
  {Mediano}}, \bibinfo {author} {\bibfnamefont {M.}~\bibnamefont {Gastpar}},\
  and\ \bibinfo {author} {\bibfnamefont {H.~J.}\ \bibnamefont {Jensen}},\
  }\bibfield  {title} {\bibinfo {title} {Quantifying {High}-order
  {Interdependencies} via {Multivariate} {Extensions} of the {Mutual}
  {Information}},\ }\href {https://doi.org/10.1103/PhysRevE.100.032305}
  {\bibfield  {journal} {\bibinfo  {journal} {Physical Review E}\ }\textbf
  {\bibinfo {volume} {100}},\ \bibinfo {pages} {032305} (\bibinfo {year}
  {2019})},\ \bibinfo {note} {number: 3 arXiv: 1902.11239}\BibitemShut
  {NoStop}%
\bibitem [{\citenamefont {James}\ \emph {et~al.}(2011)\citenamefont {James},
  \citenamefont {Ellison},\ and\ \citenamefont
  {Crutchfield}}]{james_anatomy_2011}%
  \BibitemOpen
  \bibfield  {author} {\bibinfo {author} {\bibfnamefont {R.~G.}\ \bibnamefont
  {James}}, \bibinfo {author} {\bibfnamefont {C.~J.}\ \bibnamefont {Ellison}},\
  and\ \bibinfo {author} {\bibfnamefont {J.~P.}\ \bibnamefont {Crutchfield}},\
  }\bibfield  {title} {\bibinfo {title} {Anatomy of a bit: {Information} in a
  time series observation},\ }\href {https://doi.org/10.1063/1.3637494}
  {\bibfield  {journal} {\bibinfo  {journal} {Chaos: An Interdisciplinary
  Journal of Nonlinear Science}\ }\textbf {\bibinfo {volume} {21}},\ \bibinfo
  {pages} {037109} (\bibinfo {year} {2011})},\ \bibinfo {note} {publisher:
  American Institute of Physics}\BibitemShut {NoStop}%
\bibitem [{\citenamefont {Tononi}\ \emph {et~al.}(1994)\citenamefont {Tononi},
  \citenamefont {Sporns},\ and\ \citenamefont {Edelman}}]{tononi_measure_1994}%
  \BibitemOpen
  \bibfield  {author} {\bibinfo {author} {\bibfnamefont {G.}~\bibnamefont
  {Tononi}}, \bibinfo {author} {\bibfnamefont {O.}~\bibnamefont {Sporns}},\
  and\ \bibinfo {author} {\bibfnamefont {G.~M.}\ \bibnamefont {Edelman}},\
  }\bibfield  {title} {\bibinfo {title} {A measure for brain complexity:
  relating functional segregation and integration in the nervous system},\
  }\href {https://doi.org/10.1073/pnas.91.11.5033} {\bibfield  {journal}
  {\bibinfo  {journal} {Proceedings of the National Academy of Sciences}\
  }\textbf {\bibinfo {volume} {91}},\ \bibinfo {pages} {5033} (\bibinfo {year}
  {1994})},\ \bibinfo {note} {number: 11}\BibitemShut {NoStop}%
\bibitem [{\citenamefont {Van~Essen}\ \emph {et~al.}(2013)\citenamefont
  {Van~Essen}, \citenamefont {Smith}, \citenamefont {Barch}, \citenamefont
  {Behrens}, \citenamefont {Yacoub},\ and\ \citenamefont
  {Ugurbil}}]{van_essen_wu-minn_2013}%
  \BibitemOpen
  \bibfield  {author} {\bibinfo {author} {\bibfnamefont {D.~C.}\ \bibnamefont
  {Van~Essen}}, \bibinfo {author} {\bibfnamefont {S.~M.}\ \bibnamefont
  {Smith}}, \bibinfo {author} {\bibfnamefont {D.~M.}\ \bibnamefont {Barch}},
  \bibinfo {author} {\bibfnamefont {T.~E.~J.}\ \bibnamefont {Behrens}},
  \bibinfo {author} {\bibfnamefont {E.}~\bibnamefont {Yacoub}},\ and\ \bibinfo
  {author} {\bibfnamefont {K.}~\bibnamefont {Ugurbil}},\ }\bibfield  {title}
  {\bibinfo {title} {The {WU}-{Minn} {Human} {Connectome} {Project}: {An}
  overview},\ }\href {https://doi.org/10.1016/j.neuroimage.2013.05.041}
  {\bibfield  {journal} {\bibinfo  {journal} {NeuroImage}\ }\bibinfo {series}
  {Mapping the {Connectome}},\ \textbf {\bibinfo {volume} {80}},\ \bibinfo
  {pages} {62} (\bibinfo {year} {2013})}\BibitemShut {NoStop}%
\bibitem [{\citenamefont {Sporns}\ \emph {et~al.}(2021)\citenamefont {Sporns},
  \citenamefont {Faskowitz}, \citenamefont {Teixeira}, \citenamefont {Cutts},\
  and\ \citenamefont {Betzel}}]{sporns_dynamic_2021}%
  \BibitemOpen
  \bibfield  {author} {\bibinfo {author} {\bibfnamefont {O.}~\bibnamefont
  {Sporns}}, \bibinfo {author} {\bibfnamefont {J.}~\bibnamefont {Faskowitz}},
  \bibinfo {author} {\bibfnamefont {A.~S.}\ \bibnamefont {Teixeira}}, \bibinfo
  {author} {\bibfnamefont {S.~A.}\ \bibnamefont {Cutts}},\ and\ \bibinfo
  {author} {\bibfnamefont {R.~F.}\ \bibnamefont {Betzel}},\ }\bibfield  {title}
  {\bibinfo {title} {Dynamic expression of brain functional systems disclosed
  by fine-scale analysis of edge time series},\ }\href
  {https://doi.org/10.1162/netn_a_00182} {\bibfield  {journal} {\bibinfo
  {journal} {Network Neuroscience}\ }\textbf {\bibinfo {volume} {5}},\ \bibinfo
  {pages} {405} (\bibinfo {year} {2021})}\BibitemShut {NoStop}%
\bibitem [{\citenamefont {Onnela}\ \emph {et~al.}(2005)\citenamefont {Onnela},
  \citenamefont {Saramäki}, \citenamefont {Kertész},\ and\ \citenamefont
  {Kaski}}]{onnela_intensity_2005}%
  \BibitemOpen
  \bibfield  {author} {\bibinfo {author} {\bibfnamefont {J.-P.}\ \bibnamefont
  {Onnela}}, \bibinfo {author} {\bibfnamefont {J.}~\bibnamefont {Saramäki}},
  \bibinfo {author} {\bibfnamefont {J.}~\bibnamefont {Kertész}},\ and\
  \bibinfo {author} {\bibfnamefont {K.}~\bibnamefont {Kaski}},\ }\bibfield
  {title} {\bibinfo {title} {Intensity and coherence of motifs in weighted
  complex networks},\ }\href {https://doi.org/10.1103/PhysRevE.71.065103}
  {\bibfield  {journal} {\bibinfo  {journal} {Physical Review E}\ }\textbf
  {\bibinfo {volume} {71}},\ \bibinfo {pages} {065103} (\bibinfo {year}
  {2005})},\ \bibinfo {note} {publisher: American Physical Society}\BibitemShut
  {NoStop}%
\bibitem [{\citenamefont {Jeub}\ \emph {et~al.}(2018)\citenamefont {Jeub},
  \citenamefont {Sporns},\ and\ \citenamefont
  {Fortunato}}]{jeub_multiresolution_2018}%
  \BibitemOpen
  \bibfield  {author} {\bibinfo {author} {\bibfnamefont {L.~G.~S.}\
  \bibnamefont {Jeub}}, \bibinfo {author} {\bibfnamefont {O.}~\bibnamefont
  {Sporns}},\ and\ \bibinfo {author} {\bibfnamefont {S.}~\bibnamefont
  {Fortunato}},\ }\bibfield  {title} {\bibinfo {title} {Multiresolution
  {Consensus} {Clustering} in {Networks}},\ }\href
  {https://doi.org/10.1038/s41598-018-21352-7} {\bibfield  {journal} {\bibinfo
  {journal} {Scientific Reports}\ }\textbf {\bibinfo {volume} {8}},\ \bibinfo
  {pages} {3259} (\bibinfo {year} {2018})}\BibitemShut {NoStop}%
\bibitem [{\citenamefont {Kumar}\ \emph {et~al.}(2020)\citenamefont {Kumar},
  \citenamefont {Vaidyanathan}, \citenamefont {Ananthapadmanabhan},
  \citenamefont {Parthasarathy},\ and\ \citenamefont
  {Ravindran}}]{kumar_new_2020}%
  \BibitemOpen
  \bibfield  {author} {\bibinfo {author} {\bibfnamefont {T.}~\bibnamefont
  {Kumar}}, \bibinfo {author} {\bibfnamefont {S.}~\bibnamefont {Vaidyanathan}},
  \bibinfo {author} {\bibfnamefont {H.}~\bibnamefont {Ananthapadmanabhan}},
  \bibinfo {author} {\bibfnamefont {S.}~\bibnamefont {Parthasarathy}},\ and\
  \bibinfo {author} {\bibfnamefont {B.}~\bibnamefont {Ravindran}},\ }\bibfield
  {title} {\bibinfo {title} {A {New} {Measure} of {Modularity}
  in {Hypergraphs}: {Theoretical} {Insights} and {Implications} for
  {Effective} {Clustering}},\ }in\ \href
  {https://doi.org/10.1007/978-3-030-36687-2_24} {\emph {\bibinfo {booktitle}
  {Complex {Networks} and {Their} {Applications} {VIII}}}},\ \bibinfo {series
  and number} {Studies in {Computational} {Intelligence}},\ \bibinfo {editor}
  {edited by\ \bibinfo {editor} {\bibfnamefont {H.}~\bibnamefont {Cherifi}},
  \bibinfo {editor} {\bibfnamefont {S.}~\bibnamefont {Gaito}}, \bibinfo
  {editor} {\bibfnamefont {J.~F.}\ \bibnamefont {Mendes}}, \bibinfo {editor}
  {\bibfnamefont {E.}~\bibnamefont {Moro}},\ and\ \bibinfo {editor}
  {\bibfnamefont {L.~M.}\ \bibnamefont {Rocha}}}\ (\bibinfo  {publisher}
  {Springer International Publishing},\ \bibinfo {address} {Cham},\ \bibinfo
  {year} {2020})\ pp.\ \bibinfo {pages} {286--297}\BibitemShut {NoStop}%
\bibitem [{\citenamefont {Yeo}\ \emph {et~al.}(2011)\citenamefont {Yeo},
  \citenamefont {Krienen}, \citenamefont {Sepulcre}, \citenamefont {Sabuncu},
  \citenamefont {Lashkari}, \citenamefont {Hollinshead}, \citenamefont
  {Roffman}, \citenamefont {Smoller}, \citenamefont {Zöllei}, \citenamefont
  {Polimeni}, \citenamefont {Fischl}, \citenamefont {Liu},\ and\ \citenamefont
  {Buckner}}]{yeo_organization_2011}%
  \BibitemOpen
  \bibfield  {author} {\bibinfo {author} {\bibfnamefont {B.~T.}\ \bibnamefont
  {Yeo}}, \bibinfo {author} {\bibfnamefont {F.~M.}\ \bibnamefont {Krienen}},
  \bibinfo {author} {\bibfnamefont {J.}~\bibnamefont {Sepulcre}}, \bibinfo
  {author} {\bibfnamefont {M.~R.}\ \bibnamefont {Sabuncu}}, \bibinfo {author}
  {\bibfnamefont {D.}~\bibnamefont {Lashkari}}, \bibinfo {author}
  {\bibfnamefont {M.}~\bibnamefont {Hollinshead}}, \bibinfo {author}
  {\bibfnamefont {J.~L.}\ \bibnamefont {Roffman}}, \bibinfo {author}
  {\bibfnamefont {J.~W.}\ \bibnamefont {Smoller}}, \bibinfo {author}
  {\bibfnamefont {L.}~\bibnamefont {Zöllei}}, \bibinfo {author} {\bibfnamefont
  {J.~R.}\ \bibnamefont {Polimeni}}, \bibinfo {author} {\bibfnamefont
  {B.}~\bibnamefont {Fischl}}, \bibinfo {author} {\bibfnamefont
  {H.}~\bibnamefont {Liu}},\ and\ \bibinfo {author} {\bibfnamefont {R.~L.}\
  \bibnamefont {Buckner}},\ }\bibfield  {title} {\bibinfo {title} {The
  organization of the human cerebral cortex estimated by intrinsic functional
  connectivity},\ }\href {https://doi.org/10.1152/jn.00338.2011} {\bibfield
  {journal} {\bibinfo  {journal} {Journal of Neurophysiology}\ }\textbf
  {\bibinfo {volume} {106}},\ \bibinfo {pages} {1125} (\bibinfo {year}
  {2011})},\ \bibinfo {note} {number: 3}\BibitemShut {NoStop}%
\bibitem [{\citenamefont {Luppi}\ \emph {et~al.}(2022)\citenamefont {Luppi},
  \citenamefont {Mediano}, \citenamefont {Rosas}, \citenamefont {Holland},
  \citenamefont {Fryer}, \citenamefont {O’Brien}, \citenamefont {Rowe},
  \citenamefont {Menon}, \citenamefont {Bor},\ and\ \citenamefont
  {Stamatakis}}]{luppi_synergistic_2022}%
  \BibitemOpen
  \bibfield  {author} {\bibinfo {author} {\bibfnamefont {A.~I.}\ \bibnamefont
  {Luppi}}, \bibinfo {author} {\bibfnamefont {P.~A.~M.}\ \bibnamefont
  {Mediano}}, \bibinfo {author} {\bibfnamefont {F.~E.}\ \bibnamefont {Rosas}},
  \bibinfo {author} {\bibfnamefont {N.}~\bibnamefont {Holland}}, \bibinfo
  {author} {\bibfnamefont {T.~D.}\ \bibnamefont {Fryer}}, \bibinfo {author}
  {\bibfnamefont {J.~T.}\ \bibnamefont {O’Brien}}, \bibinfo {author}
  {\bibfnamefont {J.~B.}\ \bibnamefont {Rowe}}, \bibinfo {author}
  {\bibfnamefont {D.~K.}\ \bibnamefont {Menon}}, \bibinfo {author}
  {\bibfnamefont {D.}~\bibnamefont {Bor}},\ and\ \bibinfo {author}
  {\bibfnamefont {E.~A.}\ \bibnamefont {Stamatakis}},\ }\bibfield  {title}
  {\bibinfo {title} {A synergistic core for human brain evolution and
  cognition},\ }\href {https://doi.org/10.1038/s41593-022-01070-0} {\bibfield
  {journal} {\bibinfo  {journal} {Nature Neuroscience}\ ,\ \bibinfo {pages}
  {1}} (\bibinfo {year} {2022})},\ \bibinfo {note} {publisher: Nature
  Publishing Group}\BibitemShut {NoStop}%
\bibitem [{\citenamefont {Esfahlani}\ \emph {et~al.}(2020)\citenamefont
  {Esfahlani}, \citenamefont {Jo}, \citenamefont {Faskowitz}, \citenamefont
  {Byrge}, \citenamefont {Kennedy}, \citenamefont {Sporns},\ and\ \citenamefont
  {Betzel}}]{esfahlani_high-amplitude_2020}%
  \BibitemOpen
  \bibfield  {author} {\bibinfo {author} {\bibfnamefont {F.~Z.}\ \bibnamefont
  {Esfahlani}}, \bibinfo {author} {\bibfnamefont {Y.}~\bibnamefont {Jo}},
  \bibinfo {author} {\bibfnamefont {J.}~\bibnamefont {Faskowitz}}, \bibinfo
  {author} {\bibfnamefont {L.}~\bibnamefont {Byrge}}, \bibinfo {author}
  {\bibfnamefont {D.~P.}\ \bibnamefont {Kennedy}}, \bibinfo {author}
  {\bibfnamefont {O.}~\bibnamefont {Sporns}},\ and\ \bibinfo {author}
  {\bibfnamefont {R.~F.}\ \bibnamefont {Betzel}},\ }\bibfield  {title}
  {\bibinfo {title} {High-amplitude cofluctuations in cortical activity drive
  functional connectivity},\ }\href {https://doi.org/10.1073/pnas.2005531117}
  {\bibfield  {journal} {\bibinfo  {journal} {Proceedings of the National
  Academy of Sciences}\ }\textbf {\bibinfo {volume} {117}},\ \bibinfo {pages}
  {28393} (\bibinfo {year} {2020})}\BibitemShut {NoStop}%
\bibitem [{\citenamefont {Betzel}\ \emph
  {et~al.}(2022{\natexlab{a}})\citenamefont {Betzel}, \citenamefont {Cutts},
  \citenamefont {Tanner}, \citenamefont {Greenwell}, \citenamefont {Varley},
  \citenamefont {Faskowitz},\ and\ \citenamefont
  {Sporns}}]{betzel_hierarchical_2022}%
  \BibitemOpen
  \bibfield  {author} {\bibinfo {author} {\bibfnamefont {R.}~\bibnamefont
  {Betzel}}, \bibinfo {author} {\bibfnamefont {S.}~\bibnamefont {Cutts}},
  \bibinfo {author} {\bibfnamefont {J.}~\bibnamefont {Tanner}}, \bibinfo
  {author} {\bibfnamefont {S.}~\bibnamefont {Greenwell}}, \bibinfo {author}
  {\bibfnamefont {T.}~\bibnamefont {Varley}}, \bibinfo {author} {\bibfnamefont
  {J.}~\bibnamefont {Faskowitz}},\ and\ \bibinfo {author} {\bibfnamefont
  {O.}~\bibnamefont {Sporns}},\ }\bibfield  {title} {\bibinfo {title}
  {Hierarchical organization of spontaneous co-fluctuations in densely-sampled
  individuals using {fMRI}},\ }\href@noop {} {\bibfield  {journal} {\bibinfo
  {journal} {bioRxiv}\ } (\bibinfo {year} {2022}{\natexlab{a}})},\ \bibinfo
  {note} {publisher: Cold Spring Harbor Laboratory}\BibitemShut {NoStop}%
\bibitem [{\citenamefont {Tanner}\ \emph {et~al.}(2022)\citenamefont {Tanner},
  \citenamefont {Faskowitz}, \citenamefont {Byrge}, \citenamefont {Kennedy},
  \citenamefont {Sporns},\ and\ \citenamefont
  {Betzel}}]{tanner_synchronous_2022}%
  \BibitemOpen
  \bibfield  {author} {\bibinfo {author} {\bibfnamefont {J.~C.}\ \bibnamefont
  {Tanner}}, \bibinfo {author} {\bibfnamefont {J.}~\bibnamefont {Faskowitz}},
  \bibinfo {author} {\bibfnamefont {L.}~\bibnamefont {Byrge}}, \bibinfo
  {author} {\bibfnamefont {D.~P.}\ \bibnamefont {Kennedy}}, \bibinfo {author}
  {\bibfnamefont {O.}~\bibnamefont {Sporns}},\ and\ \bibinfo {author}
  {\bibfnamefont {R.~F.}\ \bibnamefont {Betzel}},\ }\href
  {https://doi.org/10.1101/2022.06.30.497603} {\bibinfo {title} {Synchronous
  high-amplitude co-fluctuations of functional brain networks during
  movie-watching}} (\bibinfo {year} {2022}),\ \bibinfo {note} {pages:
  2022.06.30.497603 Section: New Results}\BibitemShut {NoStop}%
\bibitem [{\citenamefont {Pope}\ \emph {et~al.}(2021)\citenamefont {Pope},
  \citenamefont {Fukushima}, \citenamefont {Betzel},\ and\ \citenamefont
  {Sporns}}]{pope_modular_2021}%
  \BibitemOpen
  \bibfield  {author} {\bibinfo {author} {\bibfnamefont {M.}~\bibnamefont
  {Pope}}, \bibinfo {author} {\bibfnamefont {M.}~\bibnamefont {Fukushima}},
  \bibinfo {author} {\bibfnamefont {R.~F.}\ \bibnamefont {Betzel}},\ and\
  \bibinfo {author} {\bibfnamefont {O.}~\bibnamefont {Sporns}},\ }\bibfield
  {title} {\bibinfo {title} {Modular origins of high-amplitude cofluctuations
  in fine-scale functional connectivity dynamics},\ }\bibfield  {journal}
  {\bibinfo  {journal} {Proceedings of the National Academy of Sciences}\
  }\textbf {\bibinfo {volume} {118}},\ \href
  {https://doi.org/10.1073/pnas.2109380118} {10.1073/pnas.2109380118} (\bibinfo
  {year} {2021}),\ \bibinfo {note} {publisher: National Academy of Sciences
  Section: Biological Sciences}\BibitemShut {NoStop}%
\bibitem [{\citenamefont {Betzel}\ \emph
  {et~al.}(2022{\natexlab{b}})\citenamefont {Betzel}, \citenamefont {Cutts},
  \citenamefont {Greenwell}, \citenamefont {Faskowitz},\ and\ \citenamefont
  {Sporns}}]{betzel_individualized_2022}%
  \BibitemOpen
  \bibfield  {author} {\bibinfo {author} {\bibfnamefont {R.~F.}\ \bibnamefont
  {Betzel}}, \bibinfo {author} {\bibfnamefont {S.~A.}\ \bibnamefont {Cutts}},
  \bibinfo {author} {\bibfnamefont {S.}~\bibnamefont {Greenwell}}, \bibinfo
  {author} {\bibfnamefont {J.}~\bibnamefont {Faskowitz}},\ and\ \bibinfo
  {author} {\bibfnamefont {O.}~\bibnamefont {Sporns}},\ }\bibfield  {title}
  {\bibinfo {title} {Individualized event structure drives individual
  differences in whole-brain functional connectivity},\ }\href
  {https://doi.org/10.1016/j.neuroimage.2022.118993} {\bibfield  {journal}
  {\bibinfo  {journal} {NeuroImage}\ }\textbf {\bibinfo {volume} {252}},\
  \bibinfo {pages} {118993} (\bibinfo {year} {2022}{\natexlab{b}})}\BibitemShut
  {NoStop}%
\bibitem [{\citenamefont {Varley}\ and\ \citenamefont
  {Sporns}(2022)}]{varley_network_2022}%
  \BibitemOpen
  \bibfield  {author} {\bibinfo {author} {\bibfnamefont {T.~F.}\ \bibnamefont
  {Varley}}\ and\ \bibinfo {author} {\bibfnamefont {O.}~\bibnamefont
  {Sporns}},\ }\bibfield  {title} {\bibinfo {title} {Network {Analysis} of
  {Time} {Series}: {Novel} {Approaches} to {Network} {Neuroscience}},\ }\href
  {https://www.frontiersin.org/article/10.3389/fnins.2021.787068} {\bibfield
  {journal} {\bibinfo  {journal} {Frontiers in Neuroscience}\ }\textbf
  {\bibinfo {volume} {15}} (\bibinfo {year} {2022})}\BibitemShut {NoStop}%
\bibitem [{\citenamefont {Hoel}\ \emph {et~al.}(2013)\citenamefont {Hoel},
  \citenamefont {Albantakis},\ and\ \citenamefont
  {Tononi}}]{hoel_quantifying_2013}%
  \BibitemOpen
  \bibfield  {author} {\bibinfo {author} {\bibfnamefont {E.~P.}\ \bibnamefont
  {Hoel}}, \bibinfo {author} {\bibfnamefont {L.}~\bibnamefont {Albantakis}},\
  and\ \bibinfo {author} {\bibfnamefont {G.}~\bibnamefont {Tononi}},\
  }\bibfield  {title} {\bibinfo {title} {Quantifying causal emergence shows
  that macro can beat micro},\ }\href {https://doi.org/10.1073/pnas.1314922110}
  {\bibfield  {journal} {\bibinfo  {journal} {Proceedings of the National
  Academy of Sciences}\ }\textbf {\bibinfo {volume} {110}},\ \bibinfo {pages}
  {19790} (\bibinfo {year} {2013})},\ \bibinfo {note} {number: 49}\BibitemShut
  {NoStop}%
\bibitem [{\citenamefont {Varley}\ \emph {et~al.}(2021)\citenamefont {Varley},
  \citenamefont {Denny}, \citenamefont {Sporns},\ and\ \citenamefont
  {Patania}}]{varley_topological_2021}%
  \BibitemOpen
  \bibfield  {author} {\bibinfo {author} {\bibfnamefont {T.~F.}\ \bibnamefont
  {Varley}}, \bibinfo {author} {\bibfnamefont {V.}~\bibnamefont {Denny}},
  \bibinfo {author} {\bibfnamefont {O.}~\bibnamefont {Sporns}},\ and\ \bibinfo
  {author} {\bibfnamefont {A.}~\bibnamefont {Patania}},\ }\bibfield  {title}
  {\bibinfo {title} {Topological analysis of differential effects of ketamine
  and propofol anaesthesia on brain dynamics},\ }\href
  {https://doi.org/10.1098/rsos.201971} {\bibfield  {journal} {\bibinfo
  {journal} {Royal Society Open Science}\ }\textbf {\bibinfo {volume} {8}},\
  \bibinfo {pages} {201971} (\bibinfo {year} {2021})},\ \bibinfo {note}
  {publisher: Royal Society}\BibitemShut {NoStop}%
\bibitem [{\citenamefont {Zalesky}\ \emph {et~al.}(2012)\citenamefont
  {Zalesky}, \citenamefont {Fornito},\ and\ \citenamefont
  {Bullmore}}]{zalesky_use_2012}%
  \BibitemOpen
  \bibfield  {author} {\bibinfo {author} {\bibfnamefont {A.}~\bibnamefont
  {Zalesky}}, \bibinfo {author} {\bibfnamefont {A.}~\bibnamefont {Fornito}},\
  and\ \bibinfo {author} {\bibfnamefont {E.}~\bibnamefont {Bullmore}},\
  }\bibfield  {title} {\bibinfo {title} {On the use of correlation as a measure
  of network connectivity},\ }\href
  {https://doi.org/10.1016/j.neuroimage.2012.02.001} {\bibfield  {journal}
  {\bibinfo  {journal} {NeuroImage}\ }\textbf {\bibinfo {volume} {60}},\
  \bibinfo {pages} {2096} (\bibinfo {year} {2012})}\BibitemShut {NoStop}%
\bibitem [{\citenamefont {Langford}\ \emph {et~al.}(2001)\citenamefont
  {Langford}, \citenamefont {Schwertman},\ and\ \citenamefont
  {Owens}}]{langford_is_2001}%
  \BibitemOpen
  \bibfield  {author} {\bibinfo {author} {\bibfnamefont {E.}~\bibnamefont
  {Langford}}, \bibinfo {author} {\bibfnamefont {N.}~\bibnamefont
  {Schwertman}},\ and\ \bibinfo {author} {\bibfnamefont {M.}~\bibnamefont
  {Owens}},\ }\bibfield  {title} {\bibinfo {title} {Is the {Property} of
  {Being} {Positively} {Correlated} {Transitive}?},\ }\href
  {https://doi.org/10.1198/000313001753272286} {\bibfield  {journal} {\bibinfo
  {journal} {The American Statistician}\ }\textbf {\bibinfo {volume} {55}},\
  \bibinfo {pages} {322} (\bibinfo {year} {2001})},\ \bibinfo {note}
  {publisher: Taylor \& Francis \_eprint:
  https://doi.org/10.1198/000313001753272286}\BibitemShut {NoStop}%
\bibitem [{\citenamefont {Cutts}\ \emph {et~al.}(2022)\citenamefont {Cutts},
  \citenamefont {Faskowitz}, \citenamefont {Betzel},\ and\ \citenamefont
  {Sporns}}]{cutts_uncovering_2022}%
  \BibitemOpen
  \bibfield  {author} {\bibinfo {author} {\bibfnamefont {S.~A.}\ \bibnamefont
  {Cutts}}, \bibinfo {author} {\bibfnamefont {J.}~\bibnamefont {Faskowitz}},
  \bibinfo {author} {\bibfnamefont {R.~F.}\ \bibnamefont {Betzel}},\ and\
  \bibinfo {author} {\bibfnamefont {O.}~\bibnamefont {Sporns}},\ }\bibfield
  {title} {\bibinfo {title} {Uncovering individual differences in fine-scale
  dynamics of functional connectivity},\ }\href
  {https://doi.org/10.1093/cercor/bhac214} {\bibfield  {journal} {\bibinfo
  {journal} {Cerebral Cortex}\ ,\ \bibinfo {pages} {bhac214}} (\bibinfo {year}
  {2022})}\BibitemShut {NoStop}%
\bibitem [{\citenamefont {Jo}\ \emph {et~al.}(2021)\citenamefont {Jo},
  \citenamefont {Faskowitz}, \citenamefont {Esfahlani}, \citenamefont
  {Sporns},\ and\ \citenamefont {Betzel}}]{jo_subject_2021}%
  \BibitemOpen
  \bibfield  {author} {\bibinfo {author} {\bibfnamefont {Y.}~\bibnamefont
  {Jo}}, \bibinfo {author} {\bibfnamefont {J.}~\bibnamefont {Faskowitz}},
  \bibinfo {author} {\bibfnamefont {F.~Z.}\ \bibnamefont {Esfahlani}}, \bibinfo
  {author} {\bibfnamefont {O.}~\bibnamefont {Sporns}},\ and\ \bibinfo {author}
  {\bibfnamefont {R.~F.}\ \bibnamefont {Betzel}},\ }\bibfield  {title}
  {\bibinfo {title} {Subject identification using edge-centric functional
  connectivity},\ }\href {https://doi.org/10.1016/j.neuroimage.2021.118204}
  {\bibfield  {journal} {\bibinfo  {journal} {NeuroImage}\ }\textbf {\bibinfo
  {volume} {238}},\ \bibinfo {pages} {118204} (\bibinfo {year}
  {2021})}\BibitemShut {NoStop}%
\bibitem [{\citenamefont {Krakauer}\ \emph {et~al.}(2020)\citenamefont
  {Krakauer}, \citenamefont {Bertschinger}, \citenamefont {Olbrich},
  \citenamefont {Flack},\ and\ \citenamefont {Ay}}]{krakauer_information_2020}%
  \BibitemOpen
  \bibfield  {author} {\bibinfo {author} {\bibfnamefont {D.}~\bibnamefont
  {Krakauer}}, \bibinfo {author} {\bibfnamefont {N.}~\bibnamefont
  {Bertschinger}}, \bibinfo {author} {\bibfnamefont {E.}~\bibnamefont
  {Olbrich}}, \bibinfo {author} {\bibfnamefont {J.~C.}\ \bibnamefont {Flack}},\
  and\ \bibinfo {author} {\bibfnamefont {N.}~\bibnamefont {Ay}},\ }\bibfield
  {title} {\bibinfo {title} {The information theory of individuality},\ }\href
  {https://doi.org/10.1007/s12064-020-00313-7} {\bibfield  {journal} {\bibinfo
  {journal} {Theory in Biosciences}\ }\textbf {\bibinfo {volume} {139}},\
  \bibinfo {pages} {209} (\bibinfo {year} {2020})},\ \bibinfo {note} {number:
  2}\BibitemShut {NoStop}%
\bibitem [{\citenamefont {Barttfeld}\ \emph {et~al.}(2015)\citenamefont
  {Barttfeld}, \citenamefont {Uhrig}, \citenamefont {Sitt}, \citenamefont
  {Sigman}, \citenamefont {Jarraya},\ and\ \citenamefont
  {Dehaene}}]{barttfeld_signature_2015}%
  \BibitemOpen
  \bibfield  {author} {\bibinfo {author} {\bibfnamefont {P.}~\bibnamefont
  {Barttfeld}}, \bibinfo {author} {\bibfnamefont {L.}~\bibnamefont {Uhrig}},
  \bibinfo {author} {\bibfnamefont {J.~D.}\ \bibnamefont {Sitt}}, \bibinfo
  {author} {\bibfnamefont {M.}~\bibnamefont {Sigman}}, \bibinfo {author}
  {\bibfnamefont {B.}~\bibnamefont {Jarraya}},\ and\ \bibinfo {author}
  {\bibfnamefont {S.}~\bibnamefont {Dehaene}},\ }\bibfield  {title} {\bibinfo
  {title} {Signature of consciousness in the dynamics of resting-state brain
  activity},\ }\href {https://doi.org/10.1073/pnas.1418031112} {\bibfield
  {journal} {\bibinfo  {journal} {Proceedings of the National Academy of
  Sciences}\ }\textbf {\bibinfo {volume} {112}},\ \bibinfo {pages} {887}
  (\bibinfo {year} {2015})},\ \bibinfo {note} {publisher: National Academy of
  Sciences Section: Biological Sciences}\BibitemShut {NoStop}%
\bibitem [{\citenamefont {Demertzi}\ \emph {et~al.}(2019)\citenamefont
  {Demertzi}, \citenamefont {Tagliazucchi}, \citenamefont {Dehaene},
  \citenamefont {Deco}, \citenamefont {Barttfeld}, \citenamefont {Raimondo},
  \citenamefont {Martial}, \citenamefont {Fernández-Espejo}, \citenamefont
  {Rohaut}, \citenamefont {Voss}, \citenamefont {Schiff}, \citenamefont {Owen},
  \citenamefont {Laureys}, \citenamefont {Naccache},\ and\ \citenamefont
  {Sitt}}]{demertzi_human_2019}%
  \BibitemOpen
  \bibfield  {author} {\bibinfo {author} {\bibfnamefont {A.}~\bibnamefont
  {Demertzi}}, \bibinfo {author} {\bibfnamefont {E.}~\bibnamefont
  {Tagliazucchi}}, \bibinfo {author} {\bibfnamefont {S.}~\bibnamefont
  {Dehaene}}, \bibinfo {author} {\bibfnamefont {G.}~\bibnamefont {Deco}},
  \bibinfo {author} {\bibfnamefont {P.}~\bibnamefont {Barttfeld}}, \bibinfo
  {author} {\bibfnamefont {F.}~\bibnamefont {Raimondo}}, \bibinfo {author}
  {\bibfnamefont {C.}~\bibnamefont {Martial}}, \bibinfo {author} {\bibfnamefont
  {D.}~\bibnamefont {Fernández-Espejo}}, \bibinfo {author} {\bibfnamefont
  {B.}~\bibnamefont {Rohaut}}, \bibinfo {author} {\bibfnamefont {H.~U.}\
  \bibnamefont {Voss}}, \bibinfo {author} {\bibfnamefont {N.~D.}\ \bibnamefont
  {Schiff}}, \bibinfo {author} {\bibfnamefont {A.~M.}\ \bibnamefont {Owen}},
  \bibinfo {author} {\bibfnamefont {S.}~\bibnamefont {Laureys}}, \bibinfo
  {author} {\bibfnamefont {L.}~\bibnamefont {Naccache}},\ and\ \bibinfo
  {author} {\bibfnamefont {J.~D.}\ \bibnamefont {Sitt}},\ }\bibfield  {title}
  {\bibinfo {title} {Human consciousness is supported by dynamic complex
  patterns of brain signal coordination},\ }\href
  {https://doi.org/10.1126/sciadv.aat7603} {\bibfield  {journal} {\bibinfo
  {journal} {Science Advances}\ }\textbf {\bibinfo {volume} {5}},\ \bibinfo
  {pages} {eaat7603} (\bibinfo {year} {2019})},\ \bibinfo {note} {number:
  2}\BibitemShut {NoStop}%
\bibitem [{\citenamefont {Shine}\ \emph {et~al.}(2016)\citenamefont {Shine},
  \citenamefont {Bissett}, \citenamefont {Bell}, \citenamefont {Koyejo},
  \citenamefont {Balsters}, \citenamefont {Gorgolewski}, \citenamefont
  {Moodie},\ and\ \citenamefont {Poldrack}}]{shine_dynamics_2016}%
  \BibitemOpen
  \bibfield  {author} {\bibinfo {author} {\bibfnamefont {J.~M.}\ \bibnamefont
  {Shine}}, \bibinfo {author} {\bibfnamefont {P.~G.}\ \bibnamefont {Bissett}},
  \bibinfo {author} {\bibfnamefont {P.~T.}\ \bibnamefont {Bell}}, \bibinfo
  {author} {\bibfnamefont {O.}~\bibnamefont {Koyejo}}, \bibinfo {author}
  {\bibfnamefont {J.~H.}\ \bibnamefont {Balsters}}, \bibinfo {author}
  {\bibfnamefont {K.~J.}\ \bibnamefont {Gorgolewski}}, \bibinfo {author}
  {\bibfnamefont {C.~A.}\ \bibnamefont {Moodie}},\ and\ \bibinfo {author}
  {\bibfnamefont {R.~A.}\ \bibnamefont {Poldrack}},\ }\bibfield  {title}
  {\bibinfo {title} {The {Dynamics} of {Functional} {Brain} {Networks}:
  {Integrated} {Network} {States} during {Cognitive} {Task} {Performance}},\
  }\href {https://doi.org/10.1016/j.neuron.2016.09.018} {\bibfield  {journal}
  {\bibinfo  {journal} {Neuron}\ }\textbf {\bibinfo {volume} {92}},\ \bibinfo
  {pages} {544} (\bibinfo {year} {2016})}\BibitemShut {NoStop}%
\bibitem [{\citenamefont {Ahmed}\ \emph {et~al.}(2016)\citenamefont {Ahmed},
  \citenamefont {Devenney}, \citenamefont {Irish}, \citenamefont {Ittner},
  \citenamefont {Naismith}, \citenamefont {Ittner}, \citenamefont {Rohrer},
  \citenamefont {Halliday}, \citenamefont {Eisen}, \citenamefont {Hodges},\
  and\ \citenamefont {Kiernan}}]{ahmed_neuronal_2016}%
  \BibitemOpen
  \bibfield  {author} {\bibinfo {author} {\bibfnamefont {R.~M.}\ \bibnamefont
  {Ahmed}}, \bibinfo {author} {\bibfnamefont {E.~M.}\ \bibnamefont {Devenney}},
  \bibinfo {author} {\bibfnamefont {M.}~\bibnamefont {Irish}}, \bibinfo
  {author} {\bibfnamefont {A.}~\bibnamefont {Ittner}}, \bibinfo {author}
  {\bibfnamefont {S.}~\bibnamefont {Naismith}}, \bibinfo {author}
  {\bibfnamefont {L.~M.}\ \bibnamefont {Ittner}}, \bibinfo {author}
  {\bibfnamefont {J.~D.}\ \bibnamefont {Rohrer}}, \bibinfo {author}
  {\bibfnamefont {G.~M.}\ \bibnamefont {Halliday}}, \bibinfo {author}
  {\bibfnamefont {A.}~\bibnamefont {Eisen}}, \bibinfo {author} {\bibfnamefont
  {J.~R.}\ \bibnamefont {Hodges}},\ and\ \bibinfo {author} {\bibfnamefont
  {M.~C.}\ \bibnamefont {Kiernan}},\ }\bibfield  {title} {\bibinfo {title}
  {Neuronal network disintegration: common pathways linking neurodegenerative
  diseases},\ }\href {https://doi.org/10.1136/jnnp-2014-308350} {\bibfield
  {journal} {\bibinfo  {journal} {Journal of Neurology, Neurosurgery \&
  Psychiatry}\ }\textbf {\bibinfo {volume} {87}},\ \bibinfo {pages} {1234}
  (\bibinfo {year} {2016})},\ \bibinfo {note} {publisher: BMJ Publishing Group
  Ltd Section: Neurodegeneration}\BibitemShut {NoStop}%
\bibitem [{\citenamefont {Damoiseaux}\ \emph {et~al.}(2012)\citenamefont
  {Damoiseaux}, \citenamefont {Prater}, \citenamefont {Miller},\ and\
  \citenamefont {Greicius}}]{damoiseaux_functional_2012}%
  \BibitemOpen
  \bibfield  {author} {\bibinfo {author} {\bibfnamefont {J.~S.}\ \bibnamefont
  {Damoiseaux}}, \bibinfo {author} {\bibfnamefont {K.~E.}\ \bibnamefont
  {Prater}}, \bibinfo {author} {\bibfnamefont {B.~L.}\ \bibnamefont {Miller}},\
  and\ \bibinfo {author} {\bibfnamefont {M.~D.}\ \bibnamefont {Greicius}},\
  }\bibfield  {title} {\bibinfo {title} {Functional connectivity tracks
  clinical deterioration in {Alzheimer}'s disease},\ }\href
  {https://doi.org/10.1016/j.neurobiolaging.2011.06.024} {\bibfield  {journal}
  {\bibinfo  {journal} {Neurobiology of Aging}\ }\textbf {\bibinfo {volume}
  {33}},\ \bibinfo {pages} {828.e19} (\bibinfo {year} {2012})},\ \bibinfo
  {note} {number: 4}\BibitemShut {NoStop}%
\bibitem [{\citenamefont {Luppi}\ \emph {et~al.}(2019)\citenamefont {Luppi},
  \citenamefont {Craig}, \citenamefont {Pappas}, \citenamefont {Finoia},
  \citenamefont {Williams}, \citenamefont {Allanson}, \citenamefont {Pickard},
  \citenamefont {Owen}, \citenamefont {Naci}, \citenamefont {Menon},\ and\
  \citenamefont {Stamatakis}}]{luppi_consciousness-specific_2019}%
  \BibitemOpen
  \bibfield  {author} {\bibinfo {author} {\bibfnamefont {A.~I.}\ \bibnamefont
  {Luppi}}, \bibinfo {author} {\bibfnamefont {M.~M.}\ \bibnamefont {Craig}},
  \bibinfo {author} {\bibfnamefont {I.}~\bibnamefont {Pappas}}, \bibinfo
  {author} {\bibfnamefont {P.}~\bibnamefont {Finoia}}, \bibinfo {author}
  {\bibfnamefont {G.~B.}\ \bibnamefont {Williams}}, \bibinfo {author}
  {\bibfnamefont {J.}~\bibnamefont {Allanson}}, \bibinfo {author}
  {\bibfnamefont {J.~D.}\ \bibnamefont {Pickard}}, \bibinfo {author}
  {\bibfnamefont {A.~M.}\ \bibnamefont {Owen}}, \bibinfo {author}
  {\bibfnamefont {L.}~\bibnamefont {Naci}}, \bibinfo {author} {\bibfnamefont
  {D.~K.}\ \bibnamefont {Menon}},\ and\ \bibinfo {author} {\bibfnamefont
  {E.~A.}\ \bibnamefont {Stamatakis}},\ }\bibfield  {title} {\bibinfo {title}
  {Consciousness-specific dynamic interactions of brain integration and
  functional diversity},\ }\href {https://doi.org/10.1038/s41467-019-12658-9}
  {\bibfield  {journal} {\bibinfo  {journal} {Nature Communications}\ }\textbf
  {\bibinfo {volume} {10}},\ \bibinfo {pages} {1} (\bibinfo {year} {2019})},\
  \bibinfo {note} {number: 1}\BibitemShut {NoStop}%
\bibitem [{\citenamefont {Zamani~Esfahlani}\ \emph {et~al.}(2022)\citenamefont
  {Zamani~Esfahlani}, \citenamefont {Byrge}, \citenamefont {Tanner},
  \citenamefont {Sporns}, \citenamefont {Kennedy},\ and\ \citenamefont
  {Betzel}}]{zamani_esfahlani_edge-centric_2022}%
  \BibitemOpen
  \bibfield  {author} {\bibinfo {author} {\bibfnamefont {F.}~\bibnamefont
  {Zamani~Esfahlani}}, \bibinfo {author} {\bibfnamefont {L.}~\bibnamefont
  {Byrge}}, \bibinfo {author} {\bibfnamefont {J.}~\bibnamefont {Tanner}},
  \bibinfo {author} {\bibfnamefont {O.}~\bibnamefont {Sporns}}, \bibinfo
  {author} {\bibfnamefont {D.~P.}\ \bibnamefont {Kennedy}},\ and\ \bibinfo
  {author} {\bibfnamefont {R.~F.}\ \bibnamefont {Betzel}},\ }\bibfield  {title}
  {\bibinfo {title} {Edge-centric analysis of time-varying functional brain
  networks with applications in autism spectrum disorder},\ }\href
  {https://doi.org/10.1016/j.neuroimage.2022.119591} {\bibfield  {journal}
  {\bibinfo  {journal} {NeuroImage}\ }\textbf {\bibinfo {volume} {263}},\
  \bibinfo {pages} {119591} (\bibinfo {year} {2022})}\BibitemShut {NoStop}%
\bibitem [{\citenamefont {Novelli}\ \emph {et~al.}(2019)\citenamefont
  {Novelli}, \citenamefont {Wollstadt}, \citenamefont {Mediano}, \citenamefont
  {Wibral},\ and\ \citenamefont {Lizier}}]{novelli_large-scale_2019}%
  \BibitemOpen
  \bibfield  {author} {\bibinfo {author} {\bibfnamefont {L.}~\bibnamefont
  {Novelli}}, \bibinfo {author} {\bibfnamefont {P.}~\bibnamefont {Wollstadt}},
  \bibinfo {author} {\bibfnamefont {P.}~\bibnamefont {Mediano}}, \bibinfo
  {author} {\bibfnamefont {M.}~\bibnamefont {Wibral}},\ and\ \bibinfo {author}
  {\bibfnamefont {J.~T.}\ \bibnamefont {Lizier}},\ }\bibfield  {title}
  {\bibinfo {title} {Large-scale directed network inference with multivariate
  transfer entropy and hierarchical statistical testing},\ }\href
  {https://doi.org/10.1162/netn_a_00092} {\bibfield  {journal} {\bibinfo
  {journal} {Network Neuroscience}\ }\textbf {\bibinfo {volume} {3}},\ \bibinfo
  {pages} {827} (\bibinfo {year} {2019})},\ \bibinfo {note} {number:
  3}\BibitemShut {NoStop}%
\bibitem [{\citenamefont {Wollstadt}\ \emph {et~al.}(2021)\citenamefont
  {Wollstadt}, \citenamefont {Schmitt},\ and\ \citenamefont
  {Wibral}}]{wollstadt_rigorous_2021}%
  \BibitemOpen
  \bibfield  {author} {\bibinfo {author} {\bibfnamefont {P.}~\bibnamefont
  {Wollstadt}}, \bibinfo {author} {\bibfnamefont {S.}~\bibnamefont {Schmitt}},\
  and\ \bibinfo {author} {\bibfnamefont {M.}~\bibnamefont {Wibral}},\
  }\bibfield  {title} {\bibinfo {title} {A {Rigorous} {Information}-{Theoretic}
  {Definition} of {Redundancy} and {Relevancy} in {Feature} {Selection} {Based}
  on ({Partial}) {Information} {Decomposition}},\ }\href
  {http://arxiv.org/abs/2105.04187} {\bibfield  {journal} {\bibinfo  {journal}
  {arXiv:2105.04187 [cs, math]}\ } (\bibinfo {year} {2021})},\ \bibinfo {note}
  {arXiv: 2105.04187}\BibitemShut {NoStop}%
\bibitem [{\citenamefont {Varley}\ and\ \citenamefont
  {Hoel}(2022)}]{varley_emergence_2022}%
  \BibitemOpen
  \bibfield  {author} {\bibinfo {author} {\bibfnamefont {T.~F.}\ \bibnamefont
  {Varley}}\ and\ \bibinfo {author} {\bibfnamefont {E.}~\bibnamefont {Hoel}},\
  }\bibfield  {title} {\bibinfo {title} {Emergence as the conversion of
  information: a unifying theory},\ }\href
  {https://doi.org/10.1098/rsta.2021.0150} {\bibfield  {journal} {\bibinfo
  {journal} {Philosophical Transactions of the Royal Society A: Mathematical,
  Physical and Engineering Sciences}\ }\textbf {\bibinfo {volume} {380}},\
  \bibinfo {pages} {20210150} (\bibinfo {year} {2022})},\ \bibinfo {note}
  {publisher: Royal Society}\BibitemShut {NoStop}%
\bibitem [{\citenamefont {Schick-Poland}\ \emph {et~al.}(2021)\citenamefont
  {Schick-Poland}, \citenamefont {Makkeh}, \citenamefont {Gutknecht},
  \citenamefont {Wollstadt}, \citenamefont {Sturm},\ and\ \citenamefont
  {Wibral}}]{schick-poland_partial_2021}%
  \BibitemOpen
  \bibfield  {author} {\bibinfo {author} {\bibfnamefont {K.}~\bibnamefont
  {Schick-Poland}}, \bibinfo {author} {\bibfnamefont {A.}~\bibnamefont
  {Makkeh}}, \bibinfo {author} {\bibfnamefont {A.~J.}\ \bibnamefont
  {Gutknecht}}, \bibinfo {author} {\bibfnamefont {P.}~\bibnamefont
  {Wollstadt}}, \bibinfo {author} {\bibfnamefont {A.}~\bibnamefont {Sturm}},\
  and\ \bibinfo {author} {\bibfnamefont {M.}~\bibnamefont {Wibral}},\
  }\bibfield  {title} {\bibinfo {title} {A partial information decomposition
  for discrete and continuous variables},\ }\href
  {http://arxiv.org/abs/2106.12393} {\bibfield  {journal} {\bibinfo  {journal}
  {arXiv:2106.12393 [cs, math]}\ } (\bibinfo {year} {2021})},\ \bibinfo {note}
  {arXiv: 2106.12393}\BibitemShut {NoStop}%
\bibitem [{\citenamefont {Varley}\ \emph {et~al.}(2020)\citenamefont {Varley},
  \citenamefont {Sporns}, \citenamefont {Puce},\ and\ \citenamefont
  {Beggs}}]{varley_differential_2020}%
  \BibitemOpen
  \bibfield  {author} {\bibinfo {author} {\bibfnamefont {T.}~\bibnamefont
  {Varley}}, \bibinfo {author} {\bibfnamefont {O.}~\bibnamefont {Sporns}},
  \bibinfo {author} {\bibfnamefont {A.}~\bibnamefont {Puce}},\ and\ \bibinfo
  {author} {\bibfnamefont {J.}~\bibnamefont {Beggs}},\ }\bibfield  {title}
  {\bibinfo {title} {Differential effects of propofol and ketamine on critical
  brain dynamics},\ }\href {https://doi.org/10.1371/journal.pcbi.1008418}
  {\bibfield  {journal} {\bibinfo  {journal} {PLOS Computational Biology}\
  }\textbf {\bibinfo {volume} {16}},\ \bibinfo {pages} {e1008418} (\bibinfo
  {year} {2020})},\ \bibinfo {note} {number: 12 Publisher: Public Library of
  Science}\BibitemShut {NoStop}%
\bibitem [{\citenamefont {Newman}\ \emph {et~al.}(2022)\citenamefont {Newman},
  \citenamefont {Varley}, \citenamefont {Parakkattu}, \citenamefont
  {Sherrill},\ and\ \citenamefont {Beggs}}]{newman_revealing_2022}%
  \BibitemOpen
  \bibfield  {author} {\bibinfo {author} {\bibfnamefont {E.~L.}\ \bibnamefont
  {Newman}}, \bibinfo {author} {\bibfnamefont {T.~F.}\ \bibnamefont {Varley}},
  \bibinfo {author} {\bibfnamefont {V.~K.}\ \bibnamefont {Parakkattu}},
  \bibinfo {author} {\bibfnamefont {S.~P.}\ \bibnamefont {Sherrill}},\ and\
  \bibinfo {author} {\bibfnamefont {J.~M.}\ \bibnamefont {Beggs}},\ }\bibfield
  {title} {\bibinfo {title} {Revealing the {Dynamics} of {Neural} {Information}
  {Processing} with {Multivariate} {Information} {Decomposition}},\ }\href
  {https://doi.org/10.3390/e24070930} {\bibfield  {journal} {\bibinfo
  {journal} {Entropy}\ }\textbf {\bibinfo {volume} {24}},\ \bibinfo {pages}
  {930} (\bibinfo {year} {2022})},\ \bibinfo {note} {number: 7 Publisher:
  Multidisciplinary Digital Publishing Institute}\BibitemShut {NoStop}%
\bibitem [{\citenamefont {Glasser}\ \emph {et~al.}(2013)\citenamefont
  {Glasser}, \citenamefont {Sotiropoulos}, \citenamefont {Wilson},
  \citenamefont {Coalson}, \citenamefont {Fischl}, \citenamefont {Andersson},
  \citenamefont {Xu}, \citenamefont {Jbabdi}, \citenamefont {Webster},
  \citenamefont {Polimeni}, \citenamefont {Van~Essen}, \citenamefont
  {Jenkinson},\ and\ \citenamefont {{WU-Minn HCP
  Consortium}}}]{glasser_minimal_2013}%
  \BibitemOpen
  \bibfield  {author} {\bibinfo {author} {\bibfnamefont {M.~F.}\ \bibnamefont
  {Glasser}}, \bibinfo {author} {\bibfnamefont {S.~N.}\ \bibnamefont
  {Sotiropoulos}}, \bibinfo {author} {\bibfnamefont {J.~A.}\ \bibnamefont
  {Wilson}}, \bibinfo {author} {\bibfnamefont {T.~S.}\ \bibnamefont {Coalson}},
  \bibinfo {author} {\bibfnamefont {B.}~\bibnamefont {Fischl}}, \bibinfo
  {author} {\bibfnamefont {J.~L.}\ \bibnamefont {Andersson}}, \bibinfo {author}
  {\bibfnamefont {J.}~\bibnamefont {Xu}}, \bibinfo {author} {\bibfnamefont
  {S.}~\bibnamefont {Jbabdi}}, \bibinfo {author} {\bibfnamefont
  {M.}~\bibnamefont {Webster}}, \bibinfo {author} {\bibfnamefont {J.~R.}\
  \bibnamefont {Polimeni}}, \bibinfo {author} {\bibfnamefont {D.~C.}\
  \bibnamefont {Van~Essen}}, \bibinfo {author} {\bibfnamefont {M.}~\bibnamefont
  {Jenkinson}},\ and\ \bibinfo {author} {\bibnamefont {{WU-Minn HCP
  Consortium}}},\ }\bibfield  {title} {\bibinfo {title} {The minimal
  preprocessing pipelines for the {Human} {Connectome} {Project}},\ }\href
  {https://doi.org/10.1016/j.neuroimage.2013.04.127} {\bibfield  {journal}
  {\bibinfo  {journal} {NeuroImage}\ }\textbf {\bibinfo {volume} {80}},\
  \bibinfo {pages} {105} (\bibinfo {year} {2013})}\BibitemShut {NoStop}%
\bibitem [{\citenamefont {Schaefer}\ \emph {et~al.}(2018)\citenamefont
  {Schaefer}, \citenamefont {Kong}, \citenamefont {Gordon}, \citenamefont
  {Laumann}, \citenamefont {Zuo}, \citenamefont {Holmes}, \citenamefont
  {Eickhoff},\ and\ \citenamefont {Yeo}}]{schaefer_local-global_2018}%
  \BibitemOpen
  \bibfield  {author} {\bibinfo {author} {\bibfnamefont {A.}~\bibnamefont
  {Schaefer}}, \bibinfo {author} {\bibfnamefont {R.}~\bibnamefont {Kong}},
  \bibinfo {author} {\bibfnamefont {E.~M.}\ \bibnamefont {Gordon}}, \bibinfo
  {author} {\bibfnamefont {T.~O.}\ \bibnamefont {Laumann}}, \bibinfo {author}
  {\bibfnamefont {X.-N.}\ \bibnamefont {Zuo}}, \bibinfo {author} {\bibfnamefont
  {A.~J.}\ \bibnamefont {Holmes}}, \bibinfo {author} {\bibfnamefont {S.~B.}\
  \bibnamefont {Eickhoff}},\ and\ \bibinfo {author} {\bibfnamefont {B.~T.~T.}\
  \bibnamefont {Yeo}},\ }\bibfield  {title} {\bibinfo {title} {Local-{Global}
  {Parcellation} of the {Human} {Cerebral} {Cortex} from {Intrinsic}
  {Functional} {Connectivity} {MRI}},\ }\href
  {https://doi.org/10.1093/cercor/bhx179} {\bibfield  {journal} {\bibinfo
  {journal} {Cerebral Cortex (New York, N.Y.: 1991)}\ }\textbf {\bibinfo
  {volume} {28}},\ \bibinfo {pages} {3095} (\bibinfo {year}
  {2018})}\BibitemShut {NoStop}%
\bibitem [{\citenamefont {Parkes}\ \emph {et~al.}(2018)\citenamefont {Parkes},
  \citenamefont {Fulcher}, \citenamefont {Yücel},\ and\ \citenamefont
  {Fornito}}]{parkes_evaluation_2018}%
  \BibitemOpen
  \bibfield  {author} {\bibinfo {author} {\bibfnamefont {L.}~\bibnamefont
  {Parkes}}, \bibinfo {author} {\bibfnamefont {B.}~\bibnamefont {Fulcher}},
  \bibinfo {author} {\bibfnamefont {M.}~\bibnamefont {Yücel}},\ and\ \bibinfo
  {author} {\bibfnamefont {A.}~\bibnamefont {Fornito}},\ }\bibfield  {title}
  {\bibinfo {title} {An evaluation of the efficacy, reliability, and
  sensitivity of motion correction strategies for resting-state functional
  {MRI}},\ }\href {https://doi.org/10.1016/j.neuroimage.2017.12.073} {\bibfield
   {journal} {\bibinfo  {journal} {NeuroImage}\ }\textbf {\bibinfo {volume}
  {171}},\ \bibinfo {pages} {415} (\bibinfo {year} {2018})}\BibitemShut
  {NoStop}%
\bibitem [{\citenamefont {Tagliazucchi}\ \emph {et~al.}(2012)\citenamefont
  {Tagliazucchi}, \citenamefont {Balenzuela}, \citenamefont {Fraiman},\ and\
  \citenamefont {Chialvo}}]{tagliazucchi_criticality_2012}%
  \BibitemOpen
  \bibfield  {author} {\bibinfo {author} {\bibfnamefont {E.}~\bibnamefont
  {Tagliazucchi}}, \bibinfo {author} {\bibfnamefont {P.}~\bibnamefont
  {Balenzuela}}, \bibinfo {author} {\bibfnamefont {D.}~\bibnamefont
  {Fraiman}},\ and\ \bibinfo {author} {\bibfnamefont {D.~R.}\ \bibnamefont
  {Chialvo}},\ }\bibfield  {title} {{\selectlanguage {English}\bibinfo {title}
  {Criticality in {Large}-{Scale} {Brain} {fMRI} {Dynamics} {Unveiled} by a
  {Novel} {Point} {Process} {Analysis}}},\ }\bibfield  {journal} {\bibinfo
  {journal} {Frontiers in Physiology}\ }\textbf {\bibinfo {volume} {3}},\ \href
  {https://doi.org/10.3389/fphys.2012.00015} {10.3389/fphys.2012.00015}
  (\bibinfo {year} {2012})\BibitemShut {NoStop}%
\bibitem [{\citenamefont {Lancichinetti}\ and\ \citenamefont
  {Fortunato}(2012)}]{lancichinetti_consensus_2012}%
  \BibitemOpen
  \bibfield  {author} {\bibinfo {author} {\bibfnamefont {A.}~\bibnamefont
  {Lancichinetti}}\ and\ \bibinfo {author} {\bibfnamefont {S.}~\bibnamefont
  {Fortunato}},\ }\bibfield  {title} {\bibinfo {title} {Consensus clustering in
  complex networks},\ }\href {https://doi.org/10.1038/srep00336} {\bibfield
  {journal} {\bibinfo  {journal} {Scientific Reports}\ }\textbf {\bibinfo
  {volume} {2}},\ \bibinfo {pages} {336} (\bibinfo {year} {2012})},\ \bibinfo
  {note} {number: 1 Publisher: Nature Publishing Group}\BibitemShut {NoStop}%
\bibitem [{\citenamefont {Newman}(2006)}]{newman_modularity_2006}%
  \BibitemOpen
  \bibfield  {author} {\bibinfo {author} {\bibfnamefont {M.~E.~J.}\
  \bibnamefont {Newman}},\ }\bibfield  {title} {\bibinfo {title} {Modularity
  and community structure in networks},\ }\href
  {https://doi.org/10.1073/pnas.0601602103} {\bibfield  {journal} {\bibinfo
  {journal} {Proceedings of the National Academy of Sciences of the United
  States of America}\ }\textbf {\bibinfo {volume} {103}},\ \bibinfo {pages}
  {8577} (\bibinfo {year} {2006})},\ \bibinfo {note} {number: 23}\BibitemShut
  {NoStop}%
\bibitem [{\citenamefont {Lizier}(2014)}]{wibral_measuring_2014}%
  \BibitemOpen
  \bibfield  {author} {\bibinfo {author} {\bibfnamefont {J.~T.}\ \bibnamefont
  {Lizier}},\ }\bibfield  {title} {\bibinfo {title} {Measuring the {Dynamics}
  of {Information} {Processing} on a {Local} {Scale} in {Time} and {Space}},\
  }in\ \href {https://doi.org/10.1007/978-3-642-54474-3_7} {\emph {\bibinfo
  {booktitle} {Directed {Information} {Measures} in {Neuroscience}}}},\
  \bibinfo {editor} {edited by\ \bibinfo {editor} {\bibfnamefont
  {M.}~\bibnamefont {Wibral}}, \bibinfo {editor} {\bibfnamefont
  {R.}~\bibnamefont {Vicente}},\ and\ \bibinfo {editor} {\bibfnamefont {J.~T.}\
  \bibnamefont {Lizier}}}\ (\bibinfo  {publisher} {Springer Berlin
  Heidelberg},\ \bibinfo {address} {Berlin, Heidelberg},\ \bibinfo {year}
  {2014})\ pp.\ \bibinfo {pages} {161--193},\ \bibinfo {note} {series Title:
  Understanding Complex Systems}\BibitemShut {NoStop}%
\bibitem [{\citenamefont {Lizier}\ \emph {et~al.}(2014)\citenamefont {Lizier},
  \citenamefont {Prokopenko},\ and\ \citenamefont
  {Zomaya}}]{lizier_framework_2014}%
  \BibitemOpen
  \bibfield  {author} {\bibinfo {author} {\bibfnamefont {J.~T.}\ \bibnamefont
  {Lizier}}, \bibinfo {author} {\bibfnamefont {M.}~\bibnamefont {Prokopenko}},\
  and\ \bibinfo {author} {\bibfnamefont {A.~Y.}\ \bibnamefont {Zomaya}},\
  }\bibfield  {title} {\bibinfo {title} {A {Framework} for the {Local}
  {Information} {Dynamics} of {Distributed} {Computation} in {Complex}
  {Systems}},\ }in\ \href {https://doi.org/10.1007/978-3-642-53734-9_5} {\emph
  {\bibinfo {booktitle} {Guided {Self}-{Organization}: {Inception}}}},\
  \bibinfo {series and number} {Emergence, {Complexity} and {Computation}},\
  \bibinfo {editor} {edited by\ \bibinfo {editor} {\bibfnamefont
  {M.}~\bibnamefont {Prokopenko}}}\ (\bibinfo  {publisher} {Springer},\
  \bibinfo {address} {Berlin, Heidelberg},\ \bibinfo {year} {2014})\ pp.\
  \bibinfo {pages} {115--158}\BibitemShut {NoStop}%
\bibitem [{\citenamefont {Kolchinsky}(2022)}]{kolchinsky_novel_2022}%
  \BibitemOpen
  \bibfield  {author} {\bibinfo {author} {\bibfnamefont {A.}~\bibnamefont
  {Kolchinsky}},\ }\bibfield  {title} {\bibinfo {title} {A {Novel} {Approach}
  to the {Partial} {Information} {Decomposition}},\ }\href
  {https://doi.org/10.3390/e24030403} {\bibfield  {journal} {\bibinfo
  {journal} {Entropy}\ }\textbf {\bibinfo {volume} {24}},\ \bibinfo {pages}
  {403} (\bibinfo {year} {2022})},\ \bibinfo {note} {number: 3 Publisher:
  Multidisciplinary Digital Publishing Institute}\BibitemShut {NoStop}%
\bibitem [{\citenamefont {MacKay}(2003)}]{mackay_information_2003}%
  \BibitemOpen
  \bibfield  {author} {\bibinfo {author} {\bibfnamefont {D.~J.~C.}\
  \bibnamefont {MacKay}},\ }\href@noop {} {\emph {\bibinfo {title} {Information
  {Theory}, {Inference} and {Learning} {Algorithms}}}}\ (\bibinfo  {publisher}
  {Cambridge University Press},\ \bibinfo {year} {2003})\ \bibinfo {note}
  {google-Books-ID: AKuMj4PN\_EMC}\BibitemShut {NoStop}%
\end{thebibliography}%

\beginsupplement

\section*{SI 1. Mathematical Properties of $\mathcal{H}_{sx}$}

\subsection*{Partial Entropy Decomposition \& Partial Information Decomposition}

The redundant \textit{entropy} function $h_{sx}$ is closely related to the redundant \textit{information} function $i_{sx}$ proposed by Makkeh et al., \cite{makkeh_introducing_2021}. Our function $h_{sx}$ is defined:

\begin{equation}
    h_{sx}(\textbf{a}_1,\ldots,\textbf{a}_{k}) = \log\frac{1}{\prob(\textbf{a}_1\cup\ldots\cup\textbf{a}_k)}
\end{equation}

This measure is equivalent to the \textit{informative} component of the measure $i_{sx}$ proposed by Makkeh et al., \cite{makkeh_introducing_2021} in the context of single-target partial information decomposition. The local redundant information function $i_{sx}$ is defined:

\begin{align}
	i_{sx}(\textbf{a}^1,&\ldots,\textbf{a}^k;y) := \\ 
	& \log_2\frac{\prob(y) - \prob(y \cap (\bar{\textbf{a}}^1\cap\ldots\cap\bar{\textbf{a}}^k))}{1 - \prob(\bar{\textbf{a}}^1\cap\ldots\cap\bar{\textbf{a}}^k)} - \log_2\prob(y)
\end{align}

Which can be further decomposed into \textit{informative} and \textit{misinformative} components \cite{finn_probability_2018}:

\begin{align}
	i_{sx}^+(\textbf{a}_1,\ldots,\textbf{a}_k;y) &:= \log_2\frac{1}{\prob(\textbf{a}_1 \cup\ldots\cup\textbf{a}_k)} \\
	i_{sx}^-(\textbf{a}_1,\ldots,\textbf{a}_k;y) &:= \log_2\frac{\prob(y)}{\prob(y \cap (\textbf{a}_1 \cup\ldots\cup\textbf{a}_k))} \\
	i_{sx}(\textbf{a}_1,\ldots,\textbf{a}_k;y) &= i_{sx}^+(\textbf{a}_1,\ldots,\textbf{a}_k;y) - i_{sx}^-(\textbf{a}_1,\ldots,\textbf{a}_k;y)
\end{align}

Where it is clear that $h_{sx}(\cdot) = i_{sx}^+(\cdot;y)$, with the sole difference that $i_{sx}^+(\cdot;y)$ is implicitly defined with respect to some target variable $y$ (although $y$ has no actual impact on the value). Below, we show that, if the target $y$ is set to the joint state of the whole (\textbf{x}), then the partial entropy decomposition of $h(\textbf{x})$ with $h_{sx}$ as the shard entropy function becomes equivalent to the partial information decomposition $i(x_1,\ldots,x_N ; \textbf{x})$ with $i_{sx}$ as the redundant entropy function. The notion that the PED is equivalent to doing the PID of the information all the ``parts" disclose about the ``whole" was mentioned parenthetically in \cite{makkeh_introducing_2021}, although the finding that the informative component is all that is required is novel.

Given the equivalence between $h_{sx}(\cdot)$ and $i_{sx}^+(\cdot;y)$, it suffices to show that $i_{sx}^-(x_1,\ldots,x_N;\textbf{x})=0$ bit in all cases. When $y=\textbf{x}$, we can re-write the function as:

\begin{align}
    i_{sx}^-(x_1,&\ldots,x_N;\textbf{x}) = \\
    &\log_2\frac{\prob(x_1\cap\ldots\cap x_N)}{\prob((x_1 \cap\ldots\cap x_N) \cap (x_1 \cup\ldots,\cup x_N))} \nonumber
\end{align}

the union of $x^1\cup\ldots\cup x^k$ is clearly a superset of $x^1\cap\ldots\cap x^k$, so

\begin{equation}
    i_{sx}^- = \log_2\frac{\prob(x_1 \cap\ldots\cap x_N)}{\prob(x_1 \cap\ldots\cap x_N)}
\end{equation}

Which is clearly $\log_2(1)=0$ bit $\Box$

We can understand the partial entropy decomposition using $h_{sx}$ as being equivalent to the decomposition of $i(x_1,\ldots,x_N;\textbf{x})$. Intuitively, this is consistent with the identity for discrete variables that $\mi(\textbf{X},\textbf{X})=\entropy(\textbf{X})$.

This has curious implications: for example, while $x_1$ may misinform about $x_2$, neither variable can misinform about their contribution to the state of the whole \textbf{x}.

\subsection*{Example: Logical Exclusive-OR (XOR) Gate}

\begin{table}[!ht]
	\centering
	\begin{tabular}{cccccccccccccccc}
		\multicolumn{7}{c}{\textbf{XOR}}                     &  &  & \multicolumn{7}{c}{\textbf{AND}}                     \\ \cline{1-7} \cline{10-16} 
		$\prob$ &  & $X_1$ & $\oplus$ & $X_2$ & = & $T$ &  &  & $\prob$   &  & $X_1$ & $\wedge$ & $X_2$ & = & $T$ \\ \cline{1-7} \cline{10-16} 
		1/4 &  & 0     &          & 0     &   & 0   &  &  & 1/4 &  & 0     &          & 0     &   & 0   \\
		1/4 &  & 0     &          & 1     &   & 1   &  &  & 1/4 &  & 0     &          & 1     &   & 0   \\
		1/4 &  & 1     &          & 0     &   & 1   &  &  & 1/4 &  & 1     &          & 0     &   & 0   \\
		1/4 &  & 1     &          & 1     &   & 0   &  &  & 1/4 &  & 1     &          & 1     &   & 1   \\ \cline{1-7} \cline{10-16} 
	\end{tabular}
	\caption{\textbf{Logical XOR and AND gates.}}
	\label{tab:xor_and}
\end{table}

To demonstrate how partial entropy decomposition can be used to untangle higher-order interactions, consider the logical exclusive-OR (XOR) gate (for the lookup table, see Table \ref{tab:xor_and}). The XOR gate is an example of a \textit{synergistic} logic gate: the ability to predict the state of the target $T$ depends on having access to both $X_1$ and $X_2$ jointly: the pariwise marginal mutual informations are equal to 0: $\mi(X_1;T) = \mi(X_2;T) = 0$ bit, but the joint mutual information is nonzero: $\mi(X_1, X_2 ; T) = 1$ bit.

We can initially see that the triple-redundancy $\pentropy^{12T}(\{1\}\{2\}\{T\})=0$ bit. This is because any configuration of logical disjunctions does not actually rule out any states: for example, $\prob(X_1=0\cup X_2=0\cup T=0)=1$ as there is no configuration $(1,1,1)$ that can be excluded. Other results can be unintuitive. For example, most of the partial entropy is shared between the three bivariate relationships $\pentropy^{12T}(\{1\}\{2\})$, $\pentropy^{12T}(\{1\}\{T\})$, and $\pentropy^{12T}(\{2\}\{T\})$. How is this consistent with the fact that the mutual information between any pair of variables is zero? The bivariate redundancy can be non-zero in this case because, on average, knowing the local state of $x_1 \lor x_2$ reduces our uncertainty about the joint state of $\{x_1,x_2,t\}$. For example, suppose we learn that $x_1 = 1 \lor x_2 = 1$. This \textit{excludes} the joint configuration $\{x_1 = 0, x_2 = 0, t=0\}$. This exclusion of the associated probability mass is recognized by $h_{sx}(\cdot)$ as informative, in that it reduces our uncertainty about the joint-state of the whole, despite the fact that, on average, $X_1$ and $X_2$ disclose no information about $T$. There is no redundant information common to $X_1$, $X_2$ and $T$, however, and there a number of higher-order dependencies, such as $\pentropy^{12T}(\{1\}\{2,T\})$ and $\pentropy^{12T}(\{1,2\}\{1,T\}\{2,T\})$. 

\begin{table}[!ht]
	\centering
	\begin{tabular}{@{}llccccc@{}}
		\toprule
		Atom                  & $\pentropy^{12T} || $ &  XOR & & AND & & MaxEnt \\ \midrule
		$\{1\}\{2\}\{T\}$       &  & 0.0    & & 0.208   & & 0.193\\
		$\{1\}\{2\}$            &  & 0.415  & & 0.208   & & 0.222\\
		$\{1\}\{T\}$            &  & 0.415  & & 0.25    & & 0.222\\
		$\{2\}\{T\}$            &  & 0.415  & & 0.25    & & 0.222\\
		$\{1\}\{2,T\}$          &  & 0.17   & & 0.04    & & 0.041\\
		$\{2\}\{1,T\}$          &  & 0.17   & & 0.04    & & 0.041\\
		$\{T\}\{1,2\}$          &  & 0.17   & & 0.104   & & 0.041\\
		$\{1\}$                 &  & 0.0    & & 0.292   & & 0.322\\
		$\{2\}$                 &  & 0.0    & & 0.292   & & 0.322\\
		$\{T\}$                 &  & 0.0    & & 0.0     & & 0.322\\
		$\{1,2\}\{1,T\}\{2,T\}$ &  & 0.245  & & 0.0     & & 0.018\\
		$\{1,2\}\{1,T\}$        &  & 0.0    & & 0.104   & & 0.093\\
		$\{1,2\}\{2,T\}$        &  & 0.0    & & 0.104   & & 0.093\\
		$\{1,T\}\{2,T\}$        &  & 0.0    & & 0.0     & & 0.093\\
		$\{1,2\}$               &  & 0.0    & & 0.104   & & 0.17\\
		$\{1,T\}$               &  & 0.0    & & 0.0     & & 0.17\\
		$\{2,T\}$               &  & 0.0    & & 0.0     & & 0.17\\
		$\{1,2,T\}$             &  & 0.0    & & 0.0     & & 0.245\\ 
		\bottomrule
	\end{tabular}
	\caption{\textbf{The Partial Entropy Decomposition for the XOR, AND, and Maximum Entropy Gates.} }
	\label{tab:ped_xor}
\end{table}

\subsection*{S3: Independent Variables}
\label{sec:iid_vars}

\begin{figure}[!ht]
    \centering
    \includegraphics[scale=0.75]{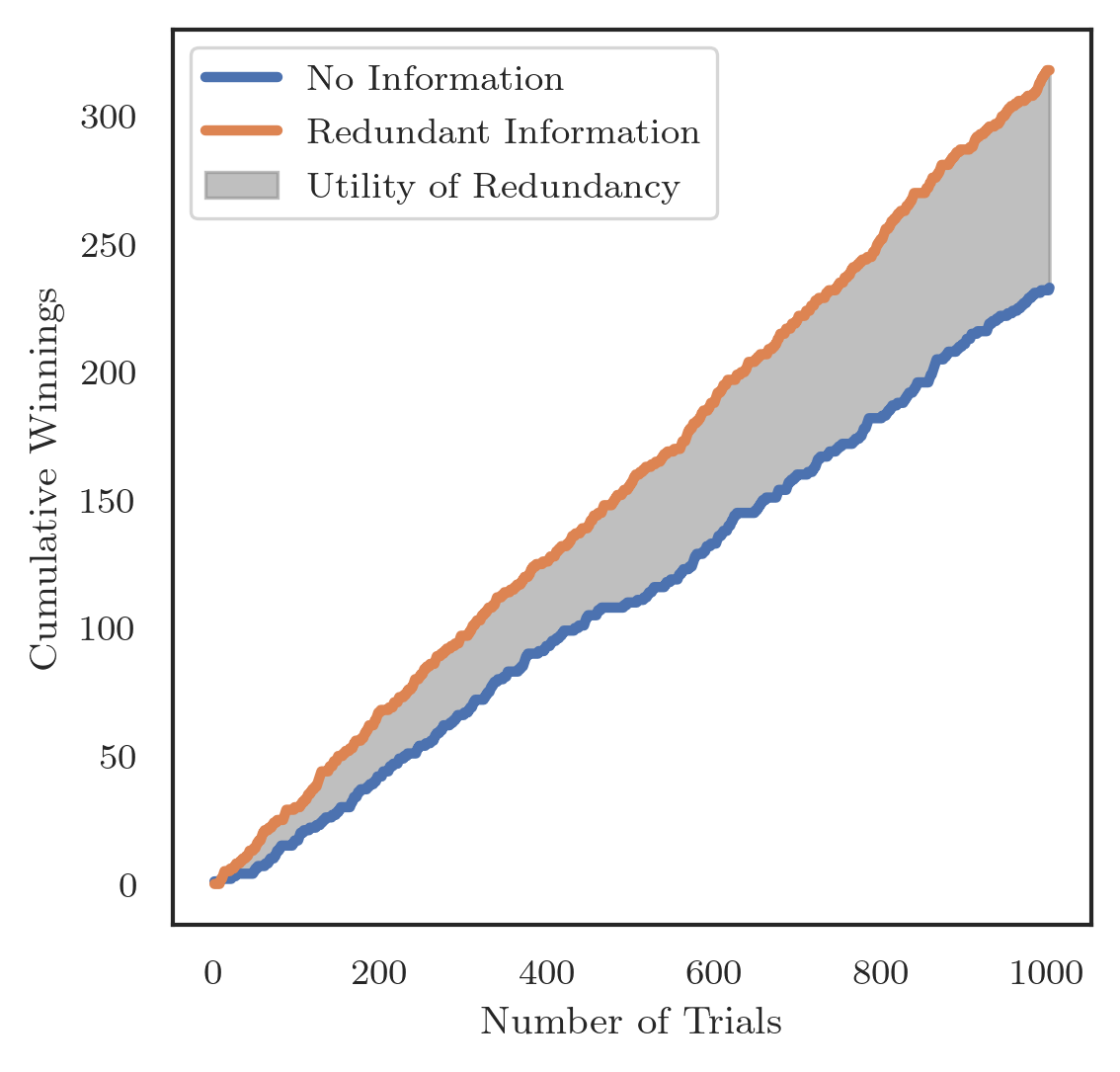}
    \caption{\textbf{Utility of redundant information.} Suppose an agent plays a gambling game, where two independent, binary variables are set at random (so all outcomes $\prob(x_1,x_2)=1/4$ for all configurations). If the agent guesses the correct variable, they win \$1 and if they guess wrong, they win nothing. Clearly, the expected value of each trial is \$0.25 (blue curve). However, if another agent learns that $X_1=x1 \lor X_2=x2$, then they can do better at the game, with an expected value of each trial of \$0.33. The difference between the two cumulative distributions of 1000 trials is the extra ``value" that can be extracted from the redundant information. This shows that, while counter-intuitive, the fact that $\pentropy^{12}(\{1\}\{2\}) > 0$ even if $X_1\bot X_2$ is interpretable in practical contexts.}
    \label{si:fig:utility}
\end{figure}

One unusual property of $h_{sx}$, as demonstrated by the logical-XOR results is that independent variables can still share entropy. This is a recognized feature of multiple measures of redundant information/entropy and is generally considered to be an issue to be excised \cite{kolchinsky_novel_2022} (some have gone so far as the suggest an axiom that such a property must be disallowed from the outset \cite{ince_measuring_2017}). While we understand that shared entropy for random variables may seem initially counter intuitive, it can be readily understood when considering the problem of inference. Let us return to our two element example (Table \ref{tab:biv}), and this time specify that $X_1\bot X_2$. We know then that $\mi(X_1;X_2)=0$ bit, however, $h_{sx}(\{X_1\}\{X_2\}) \approx 0.415$ bit. Why? The answer is that, while the two variables are independent, in all cases learning either $X_1=x_1 \lor X_2=x_2$ is sufficient to exclude a single possible state: the case where $X_1 = \neg x_1 \land X_2 = \neg x_2$. If we were to formalize this in terms of a gambling problem, we would find that, despite the independence of both variables, a player is, in fact, more likely to win with a correct guess after learning $X_1 \lor X_2$. See Figure \ref{si:fig:utility}.

Furthermore, we can see that, while $h_{sx}$ will be greater than zero for small, maximum entropy systems, as the system gets larger, the redundancy will logarithmically trend towards zero. The proof for binary systems is straightforward. For a discrete, maximum entropy system with $k$ elements, learning the state of $X_1=x_1 \lor \ldots \lor X_k = x_k$ will always exclude a single state: the state where $X_1 \not= x_1 \land \ldots \land X_k \not= x_k$. This single state $\textbf{x}^*$ will have $\prob(\textbf{x}^*) = 1/k$ (as all states have the same probability by the maximum entropy constraint). The union of all surviving configurations will be $1-\prob(\textbf{x}^*)$. Since $\lim_{k\to\infty}1/k = 0$, then the union probability will $\to 1$ and consequently $h_{sx}\to0$ bit. This suggests that, for very large, idealized systems (such as an ideal gas), the redundancy \textit{does} go to 0 bit for maximum entropy systems. How other values (such as the redundant and synergistic structure) behave remains an area of further study, although we conjecture that, as $k\to\infty$, redundancies and synergies will vanish faster than unique terms. 

\section*{S3. Basic Information Theory Review}

Here we will provide a basic overview of information theory for unfamiliar readers. For a more comprehensive treatment of the subject, see the textbooks by Cover \& Thomas \cite{cover_elements_2012} and/or MacKay \cite{mackay_information_2003}.

The basic object of study in information theory is the \textit{entropy}, which quantifies the total uncertainty that we, as observers, have about the state of some variable $X$. For the purposes of this paper, we will assume that $X$ is discrete, with a finite number of possible states that can be pulled from the support set $\mathcal{X}$. For every particular state $x\in\mathcal{X}$, there is an associated probability $P(x)$. The entropy of $X$ is given by:

\begin{equation}
    H(X) = -\sum_{x\in\mathcal{X}}P(x)\log P(x)
\end{equation}

For multiple variables, we can define the joint entropy as:

\begin{equation}
    H(X_1,X_2) = -\sum_{\substack{x_1\in\mathcal{X}_1\\ x_2\in\mathcal{X}_2}}P(x_1,x_2)\log P(x_1,x_2)
\end{equation}

We can also define the conditional entropy as the uncertainty about $X_1$ \textit{left over} after accounting for the knowledge that $X_2=x_2$:

\begin{equation}
    H(X_1|X_2) = -\sum_{\substack{x_1\in\mathcal{X}_1\\x_2\in\mathcal{X}_2}}P(x_1,x_2)\log P(x_1|x_2)
\end{equation}

From these basic components, we can define the \textit{mutual information} as the difference between our initial uncertainty about the state of $X_1$ the the remaining uncertainty about $X_1$ that is not resolved by learning the state of $X_2$:

\begin{equation}
    I(X_1 ; X_2) = H(X_1) - H(X_1 | X_2)
\end{equation}

The mutual information is symmetric in it's arguments: $I(X_1 ; X_2) = I(X_2 ; X_1)$. If we have multiple $X$s disclosing information about a single target $T$, the joint mutual information has the same form:

\begin{equation}
    I(X_1, X_2 ; T) = H(T)
    - H(T|X_1, X_2)
\end{equation}

The mutual information can also be written in terms of probabilities:

\begin{equation}
    I(X_1 ; X_2) = \sum_{\substack{x_1\in\mathcal{X}_1 \\ x_2\in\mathcal{X}_2}}P(x_1, x_2)\log \frac{P(x_1| x_2)}{P(x_1)}
\end{equation}

\subsubsection*{Local Information Theory}

Both the entropy and the mutual information can be understood as ``expected values" over some (potentially multivariate) distribution):

\begin{equation}
    H(X) = \mathbb{E}[-\log P(x)]
\end{equation}

The term $-\log P(x)$ is known as the \textit{local entropy} or the \textit{Shannon information content} and it quantifies how ``surprised" we, as observers are to see that $X=x$. It is typically denoted as $h(x)$.

\begin{equation}
    I(X_1;X_2) = \mathbb{E}\bigg[\log_2\frac{P(x_1|x_2)}{P(x_1)}\bigg]
\end{equation}

The term $\frac{P(x_1|x_2)}{P(x_1)}$ is known as the \textit{local mutual information} and it quantifies the divergence between the ``prior" probability $X_1=x_1$ and the posterior probability $X_1=x_1$ \textit{after accounting for the fact that $X_2=x_2$}. It is typically denoted as $i(x_1;x_2)$. Unlike the expected mutual information, which is strictly non-negative, the local mutual information can be either greater than, or less than, zero. If $P(x_1 | x_2) < P(x_1)$, then $i(x_1;x_2)>0$, and if $P(x_1|x_2)<P(x_1)$, then $i(x_1;x_2)<0$. In the latter case, we say that $x_1$ \textit{misinforms} on the state of $x_2$.

\end{document}